\documentclass[twocolumn]{aastex63} % preprint
\newcommand{\about}{\mbox{$\sim$}}               % ~ (about) 
\newcommand{\Av}{\mbox{$A_{\rm V}$}}                  % Av
\newcommand{\frest}{\mbox{$f_{\rm rest}$}}      % f_rest
\newcommand{\fobs}{\mbox{$f_{\rm obs}$}}      % f_obs
\newcommand{\Lbol}{\mbox{$L_{\rm bol}$}}                    % Lbol
\newcommand{\Lir}{\mbox{$L_{\rm 8-1000\; \mu m}$}}  % Lir
\newcommand{\Lsun}{\mbox{$L_\odot$}}            % Lsun
\newcommand{\NH}{\mbox{$N_{\rm H}$}}             % N_H
\newcommand{\NHH}{\mbox{$N_{\rm H_2}$}}     % N_H2
\newcommand{\NHHprime}{\mbox{$N_{\rm H_2}^\prime$}}     % N_H2
\newcommand{\nHH}{\mbox{$n_{\rm H_2}$}}       % n_H2
\newcommand{\Tb}{\mbox{$T_{\rm b}$}}               % Tb
\newcommand{\uv}{\mbox{$u$--$v$}}                     % u-v
\newcommand{\amin}{\mbox{$'$}}                          % '     (arcmin)
\newcommand{\asec}{\mbox{$''$}}                          % "    (arcsec)
\newcommand{\hr}{\mbox{$^{\rm h}$}}                   % ^h
\newcommand{\mn}{\mbox{$^{\rm m}$}}                % ^m
\newcommand{\s}{\mbox{$^{\rm s}$}}                     % ^s
\newcommand{\perbeam}{\mbox{beam$^{-1}$}}                         % beam^{-1}
\newcommand{\persquarecm}{\mbox{cm$^{-2}$}}                      % cm^{-2}
\newcommand{\squarecm}{\mbox{cm$^{2}$}}                                 % cm^{2}
\newcommand{\pergram}{\mbox{g$^{-1}$}}                                % g^{-1}
\newcommand{\kms}{\mbox{km s$^{-1}$}}                                    % km/s
\newcommand{\persquarepc}{\mbox{pc$^{-2}$}}                         % pc^{-2}
\newcommand{\persquare}[1]{\mbox{{#1}$^{-2}$}} 
\newcommand{\percubic}[1]{\mbox{{#1}$^{-3}$}} 

\newcommand{\plus}{\mbox{$+$}}    % +
\newcommand{\m}{\mbox{$-$}}
\newcommand{\thirteenCO}{\mbox{$^{13}$CO}}                % 13CO
\newcommand{\CeighteenO}{\mbox{C$^{18}$O}}              % C18O
\newcommand{\HH}{\mbox{H$_2$}}                                      % H2 (hydrogen molecule)
\newcommand{\HtwoS}{\mbox{H$_2$S}}                             % H2S
\newcommand{\CtwoH}{\mbox{C$_2$H}}                             % C2H
\newcommand{\HthirteenCN}{\mbox{H$^{13}$CN}}          % H13CN
\newcommand{\HCfifteenN}{\mbox{HC$^{15}$N}}             % HC15N
\newcommand{\HNthirteenC}{\mbox{HN$^{13}$C}}          % HN13C
\newcommand{\HCthreeN}{\mbox{HC$_{3}$N}}                 % HC3N (cyanoacetylene)
\newcommand{\HtwoCO}{\mbox{H$_2$CO}}                       % H2CO
\newcommand{\HCOplus}{\mbox{HCO$^{+}$}}                    % HCO+
\newcommand{\HthirteenCOplus}{\mbox{H$^{13}$CO$^{+}$}}     % H13CO+
\newcommand{\HOCplus}{\mbox{HOC$^{+}$}}                    % HOC+
\newcommand{\NtwoDplus}{\mbox{N$_{2}$D$^{+}$}}       % N2D+
\newcommand{\thirteenCS}{\mbox{$^{13}$CS}}                 % 13CS
\newcommand{\CHthreeOH}{\mbox{CH$_3$OH}}                 % CH3OH (methanol)
\newcommand{\vib}{\mbox{$^{\ast}$}}

\newcommand{\nd}{\nodata}
\newcommand{\tnm}[1]{\tablenotemark{\tiny #1}}
\newcommand{\citest}[1]{\citeauthor*{#1}}
\newcommand{\citesp}[1]{(\citeauthor*{#1})}

\newcommand{\arpE}{Arp~220E}
\newcommand{\arpW}{Arp~220W}

\newcommand{\alphap}{\mbox{$\alpha_\mathrm{p}$}} % alpha_p, plasma alpha
\newcommand{\alphad}{\mbox{$\alpha_\mathrm{d}$}} % alpha_d, dust alpha

\shorttitle{ALMA 1.4--0.4 mm Spectral Scan on NGC 4418 and Arp 220: I. Continuum}
\shortauthors{SAKAMOTO et al.}

\graphicspath{{./}{figures/}}

\begin{document}

\title{Deeply Buried Nuclei in the Infrared-Luminous Galaxies NGC 4418 and Arp 220 \\
I.  ALMA Observations at $\lambda = $1.4--0.4 mm and Continuum Analysis}

\author{Kazushi Sakamoto}
\affiliation{Academia Sinica, Institute of Astronomy and Astrophysics, Taipei, Taiwan}

\author{Eduardo Gonz\'{a}lez-Alfonso}
\affiliation{Universidad de Alcal\'{a}, Departamento de F\'{i}sica y Matem\'{a}ticas, 
Campus Universitario, E-28871 Alcal\'{a} de Henares, Madrid, Spain}

\author{Sergio Mart\'{i}n}
\affiliation{European Southern Observatory, Alonso de C\'{o}rdova 3107, Vitacura Casilla 763 0355, Santiago, Chile}
\affiliation{Joint ALMA Observatory, Alonso de C\'{o}rdova 3107, Vitacura 763 0355, Santiago, Chile}

\author{David J. Wilner}
\affiliation{Harvard-Smithsonian Center for Astrophysics, 60 Garden Street, Cambridge, MA 02138, USA}

\author{Susanne Aalto}
\affiliation{Department of Earth and Space Sciences, Chalmers University of Technology, Onsala Observatory, 439 92 Onsala, Sweden}

\author{Aaron S. Evans}
\affiliation{Department of Astronomy, University of Virginia, P.O. Box 400325, Charlottesville, VA 22904, USA}
\affiliation{National Radio Astronomy Observatory, 520 Edgemont Road, Charlottesville, VA 22903, USA}

\author{Nanase Harada}
\affiliation{Academia Sinica, Institute of Astronomy and Astrophysics, Taipei, Taiwan}
\affiliation{National Astronomical Observatory of Japan, Mitaka, Tokyo, 181-8588, Japan}

\begin{abstract}
We observed with ALMA three deeply buried nuclei in two galaxies, NGC~4418 and Arp~220,
at \about0\farcs2 resolution over a total bandwidth of 67 GHz in \frest\ = 215--697 GHz.
Here we (1) introduce our program,
(2) describe our data reduction method for wide-band, high-resolution imaging spectroscopy,
(3) analyze in visibilities the compact nuclei with line forests,
(4) develop a continuum-based estimation method of dust opacity and gas column density in heavily obscured nuclei,
which uses the BGN (buried galactic nuclei) model and 
is sensitive to $\log(\NHH/\persquarecm) \sim $ 25--26 at $\lambda \sim 1$ mm,
and 
(5) present the continuum data and diagnosis of our targets.
The three continuum nuclei have major-axis FWHM of \about0\farcs1--0\farcs3 (20--140 pc) aligned to their rotating 
nuclear disks of molecular gas.
However, each nucleus is described better with two or three concentric components than with a single Gaussian.
The innermost cores have sizes of 0\farcs05--0\farcs10 (8--40 pc), peak brightness temperatures 
of \about100--500 K at 350 GHz, and more fractional flux at lower frequencies.
The intermediate components correspond to the nuclear disks.  
They have axial ratios of $\approx$0.5 and hence inclinations $\gtrsim60\degr$.
The outermost elements include the bipolar outflow from Arp 220W. 
We estimate 1 mm dust opacity of $\tau_{\rm d,1mm} \approx 2.2$, $1.2$, and $\lesssim 0.4$
respectively for NGC 4418, Arp 220W, and Arp 220E.
The first two correspond to $\log(\NH/\persquarecm) \sim 26$ for conventional dust-opacity laws,
and hence the nuclei are highly Compton thick.
\end{abstract}


%\NewPageAfterKeywords
\begin{flushright}
Accepted for publication in ApJ
\end{flushright}

%%%%%%%%%%%%%%%%%%%%%%%%%%%%%%%%%%%%%%%%%%%%%%%%%%%%%%%%%%%%
\section{Introduction}  
\label{s.introduction}

\begin{figure}[!bht]
\begin{center}
\includegraphics{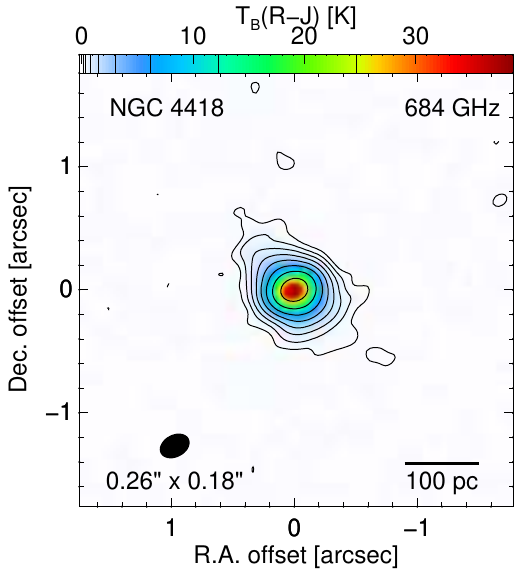} % .eps} 
\includegraphics{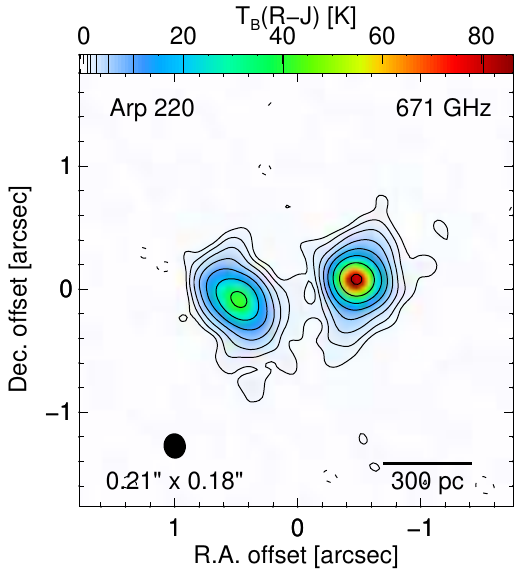} % .eps} 
\end{center}
\caption{ \label{f.B9cont}
Continuum images of NGC 4418 and Arp 220.
Contours are at $\pm 3\sigma \times 2^n$ for $n=0,1,2, \cdots$, 
where $\sigma$ is
 0.064 K (1.2 mJy \perbeam) for NGC 4418 and 
 0.21 K (2.9 mJy \perbeam) for Arp 220. 
Negative contours are dashed.
Peak intensities are 
39 K (0.70 Jy \perbeam) for NGC 4418,
46 K (0.64 Jy \perbeam) for \arpE, and 
86 K (1.19 Jy \perbeam) for \arpW. 
}
\end{figure}

NGC 4418 and Arp 220 have luminous nuclei that are among the most deeply buried in interstellar gas and dust.
The two are {ultra/luminous infrared galaxies} (U/LIRGs, \citealt{SM96})
radiating predominantly in the infrared and having \Lir\ of $10^{11.2}$ \Lsun\ and $10^{12.3}$ \Lsun, respectively \citep{Armus09}.
They stand out even among U/LIRGs for their nuclear obscuration.
They were the first galaxies in which the silicate feature at 9.7 \micron\ was detected as deep absorption 
\citep[at the level of $\Av \sim 10^2$ mag, i.e., $\NH \sim 2\times10^{23}$ \persquarecm;][]{Rieke85,Roche86,Smith89},
and their silicate absorption remains conspicuously large among local U/LIRGs \citep{Spoon07,Stierwalt13}.
Furthermore, data covering longer wavelengths (to sub/millimeter) suggest 
that the column densities to the centers of their nuclei are as large as 
$\gtrsim 10^{25}$ -- $10^{26}$ H \persquarecm\ \citep{GA12, Sakamoto13, Scoville17, GS19, Dwek20};
the mid-IR silicate feature traces only a (relatively) diffuse foreground gas \citep{Roche15}.
The nuclei of NGC 4418 and Arp 220 are 
local prototypes of luminous galactic nuclei ($L_{\rm IR} \gtrsim 10^{11}$ \Lsun)
obscured to the level of $\NH \gtrsim 10^{25}$ \persquarecm.  
As such, they deserve close examination for the study of this type of heavily obscured nuclei,
which are sometimes referred to as CONs (Compact Obscured Nuclei: \citealt{Costagliola13}; \citealt{Falstad21})
or BGNs (Buried Galactic Nuclei; \citealt{GS19}). 
The BGN models in the latter simulated dusty opaque nuclei for their thermal structure and emerging emission
by numerically solving radiative transfer and local energy balance. 
To guide readers unfamiliar with the nuclei of NGC 4418 and Arp 220, Figure~\ref{f.B9cont} shows 
the continuum images of the three compact and bright nuclei from our observations at $\lambda=0.44$ mm.

NGC 4418 ($D = 34$ Mpc; 1\arcsec=165 pc) is an Sa-type galaxy with no evident distortion in the optical, though 
a past interaction with a dwarf companion 30 kpc away is suggested from 
an \ion{H}{1} bridge \citep{Varenius17} and a probable tidal arm \citep{Boettcher20}. 
It has the deepest 9.7 \micron\ silicate absorption in the 180 local luminous infrared galaxies studied by \citet{Stierwalt13}.
The central few-10 pc area radiates most of the infrared luminosity of the entire galaxy
\citep{GS19}. See \citet{Sakamoto13} for a summary of  early works on this galaxy.
A hidden active galactic nucleus (AGN) has been suggested since \citet{Roche86}.
Signs of a nuclear starburst have also been found 
through radio detection of possible super star clusters in the nucleus \citep{Varenius14} 
and from optical population synthesis, although the latter
starburst bears only a minor fraction of the total IR luminosity \citep{Ohyama19}.
This nucleus has a kpc-scale outflow seen in dust and gas along the minor axis of the galaxy \citep{Sakamoto13,Ohyama19,Fluetsch19,Boettcher20}. 
In addition, a gas inward motion along our sight-line has been suggested from redshifted line absorption 
toward the nucleus \citep{GA12, Sakamoto13}.

Arp 220 ($D = 85$ Mpc; 1\arcsec=412 pc) is a late-stage merger with two nuclei separated by $\approx$1\arcsec\ on the sky \citep{Norris88,Graham90}.
We refer to the eastern nucleus as \arpE\ and the western nucleus as \arpW.
Both exhibit deep silicate absorption at 9.7 \micron\ \citep{Soifer99}.
Each shows its rotation in molecular gas, atomic gas, and stars,
with the two nuclei having misaligned rotational axes and apparently counter-rotating velocities \citep{Sakamoto99,Mundell01,Genzel01,Engel11}.
The nuclei are embedded in a large rotating structure of molecular gas \citep{Scoville97}. 
Vigorous star formation is evident in the two nuclear molecular disks from dozens of 
supernovae and their remnants \citep{Smith98,Varenius19}.
In addition to a large-scale superwind \citep{Heckman90,McDowell03}, 
outflows from both individual nuclei 
have been known from P-Cygni profiles of (sub)millimeter lines \citep{Sakamoto09}
and elongated structures of synchrotron emission \citep{Varenius16}.

These heavily-obscured luminous galactic nuclei are probably in a phase of rapid evolution,
and those in the major merger must have a link with the transformation of the galaxy.
The evolution must be fast, judging from the compactness of the gas concentrations in these buried nuclei
and the intensities of their nuclear activities.
The dynamical time scales are on the order of 1 Myr and 10 Myr, respectively, 
for nuclear gas concentrations of $\lesssim 100$ pc
and the close pair of merger nuclei of $\leq 1$ kpc separation.
Starbursts in U/LIRGs have gas consumption time scales on the order of 100 Myr \citep{SSS91,Herrero-Illana19}.
The link to galaxy transformation is a part of the long-held observational and theoretical framework.
It posits that the rapid supply of interstellar gas to a galactic center through strong dynamical perturbation 
(such as a major merger) triggers vigorous star formation and feeding to the central black hole (i.e., a luminous-AGN phase). 
A gas blowout due to the starburst/AGN reveals a visible quasar 
and eventually makes a gas-poor galaxy or elliptical galaxy in the case of a major merger \citep{TT72,Sanders88}.
The time scale for the most intense starburst and luminous buried AGN is 
on the order of 100 Myr in numerical simulations of this evolutionary sequence \citep{Hopkins08}.
For major mergers like Arp 220, this short phase is a turning point of galaxy evolution in terms of the overall morphology, 
dynamics, gaseous ingredients, star formation, and nuclear activities.
For a less disturbed object like NGC 4418, this phase may be recurrent without transforming the entire galaxy, 
but the nucleus still evolves rapidly in this high-luminosity phase
in terms of its stellar mass or mass of the central black hole or both.

Besides the interest in evolution, more reasons why these nuclei are worth attention.
Namely, they may represent the most intense starbursts, 
a hidden population of AGN, and sites of a starburst-AGN interplay. 
Regarding the hidden AGNs, \citet{Ricci15} found that only four out of 834 AGNs cataloged with 14--195 keV data had 
the best-fit absorbing column density $\log (\NH/\persquarecm)$ of 25--26 and that none had $\log \NH > 26$.
The intrinsic frequency of such deeply obscured AGNs has high uncertainty due to the severe selection bias against them.
It is not trivial at all to identify AGNs and estimate their luminosities in such heavily obscured cases 
because conventional AGN diagnostics are for much less extinguished systems \citep{Hickox18}. 
For example, hard-X ray observations with NuSTAR \citep{Teng15} and Swift/BAT \citep{Koss13}
did not detect AGN in our targets.
VLBI observations detected no bright non-thermal radio cores to indicate AGN either \citep{Varenius14,Varenius19}. 
However, the lack of evidence is not evidence of absence.
An AGN could be buried in a Compton-thick absorber (i.e., $\NH > 1.5\times10^{24}$ \persquarecm). 
It can also be covered by opaque plasma, or it may be without a jet and radio-quiet.
As prototypes of deeply-buried and luminous galactic nuclei, the three nuclei in NGC 4418 and Arp 220 
are potential buried-luminous AGNs accompanied by the intense starbursts seen in radio.

Millimeter-to-submillimeter emission tells us much about these hidden galactic nuclei, 
for physical scales of several pc to several 100 pc in this work.
First, these wavelengths contain various emission lines tracing the dominant (by mass) interstellar medium in these systems,
i.e., cold-to-warm molecular gas (\about10 to a few 100 K). The lines tell us the physical conditions and kinematics of the gas and thereby tell us about the fueling and feedback of the nuclear activities as well as the mass distribution in the nuclei.
Second, the short-millimeter and submillimeter continuum is usually dominated by thermal emission from dust.
It is this dust continuum that carries the vast majority of the bolometric luminosity of these nuclei.
Because of that, high-resolution information available at sub/millimeter is very useful to characterize 
the nuclei, complementing spatially unresolved observations at far-IR where the continuum spectrum peaks.
Third, the dust in these nuclei is moderately opaque (around the order of unity) at sub/millimeter wavelengths 
to produce bright emission and yet to make the nuclei penetrable except at the very center in extreme cases. 
As a guide, $\tau_{\rm 1\;mm}=1$ corresponds to $\NHH \sim 10^{25.5}$ \persquarecm\ 
for an environment similar to the Galactic disk \citep{Planck11,Galliano18}.
The structure of the nuclei can thus be probed more easily in the sub/millimeter than at far-IR and shorter wavelengths.
At the same time, the continuum emission carries information about the dust opacity and hence the obscuring column density.
Fourth, rotational lines from vibrationally excited HCN at sub/millimeter wavelengths can be an indicator and probe of
the NGC 4418/Arp 220-type nuclei with significant obscuration \citep{Sakamoto09,Aalto15,Aalto19,Imanishi19}.
These nuclei are opaque enough to trap infrared photons to increase their inner temperature 
(i.e., the greenhouse effect), and the enhanced infrared radiation vibrationally excites molecules such as HCN
\citep{GS19}.
Their rotational lines in the sub/millimeter have a chance to leak out of the dusty nuclei
and indicate their hot interior since dust opacity is lower at longer wavelengths.
And finally, the Atacama Large Millimeter-submillimeter Array (ALMA) has drastically enhanced our capability for
high-resolution and high-sensitivity observations at sub/millimeter wavelengths.
A characteristic ALMA resolution of 0\farcs1 is on the order of 10 pc at nearby hidden galactic nuclei at tens of Mpc.
While the resolution does not match the accretion disks around supermassive black holes ($r \ll 1$ pc),
it is only an order of magnitude larger than young massive star clusters \citep{Longmore14} and 
corresponds to the outer radii of molecular tori around AGN \citep{Combes19}. 
Therefore, ALMA observations are good at probing the obscuring interstellar medium around AGN (if any),
circumnuclear star formation, and their effects on the circumnuclear material.

We have conducted ALMA high-resolution spectro-imaging of the three nuclei in NGC 4418 and Arp 220 
at 1.4--0.4 mm wavelengths to probe their spatial, kinematical, and thermal structure as well as their gas and dust properties.
We covered a total bandwidth of 66 GHz for each target for key molecular lines and continuum information.
The wide bandwidth allows us 
to observe continuum at frequencies least contaminated by line emission as well as 
to constrain variation of continuum properties over frequency. 
This spectro-imaging project follows up 
our previous interferometric spectral scans on these galaxies at lower resolution \citep{Costagliola15, Martin11} and
our high-resolution sub/millimeter studies of these nuclei in smaller bandwidths 
\citep[and those already mentioned]{Sakamoto08,Costagliola13,Aalto15,Martin16}. 
\citet{Sakamoto17} was prompted by the continuum analysis in this work and analyzed 3 mm data 
of Arp 220 at resolutions up to 0\farcs05.

We present our wide-band spectro-imaging of the three nuclei
in a pair of papers to accommodate a large amount of information from ALMA.
This first paper provides an overview of the observations, details our data reduction, and presents continuum results
and analysis.
In Section \ref{s.obs}, we describe our ALMA observations.
We detail in Section \ref{s.cal} our data calibration techniques developed to handle
wide-band, high-resolution, and line-filled spectral scans with ALMA.
We then describe in Section \ref{s.cont} our handling of continuum in such a spectral scan, 
our techniques to determine the size and shape of the compact nuclei,
and the continuum parameters obtained for the three nuclei.
We present in Section \ref{s.model} models to interpret the continuum observations. 
In particular, we develop a model-based formula to estimate the dust opacity and column density from 
the spectral slope of continuum emission and apply it to the three nuclei.
Our continuum analysis takes into account the extreme opaqueness of the nuclei and their resulting
temperature structures.
In Section \ref{s.spatial}, we explore the variation of continuum structure over frequency.
We contrast our models with our compilation of spatially-resolved measurements
of the three nuclei to further assess their opacities.
We revisit in Section \ref{s.contStructure} the structure of the three nuclei, and Arp 220W in particular, in light of 
the high-quality structural data and the understanding of their frequency-dependent appearance.
We discuss our results in Section \ref{s.discussion} and summarize our findings in Section \ref{s.summary}. 
Our companion paper presents observations of molecular lines in the same data set \citep[hereafter Paper II]{Paper2}.

%%%%%%%%%%%%%%%%%%%%%%%%%%%%%%%%%%%%%%%%%%%%%%%%%%%%%%%%%%%%
\section{Observing Parameters and Log}  
\label{s.obs}
Our program consists of interferometric spectral scans on compact targets having numerous lines and bright continuum radiation.
We used an observing setup to facilitate accurate data calibration of such scans.

%%%%%%%%%%%%%%%%%%%%
\subsection{Observing Parameters}
\label{s.obs.parameters_log}

We observed the two galaxies at about 0\farcs2 resolution between 210 and 700 GHz\footnote{At ALMA, 
the frequency ranges 84--116, 211--275, 275--373, and  602--720 GHz are called Bands 3, 6, 7, and 9, respectively.
We observed in Bands 6, 7, and 9. We also use observations in Band 3.}
from June 2013 to August 2015 in the ALMA Cycles 1 and 2
through projects 2012.1.00377.S (NGC 4418) and 2012.1.00317.S (Arp 220).
We observed a single position in each galaxy with about forty 12~m-diameter antennas in ALMA.
The observed positions (ICRS) are 
R.A.=12\hr26\mn54\fs612, Dec.=$-0\arcdeg52\arcmin39\farcs41$ for NGC 4418
and 
R.A.=15\hr34\mn57\fs250, Dec.=$+23\arcdeg30\arcmin11\farcs36$ for Arp 220.
The full width at half maximum (FWHM) of the primary beam is 
24\arcsec--19\arcsec, 15\arcsec--14\arcsec, and \about7\farcs6, respectively,
for our observing frequencies in Bands 6, 7, and 9.\footnote{
Primary beam FWHM is measured to be \about$1.13 \lambda/D$, where $\lambda$ is the observing
wavelength and $D$ is the antenna diameter of 12 m (ALMA Cycle 2 Technical Handbook. \S6.2).}

Table \ref{t.tunings} lists, for each galaxy, our ten frequency setups (i.e., tunings) 
and major spectral lines in each. 
The brightest lines are CO(2--1), (3--2), and (6--5), HCN and \HCOplus(3--2) and (4--3), and HNC(4--3).\footnote{
We denote rotational transitions mostly in a simplified form: e.g., CO(2--1) instead of CO(J=2$\rightarrow$1).
}
Each frequency setup employed four 1.875~GHz-wide spectral windows, two in each sideband.
To help calibration, we overlapped adjacent tunings and the two spectral windows in each sideband 
by a few tenths of GHz (see Table \ref{t.tunings}).
We observed the two galaxies in almost the same rest-frequency range of 341--367 GHz in Band~7
but used slightly different frequency tunings in Bands~6 and 9. 
This choice was because Arp 220 has much broader lines and hence fewer continuum channels than NGC 4418.
For Arp 220, we tied all the four Band~6 tunings through their overlaps for relative flux-scaling,
and we aimed to avoid line-crowded frequencies for continuum measurements in Band 9.
Our total observing bandwidth excluding overlaps was 66~GHz in sky frequency for both galaxies,
covering 27, 25, and 14 GHz in Bands 6, 7, and 9, respectively.
We recorded the data with the spectral resolution of 0.976~MHz (and 0.488 MHz in channel spacing).
We later binned the data to lower spectral resolutions, e.g.,
20 MHz for spectra and 50 \kms\ for imaging of individual lines in \citest{Paper2};
the latter resolves the gas motion around each nucleus and corresponds to 36--116 MHz for our observing frequencies.

%%%%%%%%%%%%%%%%%%%%
\subsection{Observing Log}
\label{s.obs.obslog}
Table~\ref{t.obslog} is the log of our observations.
We aimed at spectroscopy limited by line confusion and 
observed each frequency setup for total on-source integration of 5--16 minutes in one or two sessions.
The maximum projected baseline toward the target galaxy was typically about 1.5~km for Bands 6 and 7
and 0.6 km for Band 9. 
With our minimum baselines less than 40 m, our observations had the maximum recoverable scale (MRS) of 3\arcsec--11\arcsec\ in Bands 6 and 7 and \about3\arcsec\ in Band 9.
These are adequate for the known compactness of the continuum emission in the three nuclei and the line emission
around the nuclei. 
Because of the large-scale gas motion around each nucleus (e.g., rotation), 
line emission in individual channel maps is more compact than its full extent in the velocity-integrated intensity map.
Thus the effective MRS for an emission line is larger than the continuum MRS given above.
The line MRS depends on the spatial and spectral structures of the line and the source,
and it cannot be expressed in a general form.

We also conducted a companion spectral scan of Arp 220 at \about0\farcs7 resolution through
211--357 GHz except around the 325 GHz absorption by atmospheric water vapor 
(ALMA Project ID. 2012.1.00453.S).
\citet{Martin16} already reported it in part, and full results will be reported elsewhere.
These programs shared observing techniques and the data reduction method.

%%%%%%%%%%%%%%%%%%%%%%%%%%%%%%%%%%%%%%%%%%%%%%%%%%%%%%%%%%%%
\section{Data Calibration and Imaging}  
\label{s.cal}

%%%%%%%%%%%%%%%%%%%%
\subsection{Overview}
\label{s.cal.overview}
We calibrated ALMA raw data with CASA ver.~4.6 \citep{CASA07} consulting the calibration, 
such as the rejection of bad scans, conducted by the observatory.
Our flux calibration is based on either the flux calibrator observed in each observing session 
or the quasar monitoring data in the ALMA Calibrator Source Catalogue\footnote{\sf https://almascience.nao.ac.jp/sc/}.
We used the ``Butler-JPL-Horizons 2012" model in CASA for our flux calibrators in the solar system.

Our positional reference frame is the ICRS\footnote{ALMA observations prior to ALMA Cycle 3, including ours, 
were mislabeled to be in the J2000 system but are in the ICRS.},
and we report velocities in the radio definition and in the LSRK frame.
Positions and velocities in other reference frames or conventions are converted when used in this paper.

We performed the extra calibrations in Sections \ref{s.cal.bandpass}--\ref{s.cal.B9calibration}
to improve spectral baseline, astrometry, and amplitude scales.
This additional effort is to use our data as a wide-band spectral scan 
rather than a mere collection of inhomogeneous data sets from multiple frequency tunings.

%%%%%%%%%%%%%%%%%%%%
\subsection{Bandpass Calibration}
\label{s.cal.bandpass}
We assumed that the bandpass and gain calibrators have continuous power-law spectra within each of our tunings.
Specifically, we estimated their spectra from our flux calibration or the Calibrator Catalogue. 
We then used the CASA task {\tt setjy} to enforce, for each calibrator, the same power-law model to all the four spectral windows.
It improves the original calibration by the observatory, 
in which calibrators are either assumed to have zero spectral slopes or allowed to have spectra discontinuous at the boundaries of spectral windows.\footnote{
For example, in some CASA pipeline settings, 
the bandpass calibrator has a spectrum in the form of
$S_\nu^{(i)} = a^{(i)} \nu^\alpha$, 
where $i$ is the index for spectral windows and $\alpha$  is a non-zero  spectral index common to all spectral windows.
The scale parameter $a^{(i)}$ is not required to be the same for all spectral windows. 
Instead, it usually varies among the spectral windows.
The reason is that the scale is measured independently for each spectral window and always has a measurement error.
When the model spectrum for the bandpass calibrator has amplitude gaps like this, 
the bandpass-calibrated data inherit the scaling gaps between spectral windows. 
One could avoid it by using a constant scale $a$ for all spectral windows by, for example, averaging the $a^{(i)}$s.
The same is true for gain calibrators used to calibrate amplitude gain.
}
The difference between the default calibration and ours is  illustrated in the left panel of Figure \ref{f.calib}.
Poor calibrator models would affect the target data and cause discontinuity in its calibrated spectrum.
The calibrator spectral indices that we used are in Appendix \ref{a.calSpix}.

Our bandpass calibration within each spectral window is precise enough to make the noise in our spectra limited by the thermal noise.
This non-trivial feature is thanks to ALMA being an interferometer with many antennas (to average the cross-correlation signals from
\about$10^3$ antenna-pairs), availability of power-law continuum sources without lines (i.e., quasars) to measure and correct for the instrumental response, and the virtual absence of environmental narrow-line interference in the sub/millimeter wavelengths unlike at longer wavelengths. 
The spectral dynamic range achieved by the ALMA bandpass calibration is about 800 
for our default 20 MHz spectral resolution \cite[section 10.4.6]{Cortes20}. 
We confirmed the high precision in our data by checking bandpass-calibrated spectra of bright gain calibrators.
For comparison, the signal-to-noise ratio of the continuum emission of our target nuclei is about 30--100 in each 20 MHz channel.
Therefore, the noise in our nuclear spectra is mainly thermal noise; any noise due to bandpass calibration should 
be an order of magnitude smaller. 
Any wiggles in the spectra, if they are above the thermal noise with enough margin, must be due to lines of the target nuclei.

%%%%%%%%%%%%%%%%%%%%
\subsection{Spatial Alignment}
\label{s.cal.spatialAlignment}
We improved our calibration through phase-only self-calibration and by spatially aligning datasets from different tunings. 
We first measured the positions of our target nuclei in individual tunings and each of the upper and lower sidebands.
We made the images used for this measurement without self-calibration and by adding all channels. 
We then determined our fiducial nuclear positions and used them in phase-only self-calibration to all the data. 
This procedure aligned our data to the same nuclear positions. 
It ensured that the spectral sampling and subsequent analyses in each nucleus were for the same location at any frequency.
We then performed another round of phase-only self-calibration for each tuning without a forced model 
to further reduce the residual phase noise.

For NGC 4418, the mean position of the nucleus was 
R.A. = 12\hr26\mn54.613\s\ and Dec.~= $-0$\arcdeg 52\arcmin 39\farcs41.
The deviations from the mean have an rms of 0\farcs03 and are  0\farcs11 at most.
We determined that the effect of line contamination is small in the positional scatter 
because the two sidebands in each tuning showed about the same offset even though they contained different lines.
We, therefore, adopted the mean position above for the continuum nucleus, 
assuming the scatter to be extrinsic (e.g., atmospheric or instrumental).

On Arp 220, we observed systematic error in astrometry.
Figure~\ref{f.a220position} shows it in our initial measurements before self-calibration. 
Both nuclei changed their positions as a function of hour-angle while keeping their relative position virtually constant.
The two nuclei also drifted together as a function of hour angle in our companion spectral scan of Arp 220 at lower resolution.
We, therefore, aligned data from our two scans using
R.A. = 15\hr34\mn57.295\s\ and Dec.~= $+23$\arcdeg 30\arcmin 11\farcs36  for the eastern nucleus
and
R.A. = 15\hr34\mn57.225\s\ and Dec.~= $+23$\arcdeg 30\arcmin 11\farcs52  for the western nucleus.
These coordinates agree to a few 10 mas with the ALMA measurements at 100 GHz in
\citet[see Table 1]{Sakamoto17}. 
We do not rely on absolute astrometry beyond that level of accuracy in the scientific analysis of our spectral scans.

After all self-calibration, we made cleaned data cubes, convolved them to the resolution of 0\farcs35 (FWHM),
and sampled spectra at the positions of the three nuclei.
These initial spectra had a 20 MHz resolution and were taken from every spectral window.

%%%%%%%%%%%%%%%%%%%%
\subsection{Relative Flux Scaling --- Flux Self-calibration}
\label{s.cal.fluxScaling}
A spectral scan like ours on a source with a strong continuum emission and numerous lines needs precision in relative flux calibration
to minimize amplitude jumps in the final concatenated spectrum and for accurate ratios among the lines in different tunings.
For example, if line and continuum emission is at the level of 100$\sigma$ per channel in the spectra from individual tunings, 
then a usual 10\% accuracy of flux calibration leaves amplitude jumps on the order of 10$\sigma$ at the tuning boundaries 
when we concatenate the spectra. 
Subtraction of a smooth continuum from this spectrum will leave several $\sigma$ of artificial irregularities. 
It would severely limit the detection and flux measurement of weak lines.\footnote{There will not be the same problem 
if lines do not fill the spectrum. 
In that case, one can determine and subtract continuum first in individual tunings, use the continuum information for amplitude re-scaling, and then concatenate the continuum-subtracted spectra to achieve accurate relative flux scaling and no gaps at the boundaries \citep[e.g.,][]{Harada18}. 
}
One needs a wide-band spectrum without amplitude jumps 
to robustly identify its spectral channels having as little line emission as possible. 
Continuum measurement and subtraction depend on those channels.
Therefore, precise amplitude alignment is necessary for both continuum and line measurements.

\begin{figure}[!thb]
\centering
\includegraphics[width=0.37\textwidth]{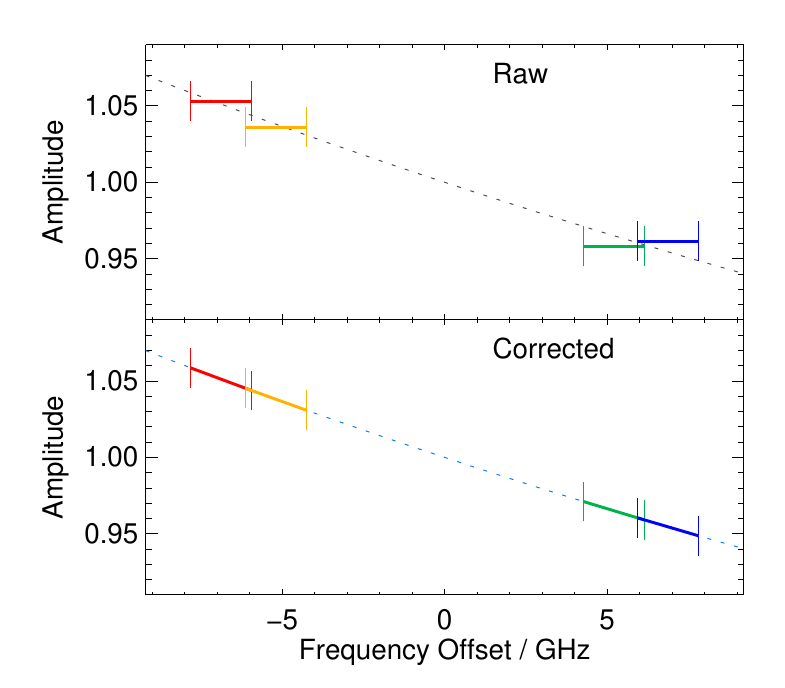} % .eps} 
\hspace{0.05\textwidth}
\includegraphics[width=0.37\textwidth]{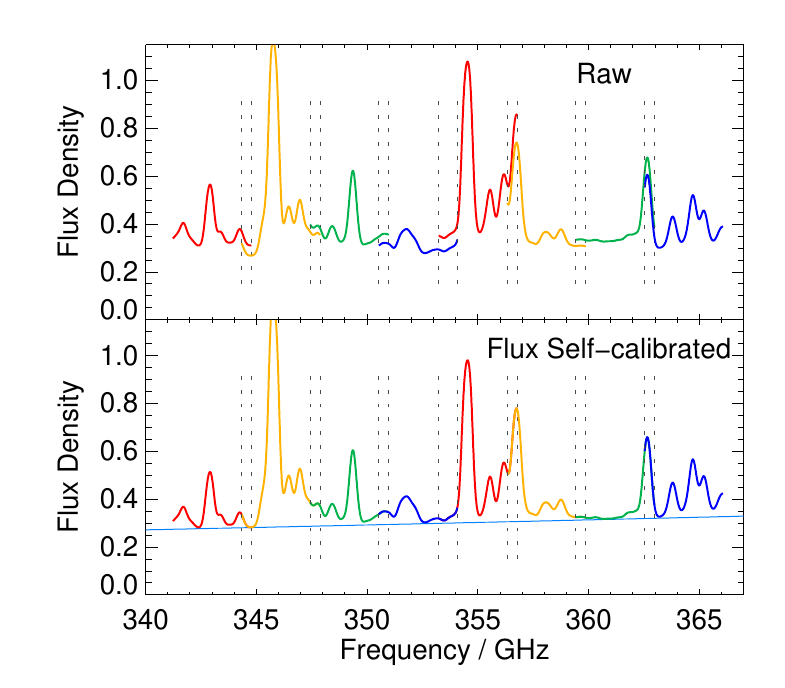} % .eps} 
\caption{ \label{f.calib}
Concepts of our ALMA calibration improvements. 
(Left) 
Illustrated are the four spectral windows in a single tuning.
The default calibration in Cycles 1--2 assumed a zero spectral slope for passband calibrators and 
did not enforce that the spectrum of a gain calibrator should be a single power-law function across all the spectral windows.
The spectrum of a gain calibrator often looks like the one in the upper sub-panel after the default calibration.
We used quasar spectral slopes measured in our observations or those taken from the ALMA calibrator archive
and enforced our passband and gain calibrators to have power-law spectra, as in the lower sub-panel.
(Right) 
Flux self-calibration among multiple tunings.
A mock spectral scan with four tunings is shown. 
A pair of spectral segments in the same color is from the two sidebands of the same tuning.
Each tuning had a flux calibration error of less than 10\%. 
It caused the visible gaps between the spectral segments at their overlapping channels, indicated by vertical dotted lines.
We compared the observed spectra there and derived a scaling-correction factor for each tuning 
to minimize the gaps. 
In this example, four scaling factors are derived via the least-square method 
from seven ratio measurements at the overlaps 
and a constraint that the mean of the scaling factors is unity.
The underlying continuum shown in light blue is better seen after this flux self-calibration.
}
\end{figure}

\begin{figure}[t]
\epsscale{0.756}
\plotone{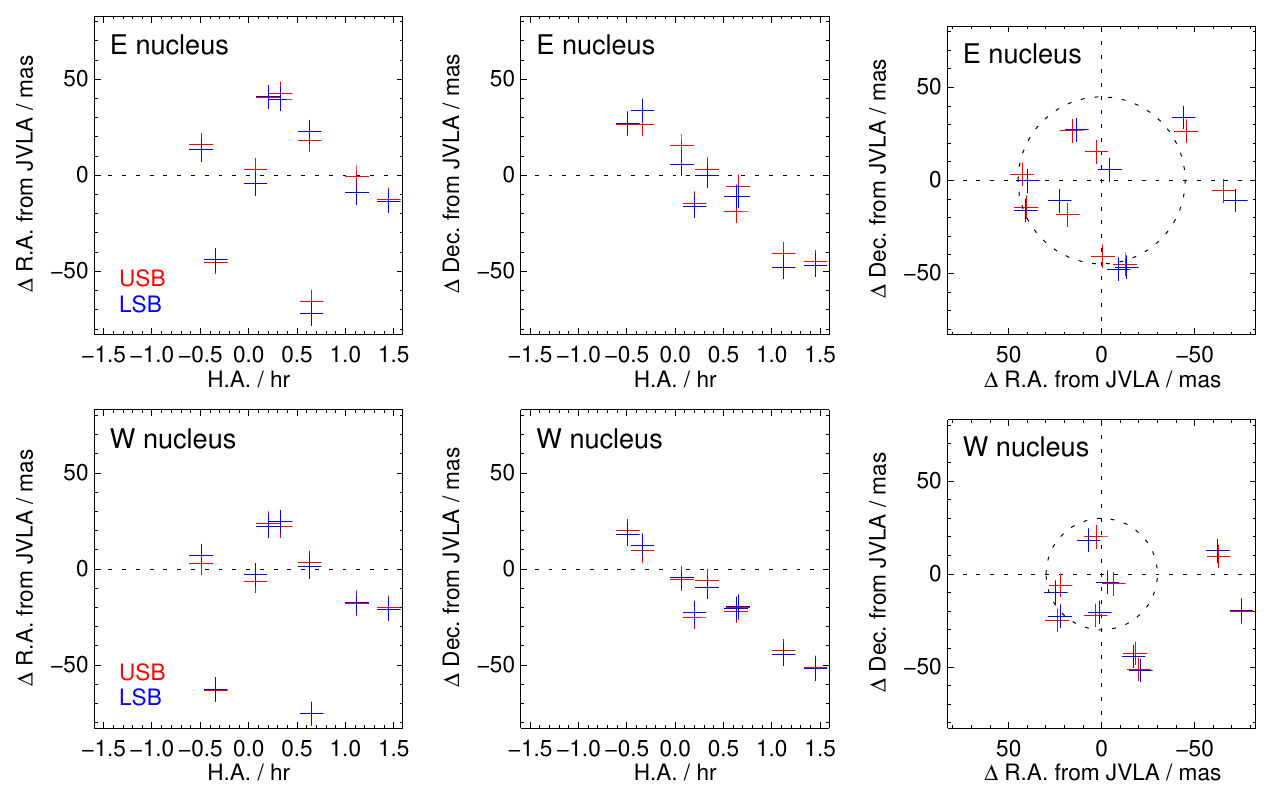} % .eps} 
\epsscale{0.244}
\plotone{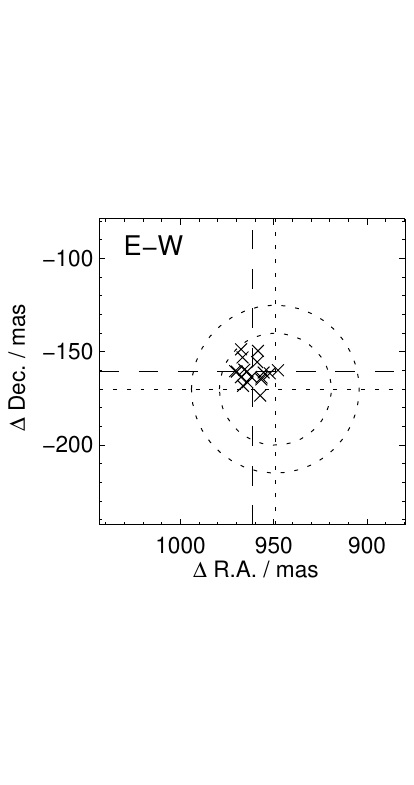} % .eps} 
\caption{ \label{f.a220position}
ALMA astrometry of the Arp 220 nuclei. 
We measured the positions of the two nuclei in each sideband in every tuning before any spatial alignment or self-calibration.
The six left panels show the offsets from the 33 GHz positions measured at the Jansky Very Large Array (JVLA) \citep{Barcos-Munoz15}. 
The top row is for the eastern nucleus and the bottom row for the western nucleus. 
The first and second columns show the offsets as a function of the hour angle of Arp 220. 
Data points in red (blue) are from the upper (lower) sideband. 
Each close pair at the same hour angle is from the same tuning.
Our typical astrometric precision is $\pm3$ and $\pm10$ mas for the western and eastern nucleus, respectively, in each
measurement.
The third column shows our measurements in the R.A.-Dec. space along with error circles of the reference JVLA positions.
The rightmost panel shows the relative position of the eastern nucleus measured from the western nucleus.
Dashed lines indicate our mean offsets; $\Delta$ R.A. = $961.6 \pm 1.6$ mas and  $\Delta$ Dec. = $-160.8 \pm 1.5$ mas.
Dotted lines and circles show the reference JVLA values and error circles. 
The data in the second column suggest some systematic error in ALMA astrometry. 
We calibrated that out.
}
\end{figure}

We refined our flux calibration by using the target signal itself as a reference. 
Accordingly, we call this procedure flux self-calibration. 
Its basic concept is in the right panel of Figure \ref{f.calib}.
Flux self-calibration bases its amplitude rescaling on
comparisons of the initial spectra at their overlaps, which we made  in our tuning setup for this purpose.
It does not matter whether the emission at an overlap is a continuum or line emission as long as it is reasonably compact and stable.
It is partly because two tunings are always compared at their overlaps, i.e., at the same frequency, using the same emission.
It is also because a spectral scan such as ours aims at a uniform spatial resolution across the scan; hence,  
the two observations to be compared should have approximately the same range of baseline \uv\ lengths at their overlaps.
Therefore, the flux self-calibration through overlapping tunings can handle targets with numerous broad lines with few
line-free channels, such as ours and the one in Figure \ref{f.calib}(right).
Our target nuclei must be stable in the sub/millimeter continuum over the timescale of our observations 
because, as we will see later, the continuum emission is predominantly thermal emission of dust 
from $\gtrsim 10$ pc regions. 
Line emission is more extended than continuum and should be more stable; 
in \citest{Paper2}, we do not detect any plausible masers in our data. 
This stability is another key of the flux self-calibration.
We used a single scaling factor for each tuning, shared by all the spectral windows in the tuning. 
The pairs of spectral windows in individual sidebands matched well at their overlaps without relative rescaling, 
thanks to our prior calibration illustrated in Figure \ref{f.calib}(left).

The actual rescaling procedure was the following.
First, for the tunings observed multiple times\footnote{ALMA observes without real-time Doppler tracking and covers slightly different sky frequencies in individual observing sessions for the same observing setup. 
We only used data in the frequency range that is common to all sessions.}, 
we compared spectra from individual observing sessions and rescaled each session 
so that its source flux agreed with that in the mean spectrum. 
Next, spectra from adjacent tunings were compared at the overlapping channels
to derive their scaling factors under a constraint that the mean of the scaling factors is unity. 
Then we used those factors to rescale the visibility data.
In other words, we assigned each tuning a scaling factor $a_i$, where $i$ is
the tuning index, and solved (with the least-squares method when necessary) 
a set of equations:
\begin{equation} \label{eq.rel_gain}
	r_{i,j} = a_i/a_j,
\end{equation}
where $r_{i,j}$ is the measured
amplitude ratio between the tunings $i$ and $j$. 
We used the constraint
\begin{equation}
	\mathrm{mean}(a_i)=1
\end{equation}
to set the overall scale of the solutions.\footnote{
Mathematically, this set of equations is similar to the one  
for interferometric gain calibration. 
For the latter, the right side of (\ref{eq.rel_gain}) is the product of the antenna gains, 
and the left side is the baseline-based amplitude of a quasar.}
The derived scaling factors were up to 6\% deviated from unity.
This flux self-calibration reduced amplitude discrepancies at the overlaps to $\lesssim1 $\% in our final data.

The flux self-calibration reduces internal amplitude inconsistency due to flux calibration errors 
in individual observing sessions. 
It also reduces the effect of any errors in the models of our reference flux calibrators by using
more than one of them.
The absolute flux scale of Cycle 2 observations should be accurate to 10\% in Bands 6 and 7
and 20\% in Band 9, according to the ALMA Cycle 2 Technical Handbook \citep{Lundgren13}.
We expect that our extra calibration reduced the random part of this error.  
After the relative scaling and concatenation of multiple observing sessions, 
each set of our concatenated observations effectively had flux calibration more than once against multiple flux calibrators. 
The error reduction should be up to a factor of $\sqrt{5}$ because we had five observing sessions
for the Band 7 observations of NGC 4418.

%%%%%%%%%%%%%%%%%%%%
\subsection{Flux Self-calibration of Non-contiguous Tunings}
\label{s.cal.fluxScaling_fullBands}
The flux rescaling described so far is only for each group of contiguous tunings. 
Therefore, it is still possible that different groups of contiguous tunings have slightly different flux scales, even though
the flux accuracy has been improved for each tuning group by their internal rescaling 
and its averaging effect.

One can reduce the remaining flux-scale biases with
another round of flux self-calibration through the following steps.
First, find continuum-dominated channels in the combined spectra of the individual tuning groups.
Those channels are easier to find in a combined spectrum than in the narrowband spectra before concatenation.
Second, fit the continuum data from multiple tuning groups with a simple function, 
such as a sum of a few power laws or a power law whose spectral index is a low-order polynomial of frequency.
Third, use the best-fit continuum spectrum as a model and determine rescaling factors, one for each tuning group,
using the deviations from the model. And last, apply the rescaling to the data.
Note that,  by replacing ``tuning group'' with ``tuning'' in the above procedure,
this method will be the same as the one used for
spectral scans on a continuum emitter with fewer lines, i.e., those having enough continuum channels in each tuning.

We took the above steps only to the third one and found the deviation of flux scale to be 3--4\%
among the non-contiguous tuning groups in Bands 6 and 7 (see Section \ref{s.cont.spix}).  
Since we do not apply the last rescaling step (to be conservative about excessive calibration), 
our data can be used (in Section \ref{s.cont.spix}) for the spectral index across our full frequency coverage. 
Also, these likely residual gaps in flux-scale among non-contiguous tunings were the reason why we subtract continuum 
individually in each contiguous tuning group (Section \ref{s.cont.def}).

Overall we expect our absolute flux scale to be accurate to about 5\% in Bands 6 and 7 
unless the ALMA primary calibrators have coherent scaling biases in these bands.
The uncertainty can be larger (up to 10\%) for the tuning groups involving only a small number of independent measurements
(i.e., those having only two overlapping tunings or sharing assumptions on secondary flux calibrators taken from the observatory database.)
Our internal flux accuracy (i.e., precision) should be on the order of 1\% in each contiguous spectral section in Bands 6 and 7.
Therefore, unless otherwise noted, we report flux-related quantities in this paper 
with enough digits to preserve the internal precision, 
and the reported errors are only for random errors.
Readers using them to compare with other observations or taking ratios across different bands in our dataset should 
add an appropriate systematic error (5--10\% for absolute flux scaling in Bands 6 and 7).

%%%%%%%%%%%%%%%%%%%%
\subsection{Band 9 Calibration} 
\label{s.cal.B9calibration}
We had two tunings in Band 9, and their calibration was most difficult.
It is partly because of the high atmospheric attenuation and system noise.
It is also because ALMA has little Band 9 flux measurements in its calibrator monitoring.
We applied the same calibration method as in the lower bands, including the flux self-calibration.
But there was little averaging effect for error reduction because the number of independent observations was small.
Therefore, we expect our flux scale to be less reliable in Band 9 than in our lower bands.
(Appendix \ref{a.b9fluxcal} has details of our Band 9 flux calibration.)
We adopt the 20\% absolute-scale uncertainty advocated in the ALMA Technical Handbook.

We had mixed results from the comparison of our Band 9 photometry with other observations. 
Our Arp 220 data have 6\% and 21\% larger continuum flux densities for the E and W nuclei, respectively,
compared to the earlier ALMA observations by \citet{Wilson14}, who adopted 15\% for their flux-scale uncertainty.\footnote{
This comparison used the same 0\farcs45-radius apertures in our continuum image smoothed to the same 0\farcs5 resolution.
We applied a minor correction for the different sky frequencies; ours is for
%and a minor correction has been applied for the different sky frequencies (ours is for 
\fobs= 671 GHz and theirs 679 GHz. }
The different ratios for the two nuclei imply that they are not solely due to flux scaling.
The total continuum flux density of Arp 220  is
$4.4 \pm 0.9$ Jy at 670 GHz in a 2\arcsec-diameter aperture in our data; the uncertainty is for flux calibration.
\citet{Dunne01} measured $6.3\pm0.8$ Jy at virtually the same frequency using the JCMT/SCUBA (\about8\asec\ resolution)
while the power-law interpolation from the 350 and 500 \micron\ measurements in the Herschel SPIRE Point Source Catalog\footnote{\url{https://irsa.ipac.caltech.edu/Missions/herschel.html}} (\about30\arcsec\ resolution) is 6.2 Jy at 670 GHz.
We detected about 70\% of these measurements.
In NGC 4418, our Band 9 flux density of the nucleus is more than 20\% larger than 
the best-fit power-law spectrum from our Band 6 and 7 measurements (Section \ref{s.cont.spix}).
Our Band 9 flux density in a 1\arcsec-diameter aperture, $1.0\pm0.2$ Jy at 680 GHz, is again about 70\%  of 
the interpolation from the Herschel 350 and 500 \micron\ measurements of 1.4 Jy.
Although ALMA detected less Band 9 flux than JCMT and Herschel, 
it must be partly due to the emission outside our small apertures
since both Arp 220 and NGC 4418 have central molecular gas structures extended to several arcsec.
In addition, any continuum emission more extended than our \about3\arcsec\ MRS (Section \ref{s.obs.obslog}) 
is filtered out in our data.
Moreover, some of the excess emission in the JCMT and Herschel measurements may be due to blended weak lines 
that were not recognized as such with the lower spectral resolutions.

Since we decided not to apply flux rescaling based on comparisons 
among tunings without overlaps (Section \ref{s.cal.fluxScaling_fullBands}),  
we present what we obtained with these cautionary notes on the flux scale and Band 9 calibration. 
Considering this elevated uncertainty in flux calibration, we will defer,  in Section \ref{s.model}, 
fully integrating our Band 9 photometry to our BGN continuum analysis that uses continuum spectral slopes. 
On the other hand, this flux uncertainty does not affect the shape and size of the Band 9 emission.
It does not affect either the relative flux scales within the Band 9 data. 
Therefore, the ratios among Band 9 lines or between Band 9 line and continuum should be accurate.

%%%%%%%%%%%%%%%%%%%%
\subsection{Note on Early ALMA Flux Calibration} 
\label{s.cal.fluxDifficulty}
There is a flux calibration issue specific to early ALMA observations that some of the flux standards have changed over time.
ALMA flux standards are solar system objects.
Their recommended models have been ``Butler-JPL-Horizons 2012'' \citep{Butler12} 
since CASA 4.0, which arrived in December 2012.
While the name remains the same, 
some of the models had updates in October 2016 in CASA 4.7.
The update includes Ceres and Pallas used in our observations of NGC 4418 (see Table \ref{t.obslog}).
The difference between the old and new models depends on the observing frequency, date, and time.
It can be more than 10\% in the worst case.

For our three observations of NGC 4418 that contained Ceres or Pallas, 
the old models in CASA 4.6 overestimate (compared to the new models) the calibrator flux densities
by 2\% (B9--1.a, Ceres), 16\% (6--1a, Ceres), and 0.9\% (B6--2, Pallas). 
We did not use Ceres for B6--1.a and instead used 3C273 to flux calibrate the observations.
During our observing program, the old models were also in effect 
for the quasar monitoring for the ALMA Calibrator Source Catalogue, which we used in our calibration.
However, we learned through ALMA Help Desk
that Ceres and Pallas were in a lower tier of flux standards and not directly used in the quasar monitoring
and that their flux densities, when needed, were bootstrapped from observations of the flux calibrators in higher tiers. 
Therefore, it must be that the old version of the flux standard models did not significantly affect flux calibration in our project.

%%%%%%%%%%%%%%%%%%%%
\subsection{Imaging and Spectrum Sampling}
\label{s.cal.imaging}
We made our final image cubes after the flux rescaling in Section \ref{s.cal.fluxScaling} 
and the self-calibration described in Section \ref{s.cal.spatialAlignment}. 
We used a spectral resolution of 20 MHz, which is 9--28 \kms\ for our  \frest\ range.
We made the cubes using {\tt robust = 0.5},\footnote{
The {\tt robust} parameter controls the relative weighting of complex visibilities in image reconstruction (i.e., Fourier transform).
It takes a value between $-2$ and $+2$. 
The former produces the smallest synthesized beam by assigning more weights to visibilities from longer baselines,
while the latter results in the largest beam by giving virtually equal weights (scaled by the data noise) to all visibilities 
\citep{Briggs95}. 
A {\tt robust} value around 0 is usually a reasonable compromise between the size and the level of side-lobes of the synthesized
beam.
} 
convolved them to the resolution of 0\farcs35 (FWHM), 
and sampled spectra at each nucleus.
Those spectra were then concatenated by averaging the overlapping channels.
We also obtained spectra at 0\farcs2 resolution in Band 7, where our data have smaller beams than in other bands. 
For this resolution, we slightly increased the visibility weights of long baselines for Arp 220 (i.e., {\tt robust = 0}).
Table \ref{t.dataparams} lists the parameters of the image cubes, including rms noise.

Figures \ref{f.spec_with_cont.ylog.N4418}--\ref{f.spec_with_cont.ylog.A220W} are the resulting spectra at
the resolution of 0\farcs35, which corresponds to 58 pc for NGC 4418 and 144 pc for Arp 220. 
The most conspicuous lines are labeled. 
They also have scale bars of widths 300, 500, and 600 \kms\ for NGC 4418, \arpE, and \arpW, respectively.
These are the full widths at about 30\% of the peak in averaged line profiles of major emission lines in the individual nuclei \citesp{Paper2}. 
Some lines, in particular those seen in Arp 220, show profiles far from Gaussian and even show absorption features.
More line identification and line analysis are in \citest{Paper2}.
But even without further line identification, 
it is evident that lines fill our spectra in at least Bands 6 and 7. 
These spectra are plotted in a logarithmic intensity scale to highlight the abundance of faint lines. 
We achieved our observational goal of ``spectroscopy limited by line confusion.''

%%%%%%%%%%%%%%%%%%%%%%%%%%%%%%%%%%%%%%%%%%%%%%%%%%%%%%%%%%%%
\section{Continuum Observations}
\label{s.cont}
We characterized the continuum in the line forests by identifying the channels that appear the least contaminated by visible lines.
We employed visibility fitting, supplemented by imaging in Band 9, for emission parameters in both spatial and spectral domains.
The continuum parameters so obtained constrain the thermal and radiative properties of the nuclei. 
The fitted continuum is also used for subtraction for line spectra and line images in \citest{Paper2}.

%%%%%%%%%%%%%%%%%%%%
\subsection{Continuum Definition and Subtraction in the Frequency Domain}
\label{s.cont.def}
We operationally define `continuum' in the spectral domain by smoothly connecting the local minima of the spectrum in consideration.
The width of a local minimum can be as small as a few channels or, in our Band 9 data showing fewer lines,
as large as tens of consecutive channels. 
The dotted lines (power-law curves) in Figures \ref{f.spec_with_cont.ylog.N4418}--\ref{f.spec_with_cont.ylog.A220W} show the continuum defined this way in our 0\farcs35 spectra.
When identifying the minima, 
we excluded evident absorption features in such lines as
CO, CN, SiO, \HCOplus, \HtwoCO,
and \HthirteenCN.
We also confirmed that the minima do not correspond to absorption lines due to the telluric atmosphere; 
the atmospheric lines are not visible in our post-calibration spectra.

For NGC 4418,  our observations have four segments at about 673--692 GHz (Band 9 tunings),
338--364 GHz (Band 7 tunings), 
247--270 GHz (B6--1 and B6--2 in Table \ref{t.tunings}), 
and 214--234 GHz (B6--3 and B6--4).
Only the tunings in the same segment had overlap with each other for flux self-calibration.
Consequently, we independently fitted a power law to the continuum data in each of the four spectral sections.
For Arp 220, there are three segments, one each in Bands 9, 7, and 6. 
In both Bands 6 and 7, all the four tunings are tied through overlaps.
We applied the same method as in NGC 4418 for both nuclei of Arp 220 in Band 7 and Arp 220E in Band 6.
For Arp 220W in Band 6, we used a single minimum at \about231 GHz in our spectrum and a spectral index 
that we estimated from the Band 6 data of our companion spectral scan. 
The reason is that while we found several minima in the companion scan, only one of them was in our frequency coverage.
Table \ref{t.cont_model-1G_parameters} has the adopted values for parameters of our local continuum function
\begin{equation} \label{eq.cont-intensity}
	I_{\nu} 
	=
	I_{\rm ref} \times \left( \frac{\nu}{\nu_{\rm ref}} \right)^{\alpha}
\end{equation}
for all three nuclei.

We obtained continuum-subtracted 0\farcs35-resolution spectra 
by subtracting the model continuum spectra from our original spectra.
It is evident, e.g., in Fig. 4, that the continuum is easier to identify 
and hence easier to subtract by having a wider bandwidth.
For example, if we had determined continuum independently in each spectral window of 1.87 GHz wide, we would have 
overestimated continuum and underestimated line flux in almost all spectral windows in Bands 6 and 7.
We will discuss the remaining limitations in our continuum measurements and subtraction in Section \ref{s.cont.Caveat}.
The continuum-subtracted spectra are presented and used for our spectral line modeling  in \citest{Paper2}.

%%%%%%%%%%%%%%%%%%%%
\subsection{Size and Shape}
\label{s.cont.sizeShape}
We estimated the size and shape of the continuum emission from the nuclei through visibility fitting. 
Visibility fitting is powerful to characterize compact sources (i.e., sources comparable in extent with the observing beam).
For them, it is more robust than fitting in the image domain \citep[e.g.,][]{Pearson99,uvmultifit14}.
Our continuum observations of NGC 4418 and Arp 220 belong to this category. 
For line emission that tends to be more extended than the continuum, visibility fitting with simple models still provides 
valuable emission parameters such as the centroid position, total flux, and equivalent size as long as
the emission is within our maximum recoverable scale.
Starting from a single Gaussian, we employ increasingly complex models in our fitting to capture the shapes of our target nuclei better.
They facilitate our model-based continuum subtraction and our analysis of the properties and structures of our subjects.

%%%%%
\subsubsection{1G fit}
\label{s.cont.sizeShape.1Gfit}
\begin{figure*}[!h]
\begin{center}
\includegraphics{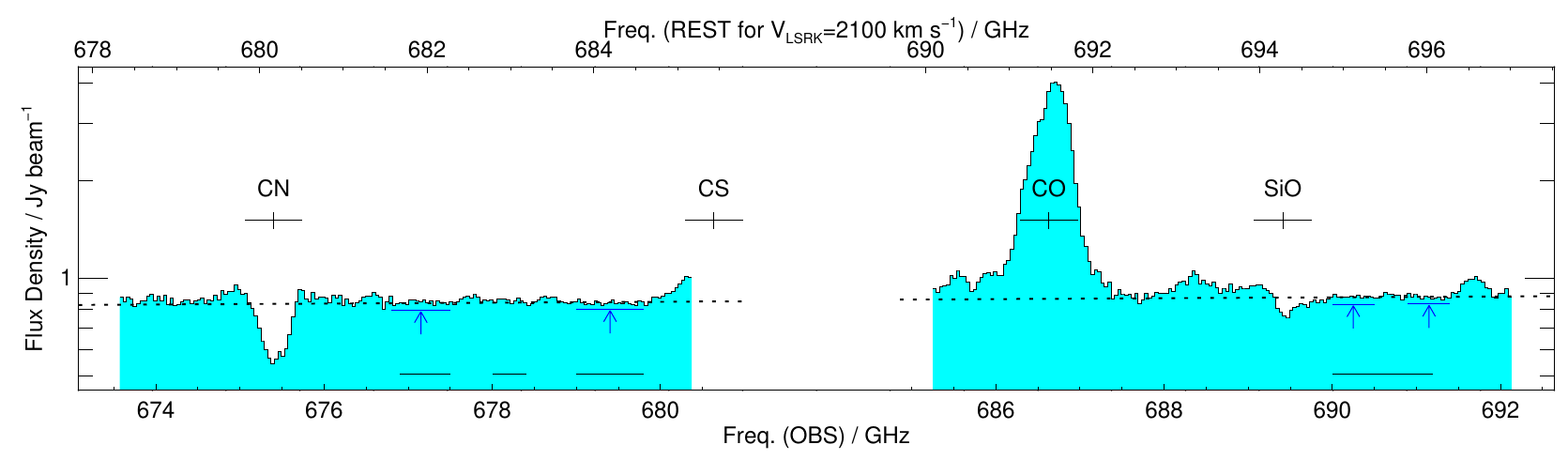} % .eps} 
\includegraphics{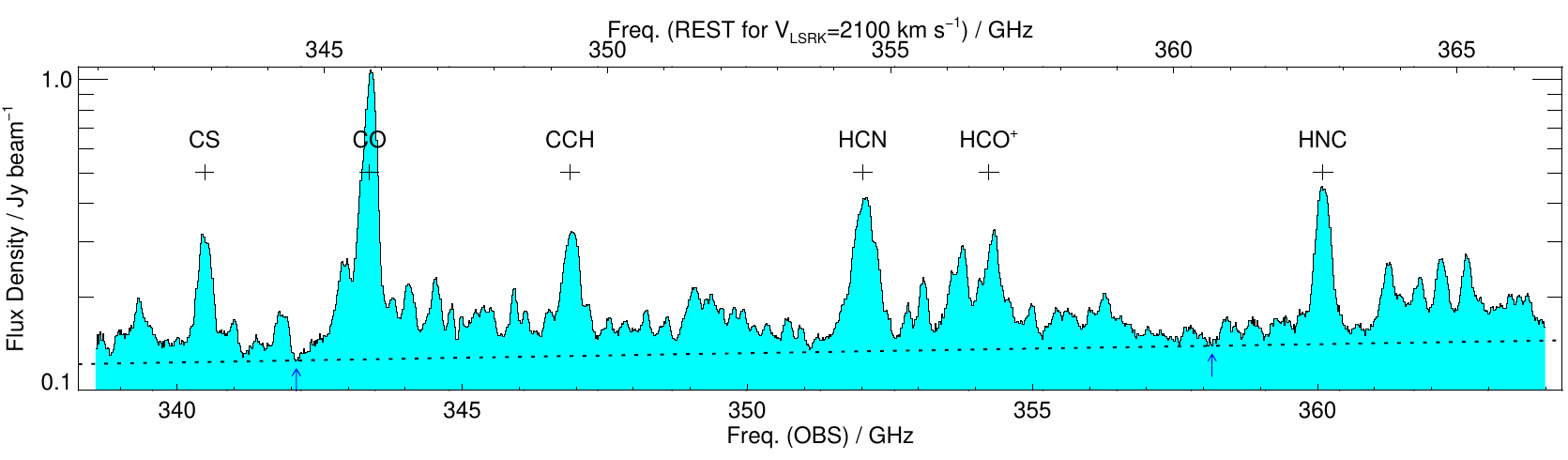} % .eps} 
\includegraphics{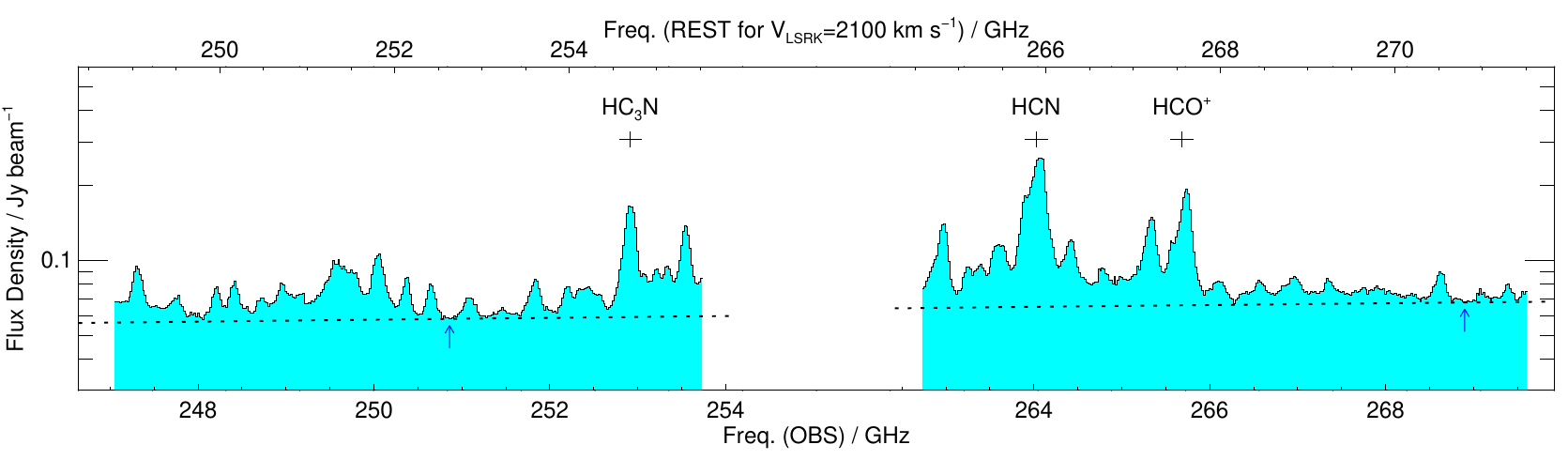} % .eps}  
\includegraphics{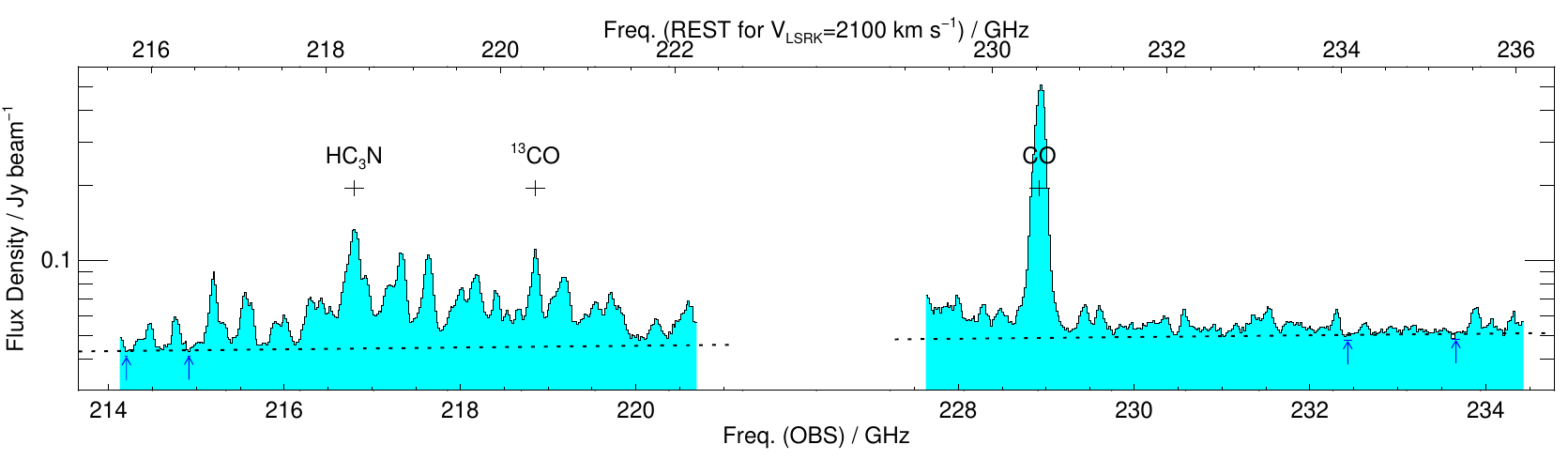} % .eps} 
\end{center}
\caption{ \label{f.spec_with_cont.ylog.N4418}
Sections of the spectrum of the NGC 4418 nucleus in a 0\farcs35 beam.
The spectral resolution is 20 MHz in all bands, but the Band 9 data are presented with 2-channel (40 MHz) binning.
Dotted lines are the power-law functions that we adopted for continuum subtraction.
Blue arrows indicate the local minima that we used to determine the continuum.
Black horizontal bars below the Band 9 spectrum indicate the channels used for our continuum image in Figure \ref{f.B9cont}.
The most prominent lines are labeled along with horizontal bars of a width of 300 \kms.
}
\end{figure*}

\begin{figure*}[!h]
\begin{center}
\includegraphics{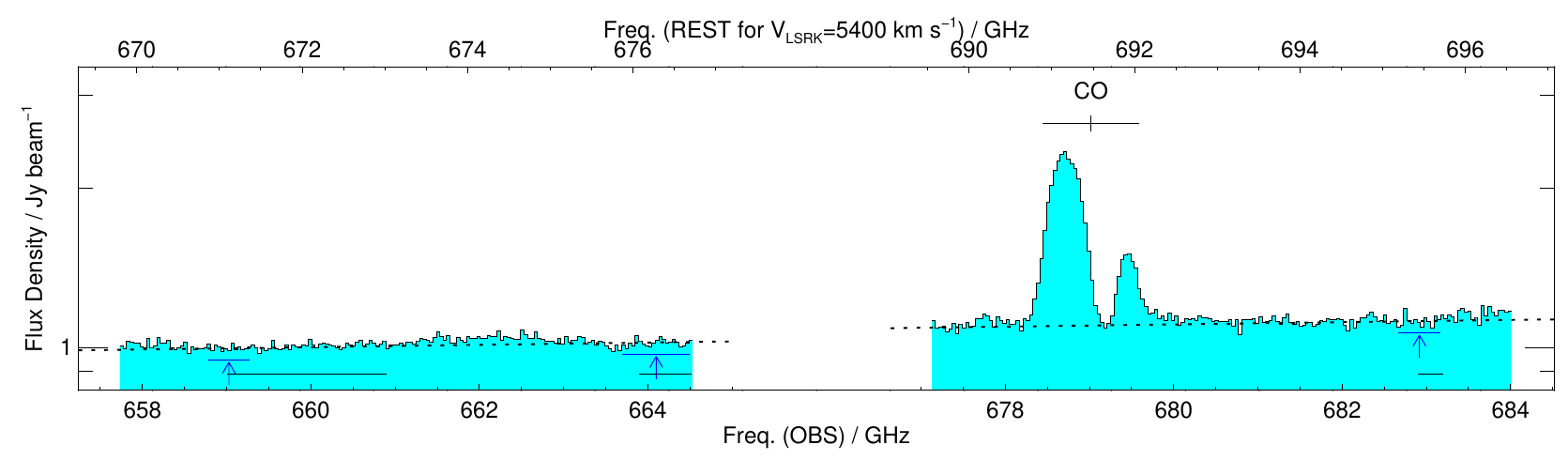} % .eps} 
\includegraphics{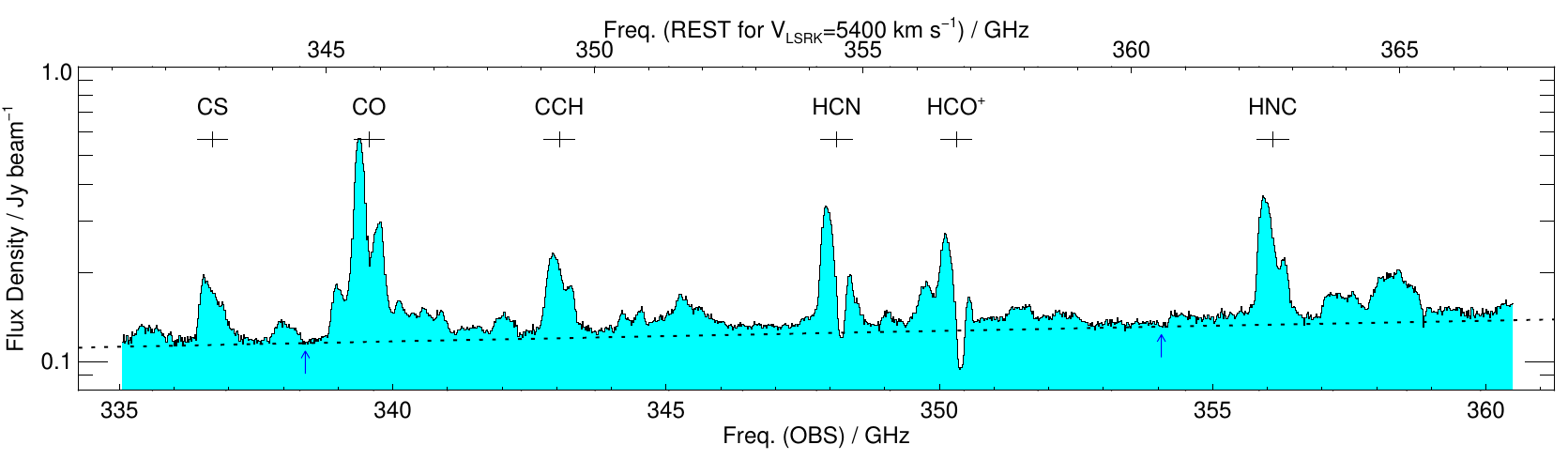} % .eps}   
\includegraphics{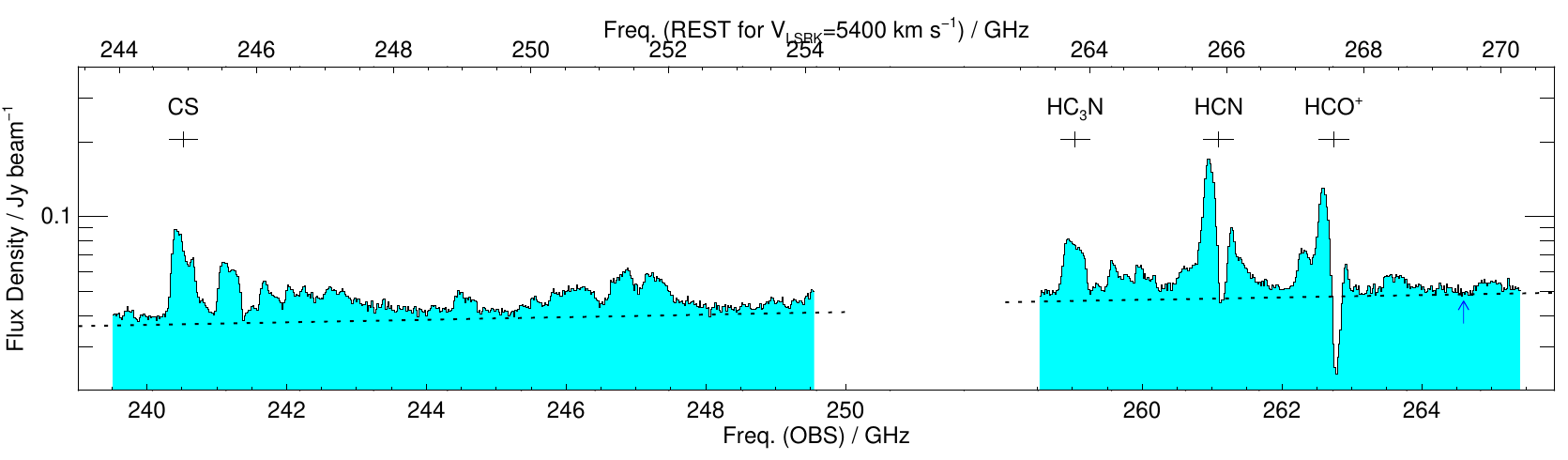} % .eps}   
\includegraphics{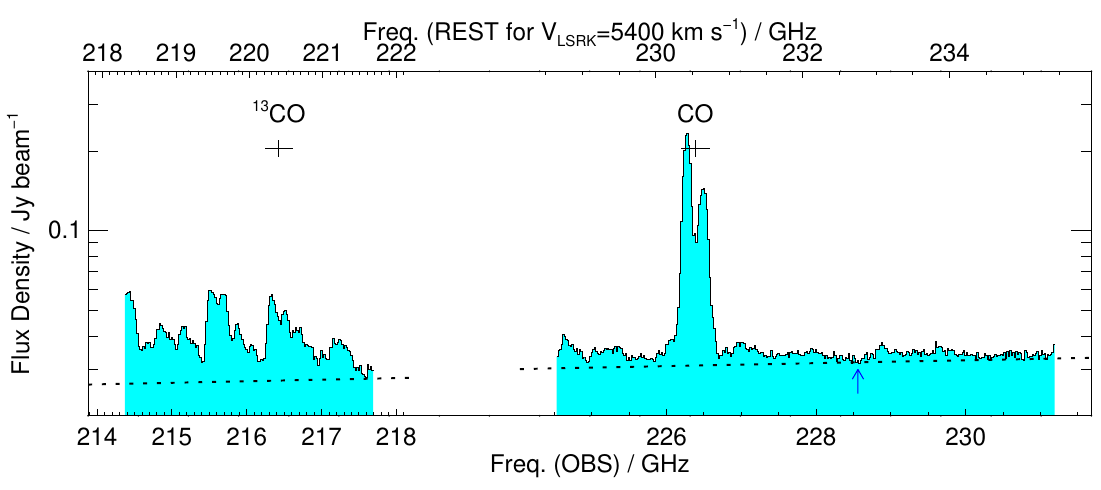} % .eps} 
\end{center}
\caption{ \label{f.spec_with_cont.ylog.A220E}
Sections of the spectrum of the Arp 220 East nucleus in a 0\farcs35 beam.
The spectral resolution is 20 MHz in all bands, but the Band 9 data are presented with 2-channel (40 MHz)  binning.
Dotted lines are the power-law functions that we adopted for continuum subtraction.
Blue arrows indicate the local minima that we used to determine the continuum.
Black horizontal bars below the Band 9 spectrum indicate the channels used for our continuum image.
The most prominent lines are labeled along with horizontal bars of a width of 500 \kms.
}
\end{figure*}

\begin{figure*}[!h]
\begin{center}
\includegraphics{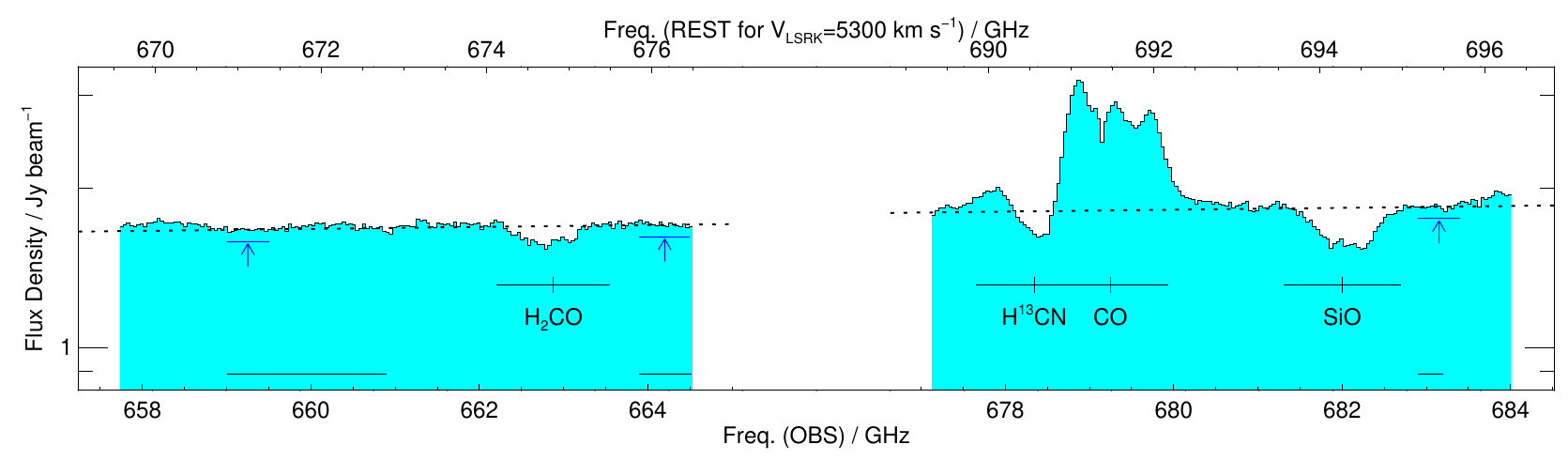} % .eps} 
\includegraphics{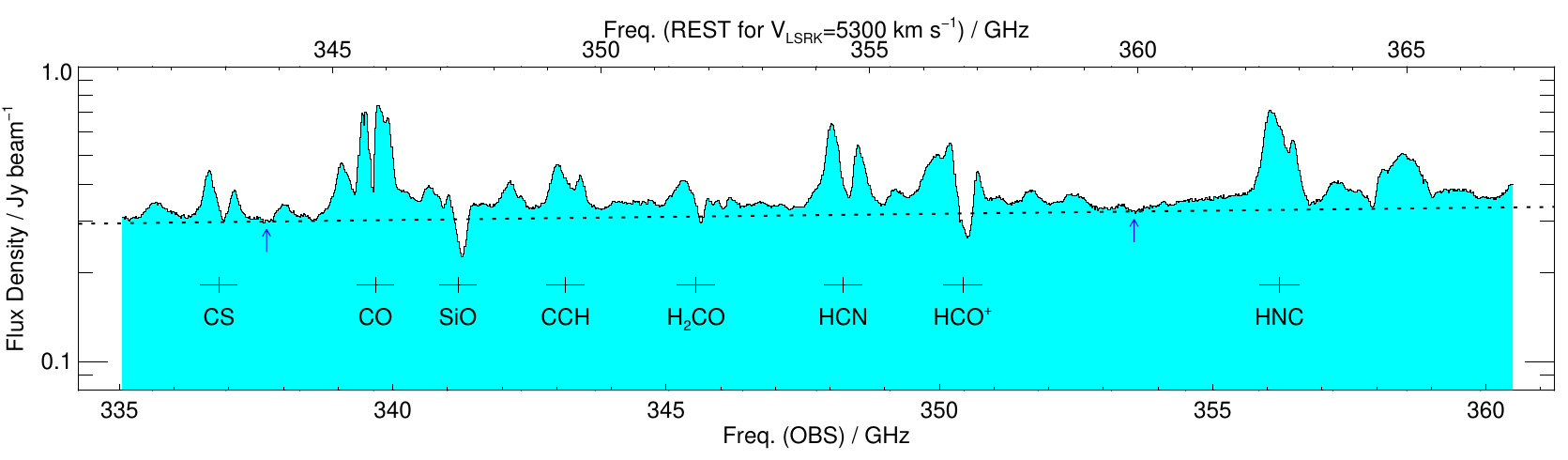} % .eps} 
\includegraphics{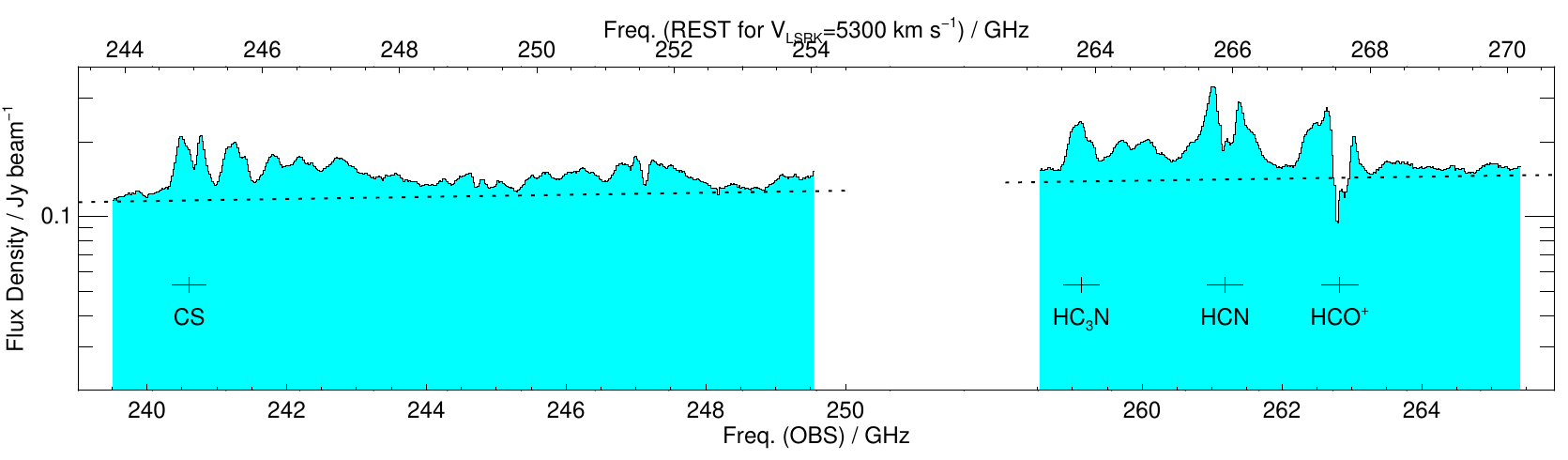} % .eps} 
\includegraphics{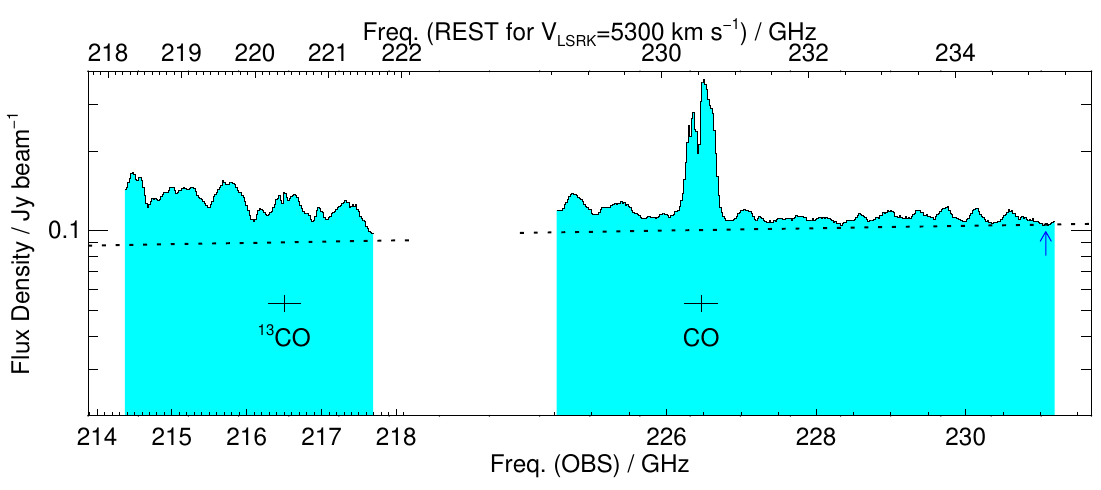} % .eps}  
\end{center}
\caption{ \label{f.spec_with_cont.ylog.A220W}
Sections of the spectrum of the Arp 220 West nucleus in a 0\farcs35 beam.
The spectral resolution is 20 MHz in all bands, but the Band 9 data are presented with 2-channel (40 MHz)  binning.
Dotted lines are the power-law functions that we adopted for continuum subtraction.
Blue arrows indicate the local minima that we used to determine the continuum.
Black horizontal bars below the Band 9 spectrum indicate the channels used for our continuum image.
The most prominent lines are labeled along with horizontal bars of a width of 600 \kms.
}
\end{figure*}

\begin{figure}[t]
\begin{center}
\includegraphics[height=65mm]{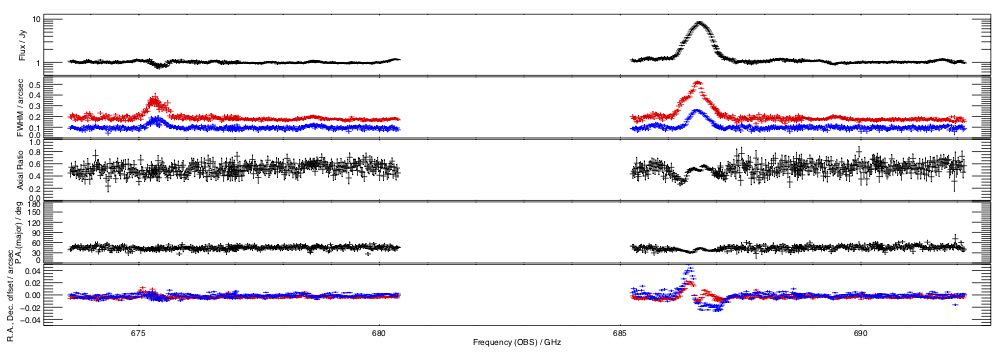} % .eps} 
\includegraphics[height=65mm]{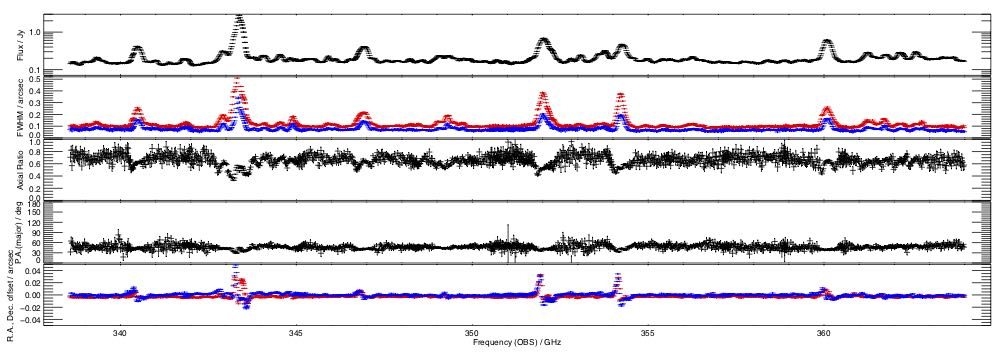} % .eps} 
\end{center}
\caption{ 
Spectra of fitted visibility parameters. 
(a; this sub Figure) NGC 4418 in Band 9 and 7, 
(b) NGC 4418 in Band 6,
(c) \arpE\ in Band 9 and 7,
(d) \arpE\ in Band 6,
(e) \arpW\ in Band 9 and 7,
and
(f) \arpW\ in Band 6. 
[Sub-figures (b)-(f) are online-only.]
Each nucleus was fitted using a single Gaussian, simultaneously for the two nuclei of Arp 220.
Five panels for each frequency segment show total flux density, 
the major (red) and minor (blue) axis FWHM, 
minor-to-major axial ratio,
the position angle of the major axis, 
and 
R.A.~(red) and Dec.~(blue) offsets from the position used for self-calibration.
Error bars are $\pm 1\sigma$.
}
\label{f.uvfit.1G}
\end{figure}

We first fitted each nucleus with an elliptical Gaussian. 
We refer to it as the 1G fitting and perform it on the calibrated visibilities. 
It provides the total flux density, centroid position, and shape of the source. 
The two nuclei of Arp 220 were simultaneously fitted.
We fitted data in every 20 MHz channel using the Levenberg-Marquardt algorithm implemented in IDL as
{\sf mpfit} \citep{More77,More93,Markwardt09} and verified the results with {\sf uvmultifit} \citep{uvmultifit14}.
Figure \ref{f.uvfit.1G} shows the fitted parameters as functions of frequency.
Since we fitted all the channels, these parameter spectra contain bright lines.
A spectrum of total flux density is similar to the 0\farcs35 spectrum of the same nucleus in
Figures~\ref{f.spec_with_cont.ylog.N4418}--\ref{f.spec_with_cont.ylog.A220W},
as it should be for a compact emitter.

There are two evident trends in the spectrum of the source size in FWHM.
The first is that all sources are larger in the line channels than in channels without lines.
In other words, these nuclei look more extended in line emission than in continuum emission.\footnote{
Absorption lines, such as the CN absorption at \fobs \about 675.4 GHz toward NGC 4418, also have larger FWHM than continuum in Fig.~\ref{f.uvfit.1G}. 
They do so because a line absorption toward a compact continuum nucleus is usually accompanied by 
circumnuclear emission of the same line (see \citest{Paper2} for line images). 
A line can have deep absorption to dominate the flux while its extended emission largely determines the fitted size.}
It means that we have to avoid line contamination as much as possible to measure the source size 
in continuum radiation.
The second interesting trend is that the spectrum of FWHM is similar to the flux spectrum for each nucleus,
indicating that brighter lines tend to be larger in spatial extent. 
We discuss this trend in the line analysis section of \citest{Paper2}. 
In addition to these two trends, our parameter spectra show that  
the position angle of the major axis and the minor-to-major axial ratio are almost constant at channels off the strong lines.
For each nucleus, the mean centroid position is the position we used for self-calibration. 
The deviations from the mean should reflect gas motion seen through the lines.

Table \ref{t.cont_visfit_1G_params} lists the average values of the fitted parameters in continuum-dominated channels.
We defined a spectral channel as being continuum-dominated 
when its intensity in our 0\farcs35 spectrum is less than a threshold.
The threshold for Bands 6 and 7 is at 105\% of our adopted continuum spectrum. 
The dotted lines in Figures~\ref{f.spec_with_cont.ylog.N4418}--\ref{f.spec_with_cont.ylog.A220W} are the continuum spectra. 
We set the threshold for Band 9 at 101\% of the continuum. 
In other words, the ``continuum-dominated'' channels should have more than 95\% or 99\% of their
signal from continuum emission.

In Bands 6 and 7, the nucleus of NGC 4418 has a major axis in FWHM of \about0\farcs10 (17 pc) and 
an equivalent (i.e., $\sqrt{\rm{major}\times\rm{minor}}$) FWHM of \about0\farcs07 (12 pc), 
with a (minor-to-major) axial ratio of \about0.6.
The nucleus of Arp 220E has a mean major axis of 0\farcs26 (110 pc) and a mean equivalent size of 0\farcs20 (80 pc), both in FWHM.
Those for the nucleus of Arp 220W are 0\farcs15 and 0\farcs14, or about 60 pc.
These are in excellent agreement with our previous measurements of the three nuclei with 
the SubMillimeter Array (SMA) at 345 GHz \citep{Sakamoto08,Sakamoto13}.

In Band 9, ours is the first direct measurement of the continuum size of the NGC 4418 nucleus around 700 GHz. 
It has an equivalent FWHM of \about0\farcs13 (21 pc) and an axial ratio of \about0.5.
The E and W nuclei of Arp 220 have equivalent sizes of 
0\farcs25 (103 pc) and 0\farcs19 (77 pc), respectively.  
Their axial ratios are similar to those in the lower bands. 
The parameters of Arp 220 are consistent with those reported by \citet{Wilson14}, 
except that we better resolved the minor axes with our higher angular resolution.

It is noteworthy that all the three nuclei appear larger in Band 9 than in Bands 6 and 7. 
It is also notable that the peak brightness temperature of the 1G fitting shows different trends among the three nuclei.
\arpE\ has its brightness temperature higher in Band 9 than in Bands 6 and 7,
while \arpW\ and NGC 4418 are less bright in Band 9 than in Band 7. 
While the $-34$\% change of peak \Tb\ from Band 7 to Band 9 in NGC 4418 might be affected much by errors in absolute flux calibration, the difference between the two Arp 220 nuclei ($+59$\% in E and $-14$\% in W from Band 7 to Band 9) is robust because the two nuclei share calibration. We interpret these observations in Section \ref{s.spatial.peakTb}.

We adopted the 1G continuum model parameters in Table \ref{t.cont_model-1G_parameters}
from the fitting results in Table \ref{t.cont_visfit_1G_params}.
We divided our frequency coverage into three or four sections, as we did to determine the local continuum spectra,
and assumed  constant structural parameters within each section.
In each section, we computed from the size and shape parameters the coupling efficiency 
of the source emission to a 0\farcs35 beam. 
To do so, we made in CASA an image of the 1G model source with a flux density of 1 Jy 
and then convolved it with the target beam. 
The peak intensity in Jy \perbeam\ is the coupling efficiency.
The efficiency $\eta_{\rm 1G}$ is as high as about 0.95 for NGC 4418
because the nucleus is small compared to the beam,
while it is as low as 0.74 for the more extended Arp 220E. 
Applying this correction to Eq.~(\ref{eq.cont-intensity}), we obtained the following 
for the source-integrated (i.e., total) continuum flux density,
\begin{equation}
	S_{\nu}  
	=
	S_{\rm ref}^{(\rm 1G)} \times \left( \frac{\nu}{\nu_{\rm ref}} \right)^{\alpha},
\end{equation}
where $S_{\rm ref}^{(\rm 1G)} = I_{\rm ref}/\eta_{\rm 1G}$ is in Table \ref{t.cont_model-1G_parameters}.

%%%%%
\subsubsection{r2G fit}
\label{s.cont.sizeShape.r2Gfit}
We further fitted each nucleus using two restricted Gaussians to describe the structure better.
The fitting procedure is as follows.
Using visibilities in the most continuum-dominated channels, 
we fit the target nucleus with two concentric and homologous Gaussians,
i.e., elliptical Gaussians sharing the center, axial ratio, and major-axis position angle.
The shared parameters are fixed at the 1G model values in Table \ref{t.cont_model-1G_parameters}.
These restrictions are appropriate for our data that marginally resolve the individual nuclei.
We refer to it as the restricted-2G fit, or r2G fit for short.
The channels we used have an intensity excess of less than 1\% over the model continuum spectrum.
We fitted the nuclei of Arp 220 one at a time. 
When fitting one, we subtracted the other by using its 1G model so that the visibility data contain a single nucleus. 
We also excluded the baselines shorter than 0.2 M$\lambda$, which corresponds to 1\arcsec. 
It suppresses subtraction residuals.
The fixed axial ratio and position angle of a nucleus define a family of similar ellipses in the \uv\ domain.
An ellipse in the family has the semi-minor axis of 
\begin{equation}
	\rho 
	\equiv
	\sqrt{
		\left(
			v \cos \phi + u \sin \phi
		\right)^2
		+
		q^2
		\left(
			-v \sin \phi + u \cos \phi
		\right)^2
	}, 
\end{equation}
where $\phi$ is the position angle of the major axis, 
and $q$ 
is the minor-to-major axial ratio, both in the image domain.
We phase-shift visibilities so that the phase center is at the target nucleus.
The visibility of the double Gaussian is then a function of $\rho$:
\begin{equation}
	V(\rho)
	=
	\sum_{i=1}^2	
	S_{\nu}^{(i)} 
	\exp \! 
	\left[ 
		- \frac{(\pi \theta_{\rm maj}^{(i)} \rho )^2}{4 \ln 2} 
	\right],
\end{equation}
where $S_{\nu}^{(i)}$ and $\theta_{\rm maj}^{(i)}$ are the total flux density and the image-domain FWHM, respectively,
of the $i$-th Gaussian. 
In the continuum-dominated channels, 
we averaged the real parts of the complex visibilities in bins of $\rho$ and fitted the outcome with this function.

Table~\ref{t.cont_visfit_r2G_params} and Figure~\ref{f.uvRealFit.2Gauss} show the r2G fitting results.
Figure~\ref{f.uvRealFit.2Gauss} shows that, in Band 7, 
none of the three nuclei is fitted well with a single Gaussian. 
Specifically, the data have too large amplitudes at large $\rho$ compared to the 1G models shown as blue dashed lines. 
It indicates the presence of a compact core besides the more extended source fitted with the single Gaussian.
(The more compact a Gaussian is, the slower its visibility amplitude declines at longer baselines.)
Indeed, each of the three nuclei is fitted much better with two Gaussians;
we show them as dotted lines in magenta and their sum as a red line in each panel.
The same appears true in Band 6 though our limited baseline lengths make it more difficult to
separate two components in each nucleus, particularly for NGC 4418.
In Band 9, the failure of the single-Gaussian fitting is again evident in NGC 4418 and Arp 220E, but not so in Arp 220W.
The smaller components have major-axis FWHM (in Band 7)  in the range of 0\farcs05--0\farcs10 (8--40 pc).
The other fitting components are \about0\farcs2--0\farcs4 in FWHM.
It is evident in the Band 7 fits that the fractional contribution of the more compact component
is higher in NGC 4418 (\about70\%) than in the Arp 220 nuclei (\about25\%). 
In other words, among the two Gaussians shown as magenta dotted lines, the one having less falloff at longer baselines
has a higher fraction in total flux (i.e., in the amplitude at $\rho=0$) in NGC 4418 than in the two nuclei of Arp 220.
We also observe in Figure~\ref{f.uvRealFit.2Gauss} and Table~\ref{t.cont_visfit_r2G_params}
that the more extended components contribute to the total flux densities more at higher frequencies.

%%%%%
\subsubsection{Continuum Images at $\lambda \sim 0.44$ mm}
\label{s.cont.B9images}
We made continuum images of the three nuclei in Band 9 for their direct observations in the image domain.
The choice of Band 9 was because our spectra in Figures~\ref{f.spec_with_cont.ylog.N4418}--\ref{f.spec_with_cont.ylog.A220W} show more continuum dominance and fewer lines in Band 9 than in lower bands.
We used the channels marked with black bars in the Band 9 spectra. 
At these channels,  the excess of intensities over our continuum spectrum (dotted lines) is 3\% or less in the 0\farcs35 spectra. 
We also checked our visibility fitting in Section \ref{s.cont.sizeShape.1Gfit}
and the low-resolution spectra in Appendix~C to decide these channels.
The total bandwidth used for the continuum imaging is 3.1 GHz for NGC 4418 and 2.9 GHz for Arp 220.
The imaging used multi-frequency synthesis with two terms.

The Band 9 continuum images are the ones in Figure~\ref{f.B9cont}. 
The continuum emission very strongly concentrates toward the three nuclei.
This compactness justifies our visibility fitting with only a few concentric components.
In addition, the high dynamic-range images reveal faint extended features around the compact peaks.
Most notable is the one in Arp 220W extending in P.A.\about$-15\arcdeg$. 
A similar elongated emission was seen in the 3 mm continuum and  was attributed to
a bipolar outflow from the nucleus by \citet{Sakamoto17}. 
Extended emission around \arpE\ 
is also similar to that seen at 3 mm.
In NGC 4418, the faint extended emission around the central peak has an extent of about 1\arcsec\ at 
our sensitivity.  
Its elongation in P.A. \about 45\arcdeg\ is similar to that of molecular line emission in 
\citet{Sakamoto08} and \citet*{Paper2}.

Our continuum(-dominated) images may well have more line contamination in the extended low-level features 
than toward the central peaks because we selected our `continuum' channels primarily on those peaks.
However, the faint and extended emission is not due to a few particular lines because it is visible in
images made with various subsets of our continuum-dominated channels.
Therefore, it is most likely that the extended features are also predominantly due to continuum emission.

We also made continuum images in other bands, 
but they were not sensitive enough to reveal extended emission or new features. 
It is because of the small bandwidths available for continuum imaging in Bands 6 and 7.
Therefore, visibility modeling with simple functions is adequate for the continuum in those bands.

%%%%%
\subsubsection{Multi-component Visibility Modeling for Band 9 and Band 3} 
\label{s.contB9visfit}
We characterized the features in the Band 9 continuum images through visibility fitting with 
more parameters, taking advantage of the high continuum sensitivity.
We used three Gaussians for NGC 4418 and \arpE. 
We assign one to the main component (similar to the one in the 1G fitting).
Another is for the core component (seen in the r2G fitting). 
The third is for the extended emission.
We used two Gaussians for \arpW\ since it did not show a significant core component in the r2G fitting.
These Gaussians, five in total for Arp 220, were simultaneously fitted to the visibilities used for our continuum imaging.
Each Gaussian had six parameters, 
one for the total flux, 
two for the position, 
two for the major-axis and minor-axis sizes, 
and one for the position angle of the major axis.

Table \ref{t.B9cont.multiGfit} has the fitted parameters.
Cautions are due against the over-interpretation of the multi-component decompositions.
The multiple components in each fitting approximately reproduce the observations only in total. 
There are many ways to fit the same data by using, for example, functions other than Gaussian. 
Nevertheless, it is notable that \arpW\ is decomposed to the main component 
elongated along the position angle of 84\degr\ and the extended outflow component elongated along 163\degr;
their flux ratio is about 2 to 1. 
Judging from the residual image of this fit, 
the core component of \arpW, if any in Band 9, should have less than 1\% of the total flux of the nucleus.
This fit also reproduced our observation in the r2G fitting that NGC 4418 has a more contribution of the core component than 
\arpE. 
The size and flux of the core components in NGC 4418 and \arpE\ are also approximately reproduced.
Interestingly, the core and main components are misaligned by about 40\degr\ and 80\degr\ in their position angles, 
respectively in NGC 4418 and \arpE; the formal uncertainties of the position angles are $\pm10\degr$ for
the cores and much less for the main components (i.e., nuclear disks).

For comparison, we extended the visibility fitting of the 3 mm (Band 3) continuum in Arp 220 by \citet{Sakamoto17}
to six Gaussians, i.e., a compact core, the main component (or nuclear disk), and an extended emission for each nucleus.
We previously used only up to four Gaussians; they were for the cores and the main components.
Table \ref{t.B3cont.multiGfit} shows the new fitted parameters. 
The core and main components did not change their shapes much from the previous Band 3 fitting \citep[see Table 1]{Sakamoto17},
and they are almost similar to the corresponding components (if any) in Band 9.
An exception is that the apparent misalignment of the core and main components is not 
in the Band 3 continuum emission of Arp 220E.
The extended emission components newly fitted in Band 3 are minor contributors to flux densities in individual nuclei,
and their shapes and orientations are similar to those in Band 9.

In particular, the extended bipolar component in Arp 220W has a major-axis FWHM and position angle 
in excellent agreement with those in Band 3. 
It has an average FHWM of about 0\farcs5 and a position angle of 165\degr, 
while its flux ratio between Band 9 and Band 3 is about 260.
This ratio suggests that the Band 3 bipolar feature is mostly plasma  (i.e., free-free and synchrotron) emission 
under a robust assumption that the dust outflow in Band 9 is optically thin. 
The bipolar structure in Band 3 is too bright if it were thermal dust emission. 
This bipolar feature must correspond to the one that \citet{Varenius16} found in their 150 MHz image 
because both have about the same position angle.
Around \arpE, both fittings in Bands 3 and 9 find an extended feature whose major-axis 
position angle is about \plus15\degr,
which makes an angle of about 40\degr\ with the major axis of the eastern nuclear disk.
Its Band 9 to Band 3 flux ratio of about 180 suggests that this 3 mm feature is also plasma emission.
Unlike in \arpW, this 3 mm plasma emission and the 0.4 mm dust emission do not align with
the plasma feature that \citet{Varenius16} saw at 150 MHz with the north-south elongation.

%%%%%%%%%%
\subsection{Model-based Continuum Subtraction in Line-Image Cubes}
\label{s.cont.contSub}
We used our continuum models to subtract continuum emission from our line image cubes.
In Bands 6 and 7, we mainly used our r2G model. 
In Band 9, our model is based on the multi-Gaussian fitting in Section \ref{s.contB9visfit}.

Table \ref{t.cont_model-r2G_parameters} lists the r2G-model parameters. 
Only in Band 6 and for NGC 4418 we used the 1G model in Table \ref{t.cont_model-1G_parameters} 
because we had little long-baseline data to be affected by the compact component in the r2G model.
For NGC 4418 in Band 7, we adopted an r2G model that averaged our r2G fits in the band.
For \arpE, we adopted the average values of the r2G fitting results above 260 GHz for the three r2G-specific parameters, 
namely the fractional contribution of the compact component and two major-axis sizes.
The frequency range is where we could distinguish the two components.
For \arpW, we used averages of the r2G fits in Band 7 and Band 6 separately. 
In Band 6, we had another compromise to average the fits using the data having up to 3\% line contamination instead of 1\%.
Again, while this made the model parameters less accurate in Band 6, 
there are fewer long-baseline data in this band and affected by it. 
For each of the adopted r2G models, we computed the source-beam coupling efficiency, $\eta_{\rm r2G}$.
We used it to obtain the model total flux density of the continuum emission as 
\begin{equation}
	S_{\nu}  
	=
	S_{\rm ref}^{(\rm r2G)} \times \left( \frac{\nu}{\nu_{\rm ref}} \right)^{\alpha},
\end{equation}
where $S_{\rm ref}^{(\rm r2G)} = I_{\rm ref}/\eta_{\rm r2G}$.  
Table \ref{t.cont_model-r2G_parameters} also has $\eta_{\rm r2G}$ and $S_{\rm ref}^{(\rm r2G)}$.

Our Band 9 model used the best-fit parameters in Table \ref{t.B9cont.multiGfit} with a minor flux scaling.
The scaling is for each nucleus to have the power-law spectrum 
in agreement (at 0\farcs35 resolution) with the observed continuum spectra,
i.e., those in Columns 3 and 4 of Table \ref{t.cont_model-1G_parameters}.

Our model-based continuum subtraction is distinct from the conventional method.
In the latter, one has many line-free channels in the same tuning and subtracts their average
or polynomial interpolation in frequency.
In contrast, the model-based method enables continuum subtraction in a spectral scan 
even when a tuning has no line-free channel.
It is made possible by importing continuum information from nearby tunings that have continuum-dominated channels.
The conventional method, when applied to data with a line forest, tends to overestimate the continuum.
Hence it tends to over-subtract emission in the `continuum-subtracted' line images.

One may wonder whether we should have used functions other than the Gaussians to decompose
the continuum nuclei for better modeling and hence subtraction of the continuum. 
Different multi-component fittings are certainly possible for the data in Figure~\ref{f.uvRealFit.2Gauss}, for example.
However, what matters for continuum subtraction is not the form of the individual components (dotted lines in magenta) 
but their sum (red line).
In this regard, there is not much room to improve the r2G models in Fig.~\ref{f.uvRealFit.2Gauss}
by using other functional forms.

Nonetheless, a drawback in our implementation of the model-based continuum subtraction is due to our continuum modeling.
All methods of continuum subtraction pass any imperfection of the adopted continuum models 
to the continuum-subtracted line data.
Ours has limited accuracy in our continuum models
because they use a small number of parameters to simplify the import of continuum information across tunings. 
An alternative, more complex method may use CLEAN components from the imaging of continuum-dominated channels 
as a conduit for continuum information.

%%%%%%%%%%%%%%%%%%%%
\subsection{Peak Intensity}
\label{s.cont.peakI}
The peak intensity of the continuum in the image domain is expressed as follows for the Gaussian models.
\begin{equation}
	I_{\nu}^{(1G)}(0) 	= 
	S_{\rm \nu}^{(0)}
	\frac{4 \ln 2}{\pi \theta_{\rm maj} \theta_{\rm min}}
\end{equation}
\begin{equation}
	I_{\nu}^{\rm (r2G)}(0) 
	= 
	\sum_{i=1}^2
	I_{\nu}^{(i)}(0) 	
	= 
	\sum_{i=1}^2
	S_{\rm \nu}^{(i)}
	\frac{4 \ln 2}{\pi \theta_{\rm maj}^{(i)} \theta_{\rm min}^{(i)}}
\end{equation}
We calculated the peak intensities and listed their equivalent Rayleigh-Jeans brightness temperatures in 
Tables \ref{t.cont_model-1G_parameters} and \ref{t.cont_model-r2G_parameters} for our 1G and r2G models.
The peak continuum brightness temperatures (\Tb) in the 1G models are \Tb\ \about\ 200 K for NGC 4418 and \arpW\
and 25--50 K for \arpE;
Table \ref{t.cont_visfit_1G_params} has the individual 1G fitting results.
These are consistent with the earlier SMA measurements at 0.86 mm in  \citet{Sakamoto13,Sakamoto08}. 
NGC 4418 and \arpW\ have even higher peak \Tb\ of about 400--500 K in our r2G model 
because of their significant compact components found in the two-component fitting.
Our analysis of ALMA 3~mm data also indicates for the core of \arpW\ a peak \Tb\ of 
about 500 K and 630 K respectively from 2 and 3-Gaussian models \citep[and Table \ref{t.B9cont.multiGfit}]{Sakamoto17}.

%%%%%%%%%%%%%%%%%%%%
\subsection{Spectral Index}
\label{s.cont.spix}
Our continuum models so far are local, as they approximate
the observed continuum emission only in individual spectral segments. 
The model spectral indices (in Column 4 of Table \ref{t.cont_model-1G_parameters}) have significant uncertainties
due to the small fractional bandwidths of the frequency sections. 
For NGC 4418 and \arpE\ having comparable continuum strengths,
we estimate the 1$\sigma$ uncertainties to be as large as 0.4 in Band 7 (while smaller at lower frequencies).
The local models are sufficient for continuum subtraction in the individual spectral segments with small fractional bandwidths.
But a more precise shape of the continuum spectrum (i.e., accurate spectral index) tells us more about each nucleus.

%%%%%
\subsubsection{Spectral Indices at $\lambda \sim 1$ mm of the Three Nuclei}
\label{s.cont.spix.at1mm}
We obtained better estimates of the spectral indices at $\lambda \approx 1$ mm by combining all our data
in Bands 6 and 7, namely, $\alpha = 2.36 \pm 0.08$ for NGC 4418, $2.67 \pm 0.10$ for \arpW,
and $3.28 \pm 0.09$ for \arpE. See Table~\ref{t.contSpectra} for other parameters.
Since the data are from multiple frequency segments not directly calibrated against each other with overlapping channels,
the estimated spectral indices are affected by the absolute errors in flux calibration.
Its magnitude for our data is poorly known except for some general descriptions from the observatory. 
 In addition, any line contamination at our local spectral minima appends error to our flux measurements there. 
 While we argue in Section \ref{s.cont.Caveat} that the degree of line contamination should be small, 
 we expect it to be variable among our continuum data points and do not know its magnitude at every data point.
Therefore, we assumed a constant fractional error for all data points 
and scaled it so that the reduced $\chi^2$ of the power-law fit is unity. 
The fractional errors estimated this way turn out to be 3--4\%. 
A part of the reason why it is better than the standard \about10\% accuracy of flux calibration must be 
our flux self-calibration in the individual spectral segments. 
The small fractional errors also suggest that line contamination is minor in our continuum measurements;
otherwise, random degrees of line contamination should cause large scatter in our measurements.

%%%%%
\subsubsection{Additional Data for NGC 4418}
In the left panel of Figure \ref{f.contSpecGlobal} is the continuum spectrum of the nucleus of NGC 4418.
ALMA data from this work (red) and our earlier measurements in \citet[blue]{Costagliola15} nicely align.
The angular resolution was \about1\arcsec\  and \about0\farcs8 in Bands 6 and 7, respectively,
for the latter measurements.
Hence they are not corrected for the source-beam coupling efficiency.
The power-law fitting to the combined data found the spectral index of $\alpha = 2.35 \pm 0.08$ at $\nu \sim 300$ GHz.

The ALMA spectral index is marginally consistent with our earlier SMA value of $2.55 \pm 0.18$ 
at virtually the same frequency  \citep{Sakamoto13}.
However, the SMA flux measurements and those from single-dish bolometer observations in Fig.~\ref{f.contSpecGlobal}(left)
have higher flux densities than the ALMA spectrum by about 25\%.
The larger flux densities and a slightly shallower spectral slope 
are probably due to line contamination and extended emission in the bolometer and SMA data, 
although the latter excluded bright lines from continuum measurements. 
Any difference in flux standards among observatories would also contribute to the discrepancy.

The 98 GHz data point in Fig.~\ref{f.contSpecGlobal}(left) is from ALMA and is above the extrapolation of 
our best power-law fit around 300 GHz.
The likely curvature of the continuum spectrum was already noted in \citet{Costagliola15} 
and is consistent with a more contribution of plasma emission at lower frequencies.

%%%%%
\subsubsection{Comparison with Band 9 Data}
The right panel of Fig.~\ref{f.contSpecGlobal} shows the continuum spectra of the three nuclei 
measured with ALMA in Bands 3, 6, 7, and 9. 
The dotted lines are the power-law spectra that best fit our measurements between 200 and 400 GHz.  
The steeper spectral slope of \arpE\ than \arpW\ in the frequency range is supported
by the Band 9 data at 670 GHz. 
However, the Band 9 flux of the NGC 4418 nucleus is not consistent with the simple extrapolation of the 200--400 GHz spectrum,
as we already noted in Section \ref{s.cal.B9calibration}.
Considering the difficulty of Band 9 calibration, 
we did not include our Band 9 data in our spectral fitting for the continuum spectral indices of the three nuclei.
We use our spectral indices at $\nu \sim 300$ GHz to compare with models in Section \ref{s.model}.

%%%%%%%%%%%%%%%%%%%%
\subsection{Limitations in Continuum Measurements}
\label{s.cont.Caveat}
Our continuum measurements in this section have at least two limitations.
One is on the continuum definition. 
The other is about the structure that is either too large or too small for our detection or modeling. 
We described our observations only in such basic shapes as up to a few concentric Gaussians 
even though no structural components in the nuclei have a priori reason to be Gaussian. 
It is related to the latter limitation.

Our operational definition of `continuum' in the spectra only ensures that the emission exceeding it is line emission.
The `continuum' may be partly from molecular line transitions
because emission lines may fill the frequency space. 
Indeed, the `continuum' defined through spectral minima has more flux density 
when we smooth the spectrum to mimic broader lines. 
It implies that the `continuum' would be weaker if the lines were narrower than they are.
In addition, we may have overestimated the `continuum' by missing 
lower local minima outside our frequency coverage. 
Despite those, Fig.~\ref{f.uvfit.1G} clearly shows
that the emission centroid stays still across frequency except on the channels with bright lines. 
It is as expected when the emission is continuum-dominated in most spectral channels (i.e., away from bright lines).
The exact magnitude of line contamination in our continuum is hard to decide through spectral modeling (e.g., the one in \citest{Paper2}) given uncertainties in chemical abundance and excitation. 
However, we estimated in Section \ref{s.cont.spix.at1mm} that it must be less than 3--4 \%. 
The actual contamination must be much smaller than this upper limit unless our flux calibration was perfect. 
For these reasons, we regard that most of the continuum we decided should be genuine.
In other words, it is an emission that changes in flux and spatial distribution very slowly over frequency, 
such as the ones from dust or plasma. 
There should be inevitable contamination by low-level line emission at every frequency in our three galactic nuclei.
As we summarized in Table \ref{t.lineFraction}, line contribution to the total flux in our spectra is significant, 
at the level of about 20--30\% in Bands 6 and 7.
This high-level bias can be in continuum measurements around $\lambda = 1$mm if one completely ignores lines.
Therefore, spectroscopy and line-removal are essential for accurate continuum measurements.
Our analysis procedures are our best attempt to minimize the bias.

We miss large-scale emission much more extended than our observing beam because
we modeled the three nuclei as objects whose continuum intensities monotonically and quickly decrease 
with an increase in radius. 
Arp 220 has a structure of at least several arcsec in size to encompass the two nuclei. 
NGC 4418 also has circumnuclear CO emission with an extent of \about5\arcsec\ \citep[and \citest{Paper2}]{Sakamoto13}.
Our correction for the source-beam coupling efficiency does not account for such extended emission. 
Indeed, the Band 9 continuum flux density in a 2\arcsec-diameter aperture toward Arp 220 
(Appendix \ref{a.lowResSpectra}) is about 10\% larger than the sum of the flux densities of the two nuclei.

We also miss small-scale structures that our simple models cannot describe.
For example, all our models have point symmetry. Hence they cannot capture any features that are not centrally symmetric.
In addition, observations at higher resolution should tell us about the radial distribution of continuum intensity
more accurately than the current data.
\citet{Sakamoto17} found in the 0\farcs05 data of the $\lambda = 3$ mm continuum of Arp 220
asymmetric features in each nucleus after removing the best-fit elliptical Gaussian.
We also found that a double-Gaussian is superior to an exponential disk 
in fitting the radial continuum distribution in \arpW, 
even though both have a sharper central peak and longer outer tail than a Gaussian.
We have been unable to confirm these features and properties in our current data.
It is because of the lower angular resolution and the smaller continuum bandwidths. 
The simplified functional forms in our spectral and spatial modeling limit our models 
to what our data warrant. 
In the future, more observations should allow similar analysis at $>$200 GHz and tell us, 
for example, spatial variation of the continuum spectral index within each nucleus.

%%%%%%%%%%%%%%%%%%%%%%%%%%%%%%%%%%%%%%%%%%%%%%%%%%%%%%%%%%%%
\section{Continuum Modeling: Opacities and Column Densities from Spectra}
\label{s.model}
We use radiative transfer models to interpret our continuum observations of the nuclei. 
Beyond a simple analytic model, we employ the BGN models of \citet{GS19},
who simulated buried galactic nuclei using internally heated spheres of gas and dust 
of various total column densities. 
We evaluate the opacity and column density of our target nuclei using their continuum spectra.

%%%%%%%%%%%%%%%%%%%%
\subsection{Spectral Index as Indicator of Opacity and Column Density}  \label{s.model.slab}
One of the fundamental parameters of a buried galactic nucleus is its column density or opaqueness to radiation.
When the opacity is due to dust, its frequency-dependent emissivity (and hence opacity) modulates
the continuum spectrum in a manner determined by the opacity of the system.

One can therefore obtain the opacity as well as column density 
from the spectrum of the dust continuum emission. 
For a simple spectrum, this is possible with the spectral index at a single frequency.
The most basic example is a uniform slab of dust (and gas) with negligible background radiation.
The source flux density at a frequency $\nu$ is
	$
	S_\nu = (1 - e^{-\tau_\nu}) B_\nu(T) \Omega_\nu,
	$
where $\tau_\nu$ is the opacity at the frequency, $B_\nu(T)$ is the Planck function for the temperature $T$,
and $\Omega_\nu$ is the source solid angle.
We assume that the source is warm enough to allow the Rayleigh-Jeans approximation at the observing frequency,
that the apparent source size is independent of $\nu$,
and that the mass opacity coefficient of the dust has a power-law index of $\beta$, i.e., $\tau_\nu \propto \nu^\beta$.
The source spectrum in this case is
	$
	S_\nu \propto (1 - e^{-\tau_\nu}) \nu^2
	$
and its spectral index is
\begin{equation} \label{eq.simple_spix}
	\alpha_\nu 
	= 
	\frac{d \log S_\nu}{d \log \nu} 
	=
	2 + \beta \frac{\tau_\nu}{\exp(\tau_\nu) -1} .
\end{equation}
It monotonically varies between $2 + \beta$ for $0 < \tau_\nu \ll 1$ and 2 for $\tau_\nu \gg 1$
as shown in Figure \ref{f.contModelSpix_slab}.
Therefore, an observed spectral index $\alpha_\nu$ informs the optical depth $\tau_\nu$ for a given $\beta$.
The mass column density of the slab is then obtained from the opacity and a mass opacity coefficient.

For galactic nuclei having \NHH $\geq 10^{24}$ \persquarecm\ of gas and dust,
spectral indices at (sub)millimeter wavelengths are the ones most sensitive to the large column densities.
Indeed, the slab model indicates that the optical depth (and hence column density) is best determined 
when the slope in Fig.~\ref{f.contModelSpix_slab} is the steepest or
when the spectral index is measured around the wavelength where the opacity is unity. 
Therefore, the spectral index at $\lambda \sim 1$~mm should be a good indicator of dust opacity and column density
for $\NHH \about 10^{25-26}$ \persquarecm.

Although conveniently simple, the uniform and isothermal slab is a crude model for a galactic nucleus.
For example, a dust cocoon covering a luminous AGN must be warm inside and cool outside
rather than isothermal. 
The source spectrum, as well as spectral index, depends on the temperature structure.
It is also unclear how the column density obtained from the slab model is related 
to the one between the central AGN and the surface of the cocoon. 
A sphere is the second simplest structure and can better reflect the physical parameters of a nucleus when
its radiative transfer and thermal structure are solved.

%%%%%%%%%%%%%%%%%%%%
\subsection{$\alpha$--$\tau$ Relation from BGN Models} \label{s.model.BGN}
The BGN models allow us to verify the analytic slab model for its basic properties
and relate a radio spectral index $\alpha$ to the opacity $\tau$ and column density 
for a spherical source having radial density and temperature structures.
We use the fiducial BGN model and refer readers to \citet{GS19} for the full description of the modeling.
In short, it solved radiative transfer and thermal energy balance for an internally heated dust sphere 
to derive its internal temperature distribution and emergent continuum radiation. 
(It also simulated line radiation for HCN.) 
The radiative transfer calculation safely ignored scattering 
since the mean free path of the energy-carrying IR photons is much smaller than the source size 
for the opaque nuclei of our interest. 
The self-similar nature of such a dust sphere \citep{IvezicElitzur97} makes the model applicable to a wide range of BGNs. 
The fiducial parameters in the BGN model, in Table 1 of \citet{GS19}, include the dust $\beta$ of 1.6 
and the radial density profile $\propto r^{-1}$ between the outer and inner radii whose ratio is 17.
The temperature structure is simulated for either
a central point source or extended source to mimic AGN and starburst, respectively.
The resulting spectra are virtually identical at $\lambda \gtrsim 30$~\micron\ for the two types of luminosity sources
buried in the same total column of gas and dust, as shown in Figure~2b of \citet{GS19}.
On the one hand, it means that one cannot determine the luminosity source of a BGN from its (sub)millimeter spectral index
of thermal dust emission.
On the other hand, it also means that such an index constrains the total optical depth of a BGN regardless of the nature of the dominant luminosity source.
We fitted the BGN model spectra for spectral indices at various (sub)millimeter wavelengths.

Figure \ref{f.contModelSpix_BGN}(left) compares in the BGN models the spectral index at $\lambda = 1$~mm with the 1 mm
optical depth and column density, both measured from the center to the surface. 
The figure indicates that the spectral index is a function of the optical depth and not the source luminosity.
The dotted curve in Fig.~\ref{f.contModelSpix_BGN}(left) is equation (\ref{eq.simple_spix}) for the analytic slab model.
It agrees with the BGN calculations rather well.
Therefore, the opacity and the column density calculated from the 1 mm spectral index with the simple uniform-slab model 
closely approximates the same quantities measured from the center to the surface in the BGN model 
when the estimated opacity is order of 1.

Figure \ref{f.contModelSpix_BGN}(right) presents the $\alpha$--$\tau$ relation in the BGN model 
at five wavelengths. 
This relation depends little on wavelength in the plotted simulations.
It corresponds to equation (\ref{eq.simple_spix}) being independent of frequency.
We obtained the following approximate formula for the relation by a shift and stretch of (\ref{eq.simple_spix}) :
\begin{equation} \label{eq.bgn_spix}
	\frac{\alpha_\nu - 1.55}{2.0}
	=
	\frac{x}{\exp(x) -1},
	\quad
	\mbox{$x \equiv 0.75\, {\tau_\nu}^{0.84}$}.
\end{equation}
The solid line in Fig.~\ref{f.contModelSpix_BGN}(right) is this function.
Unlike in the slab model, the range of $\alpha_\nu$ for $\tau_\nu \geq 0$ is larger than the $\beta$ of 1.6.
The reason is the following.
In the most opaque cases and at the shortest wavelengths, 
the photosphere (for the wavelengths under consideration) is at large radii and hence at relatively low temperatures of \about100 K.
Therefore, the Rayleigh-Jeans approximation fails, 
and $\alpha_\nu$ approaches shallower slopes of the Planck function. 
Hence, $\alpha_\nu < 2$ for such situations.\footnote{It is expected that the $\alpha$--$\tau$ relation depends
on wavelength at $\tau \gtrsim 10$ since the R-J approximation works better for longer wavelengths.
%having $\alpha$ closer to 2 at longer wavelengths. 
The BGN simulations did not cover such cases of \NHH $> 10^{26}$ \persquarecm. 
The simulation-based formula (\ref{eq.bgn_spix})  is valid only for the parameter range of the simulations. }
In contrast, when the opacity is low, the bulk of emission is from the warmer interior.
Therefore, the slope of the Planck function is close to the R-J limit of 2.
The spectral index $\alpha_\nu$ is the sum of the slope and $\beta$ for the optically thin emission.

%%%%%%%%%%%%%%%%%%%%
\subsection{$\tau$--\NHH\ Relation} \label{s.model.tauToNHH}
Optical depths estimated from spectral indices or obtained by other means can be translated to gas column densities using 
a mass opacity coefficient. 
The $\tau$--\NHH\ relation is independent of source geometry and applicable to both the slab and BGN models.
The fiducial BGN model adopted 
$\kappa_{\rm 1.1\,mm} = 1.2$ \squarecm\, \pergram\ for dust and
a gas-to-dust mass ratio of 100.
Thus the conversion relation is 
\begin{equation}  \label{eq.bgn_nhh-tau}
\frac{\NHH}{\persquarecm} 
= 
\tau_\lambda \times 10^{25.2}  
\left(
	\frac{\lambda}{\text{mm}}
\right)^{1.6}
\end{equation}
for a mass abundance $X = 0.715 \approx 1/1.40$ of hydrogen among all elements \citep{Asplund09,Przybilla08}.
In \citet{GS19}, \NHH\ denoted the mass column density of gas between the inner and outer radii of the sphere
(with a small central cavity) divided by the \HH\ mass; helium and heavy elements in the gas were disregarded for simplicity.
Renaming that quantity as \NHHprime\ in this paper, its relation to the \NHH\ here is
$\NHHprime = \NHH / X \approx 1.40 \NHH$.
For reference, the proton column density in the fiducial BGN model is 
$N_{\rm p}/\persquarecm = \tau_\lambda \times 10^{25.6} (\lambda/{\rm mm})^{1.6}$.

The uncertainty of the \NHH\ from (\ref{eq.bgn_nhh-tau}) arises from the assumed gas and dust properties 
and the error in $\tau$. 
The opacity $\tau$  inherits errors from the $\alpha$--$\tau$ relation and the BGN modeling itself.
To assess the uncertainties in the numerical constants in (\ref{eq.bgn_nhh-tau}), 
we show in Figure \ref{f.tau2NH2} some of the $\tau$--$\NHH$ relations in the literature, in particular those used for Arp 220.
The formula (\ref{eq.bgn_nhh-tau}) adopted in this paper is the red line.
It is close to the relation used in \citet{Wilson14}. 
It gives about three times less column density for the same opacity 
compared to the formulas of \citet{Scoville14} and \citet{Hildebrand83}, 
which were used for Arp 220 in \citet{Scoville17} and \citet{Sakamoto08}, respectively.
Since the relation depends on the poorly constrained dust properties and gas-to-dust mass ratio in our target nuclei, 
the choice of any one of these relations does not seem much better justified than others. 
From the spread in the plot, we caution that the \NHH\ we estimate using the formula (\ref{eq.bgn_nhh-tau})
can be lower by up to a factor of 3 (i.e., by 0.5 dex) than other estimates using a different $\tau$--\NHH\ relation.

%%%%%%%%%%%%%%%%%%%%
\subsection{Opacities and Column Densities of the Three Nuclei}  \label{s.model.tauAndColumn}

%%%%%
\subsubsection{Decomposition of Continuum Emission}   \label{s.model.tauAndColumn.decomposition}
Millimeter-submillimeter continuum emission generally consists of thermal emission from dust and
free-free and synchrotron emission from plasma. 
We need to subtract the plasma emission from the continuum to use the $\alpha$--$\tau$ relation of dust emission.

The spectral index of a multi-component emission is the weighted mean of the spectral indices 
of the components, where the weights are the fractional contributions of the individual components 
to the total flux density. In other words, a composite spectrum
\begin{equation} \label{eq.compositeSpectrum}
	S_\nu 
	= 
	\sum_{i} S^{(i)}_{\nu}
\end{equation}
has the spectral index
\begin{equation} \label{eq.compositeSpix}
	\alpha_{\nu} 
	\equiv
	\frac{d\log S_\nu}{d \log \nu} 
	=
	\frac{\nu}{S_\nu}
	\frac{d S_\nu}{d \nu}
	=
	\sum_{i} \frac{S^{(i)}_{\nu}}{S_\nu} \alpha_{i, \nu} , \\
\end{equation}
where  
$\alpha_{i, \nu} \equiv d\log S^{(i)}_{\nu}/d \log \nu $ is the spectral index of the $i$-th component
and ${S^{(i)}_{\nu}}/{S_\nu}$ is the fractional flux contribution of the same component at $\nu$.
Note that the observed emission of one component, $S^{(i)}_{\nu}$, may be modulated (e.g., extinguished and spectral slope altered) 
by another component depending on their geometry.
We adopt the linear formula (\ref{eq.compositeSpix}) with a simplification that each $S^{(i)}_{\nu}$ is 
independent and unmodulated.

We estimate the fractional contribution of plasma emission to the total $\lambda = 1$ mm flux density
to be about 5\% in the three nuclei.
Table \ref{t.contSpectralAnalysis} lists the fractions in Column 6 and explains details of their derivations.
We used for each source 30--100 GHz data to estimate the plasma continuum there.
We assumed spectral indices of $-0.7$ for synchrotron and $-0.1$ for free-free emission.
We calculated at 300 GHz the total plasma flux density in Column 4 
and the spectral index for the combined plasma emission in Column 5.
The synchrotron to free-free ratio at the lower frequency is either from an observed centimeter-wave spectral index 
or an assumption, but it makes little difference to the final dust opacity.

Spectral indices of dust emission at $\lambda = 1$ mm are in Column 7 of Table \ref{t.contSpectralAnalysis}.
We computed them with (\ref{eq.compositeSpix}) in the following way.
At any frequency $\nu$, an observed spectral index $\alpha_\nu$ is decomposed to the spectral index of dust emission
$\alpha_\mathrm{d,\nu}$ and that of plasma emission $\alpha_\mathrm{p,\nu}$ as
$
	\alpha_{\nu}
	=
	f_\mathrm{p,\nu}\alpha_\mathrm{p,\nu}
	+	
	(1 - f_\mathrm{p,\nu})\alpha_\mathrm{d,\nu}
$,
where $f_\mathrm{p,\nu}$ is the fraction of the plasma emission in the total flux density.
The spectral index of the dust emission is then calculated as
\begin{eqnarray}
	\alpha_\mathrm{d,\nu}
	& = &           \label{eq.alpha_d_exact}
	\frac{\alpha_{\nu} - f_\mathrm{p,\nu}\alpha_\mathrm{p,\nu} }{1 - f_\mathrm{p,\nu}}   \\
	& \approx & \label{eq.alpha_d_approx}
	 \alpha_{\nu} + f_\mathrm{p,\nu} (\alpha_{\nu} - \alpha_\mathrm{p,\nu} )	
	\quad 
	\mbox{for $f_\mathrm{p,\nu} \ll 1$} .
\end{eqnarray}

Our spectral indices observed around 1 mm are already good proxies for dust spectral indices
since the 1 mm emission is about 95\% dust emission in our cases. 
It is an advantage of the \about1 mm analysis over the same analysis at $\lambda \sim 3$ mm, 
where plasma emission is significant or dominant even in our deeply dust-enshrouded nuclei.
The second term on the right side of (\ref{eq.alpha_d_approx}) is the correction to the observed spectral index
to obtain the dust spectral index.
It is affected by the plasma spectral index used at a lower frequency, denoted as $\alpha_\mathrm{p,low}$ here, 
through both $f_\mathrm{p,\nu}$ and  $\alpha_{\nu} - \alpha_\mathrm{p,\nu}$. 
However, the two change in the opposite directions as $\alpha_\mathrm{p,low}$ changes.
Compared to $\alpha_\mathrm{p,low}=-0.4$ for example, 
a flatter [steeper] plasma spectrum with $\alpha_\mathrm{p,low}$ closer to $-0.1$ [$-0.7$] 
(i.e., free-free [synchrotron] dominated) makes the plasma emission at $\nu=300$ GHz stronger [weaker]
and $f_\mathrm{p,\nu}$ larger [smaller]
while it makes $(\alpha_{\nu} - \alpha_\mathrm{p,\nu})$ smaller [larger].
Combined with the small $f_\mathrm{p,\nu}$, the uncertainty in $\alpha_\mathrm{p,low}$ adds little to
the errors of dust spectral index $\alpha_\mathrm{d,\nu}$.
Therefore, the uncertainties of $\alpha_\mathrm{d,\nu}$ in Table \ref{t.contSpectralAnalysis} do not include
those from the plasma spectral indices.

%%%%%
\subsubsection{Dust Opacities}		\label{s.model.tauAndColumn.opacities}
Dust optical depths at $\lambda=1$ mm are estimated 
to be $\tau_\mathrm{d,1mm} \approx 2.2$, $\leq0.1$, and $\approx1.2$, 
for NGC 4418, Arp 220E and,  Arp 220W, respectively (Column 8 of Table \ref{t.contSpectralAnalysis}), 
from the dust-emission spectral indices using the BGN $\alpha$--$\tau$ relation (\ref{eq.bgn_spix}).
The $1\sigma$ upper limit for Arp 220E is because the dust spectral index is very close to 
its upper limit in the model.
One cannot precisely determine opacity in such a case using the $\alpha$--$\tau$ method.
The result is affected much by the (uncertainty of) dust $\beta$.

For comparison, a shortcut to the same opacity uses the raw spectral indices, a fixed $+5$\% correction to them
for plasma subtraction, and the $\alpha$--$\tau$ relation (\ref{eq.simple_spix}) of the slab model. 
It leads to $\tau_\mathrm{d,1mm}^\mathrm{(s)} \approx 2.1$, $0.2$, and $1.2$
(or 2.3, 0.4, 1.4 for $\beta=1.8$),
for NGC 4418, Arp 220E and,  Arp 220W, respectively.
Considering the reasonable agreement with the formal estimates above, 
this simplified method will be a good starting point when handling a large data set of non-uniform quality.

Our new estimates are broadly consistent with previous ones.
For example, NGC 4418 has $\tau_\mathrm{d,1mm} \sim 1$ estimated by \citet{Sakamoto13}.
Approximately translating opacities at other wavelengths to $\tau_\mathrm{d,1mm}$ with $\beta=1.8$, 
Arp 220W has estimates of \about\ 0.8 and \about\ 6, 
respectively, by \citet[converted from $\tau_\mathrm{d,0.86mm}$]{Sakamoto08}
and \citet[converted from $\tau_\mathrm{d,2.6mm}$]{Scoville17}.
Arp 220E and W also have estimates of $\tau_\mathrm{d,1mm} \sim 0.4$ and $1.2$, respectively, by
\citet[reported in $\tau_\mathrm{d,0.434mm}$]{Wilson14},
and $\sim 0.6$ and $3.2$ by \citet{Dwek20}.
\citet{Sakamoto13,Sakamoto08}, \citet{Scoville17}, and \citet{Wilson14} used the isothermal slab model to estimate the opacities.
We saw in Section \ref{s.model.BGN} that such a slab opacity approximates the radial optical depth
(i.e., the one between the center and the surface of a sphere) in the BGN model.
\citet{Dwek20} adopted a spherical model and a temperature structure for a centrally-heated, optically thin cloud.

%%%%%
\subsubsection{Column Densities}		\label{s.model.tauAndColumn.columns}
We obtained the following gas column densities for the three nuclei using the conversion formula (\ref{eq.bgn_nhh-tau}) from the
dust opacities: $\log(\NHH/\persquarecm) = 25.65\pm0.06$, $< 24.4$, and $25.4\pm0.1$ 
for NGC 4418, Arp 220E, and Arp 220W, respectively.
The uncertainties above are solely from the 1 mm optical depths of dust emission.
They do not include the uncertainties in the $\tau$--\NHH\ conversion formula, $\beta$, or the BGN modeling itself. 
The result of Arp 220E is most susceptible to the choice of $\beta$ and could be about 25.0 for $\beta=1.8$ rather 
than 1.6.
In addition, depending on the conversion formula, the \HH\ column densities could be \about0.5 dex larger,
which we noted in Section \ref{s.model.tauToNHH}.

%%%%%%%%%%%%%%%%%%%%%%%%%%%%%%%%%%%%%%%%%%%%%%%%%%%%%%%%%%%%
\section{Frequency-Dependent Continuum Structure and Its Implications}
\label{s.spatial}
The continuum information that a high-resolution spectral scan provides for a galactic nucleus
is not limited to a spectrum and the shape of the target nucleus.
It also includes any variation of the spatial structure in the continuum emission over frequency, although such information
has been hard to collect and rarely used.
Here we combine theoretical expectations and a new compilation of size and shape spectra of our
target nuclei to start constraining their opacities and structures from this new angle.

%%%%%%%%%%%%%%%%%%%%
\subsection{Variation of Apparent Source Size with Frequency as Another Indicator of Opacity}
\label{s.spatial.basic}
The apparent size of a dusty object varies much with frequency 
when the object is moderately opaque, i.e., $\tau_\nu \gtrsim 1$. 
Such size variation is minimal when the object is optically thin ($\tau_\nu \ll 1$) or 
if a large part of the object having a clear outer boundary is optically thick ($\tau_\nu \gg 1$).
Therefore, the degree of size variation with frequency is another indicator of source opacity for a dusty nucleus.
This diagnostic has an advantage over the $\alpha$--$\tau$ method in the previous section 
in that a source size (FWHM) can be measured even without flux calibration.

The reasons for the opacity-dependent size variation are simple. 
An optically thin source hardly changes its apparent size with the frequency
because the entire emitting dust is visible from the outside.
Since emissivity at all locations varies with frequency in the same way, the source shape and size (such as FWHM)
remain the same, 
while the source brightness and total flux density vary with frequency.
In contrast, an object of moderate opacity ($\tau_\nu \gtrsim 1$) at its central sightline has its emission from inner dust particles absorbed by the outer, foreground dust. 
At a higher frequency, the emission from the interior dust has even less chance to reach us 
since the foreground dust becomes more opaque with increasing frequency.
As a result, the dust emission  saturates at higher frequencies
along the central sightline but not along the optically thin sightlines through the source outskirt.
The net effect at higher frequencies is 
that the continuum intensity increases more at larger radii than near the source center.
Therefore, a moderately opaque source has larger FWHM at higher frequencies.
An object that has a sharply falling dust density distribution (e.g., a distribution with a radial cutoff) and is optically thick
($\tau_\nu \gg 1$) across its sky-projected surface would have its emission saturated almost everywhere. 
Therefore, its appearance is nearly independent of $\tau_\nu$ and $\nu$
as long as $\tau_\nu$ remains $\gg 1$.
In short, no photosphere exists in an optically thin source.
The photosphere moves much with frequency in a moderately opaque source.
And the photosphere stays near the surface if a source with a definite outer boundary is highly opaque.

\subsection{BGN Models on Frequency-dependent Structure}
\label{s.spatial.bgn}
The BGN simulations in \citet{GS19} provide sky-projected, radial profiles of continuum brightness temperature
for each model source at various observing frequencies. 
Figure \ref{f.contModelTbProfile} is an example, showing the continuum brightness profiles for a fiducial AGN model
(for $L_\mathrm{IR}=10^{11}\Lsun$ and $\NHH = 10^{26}$ \persquarecm).
Although the gas column density is a few times larger than in the two most deeply buried nuclei in our sample,
the figure illustrates a general trend that the radial brightness profile tends to be more strongly peaked toward the center 
and hence has smaller FWHM at lower frequencies.
Conversely, the source tends to have a larger half-peak size at a higher frequency because one looks at only
the outer volume of the gas sphere (i.e., down to the photosphere).
The apparent source size significantly changes over frequency when the frequency range includes where $\tau_\nu$ is unity.

%%%%%%%%%%%%%%%%%%%%
\subsection{Spectra of Structure and Brightness in the Three Nuclei} 
\label{s.spatial.obsSummary}
Figure \ref{f.contParams} presents the continuum parameters of the three nuclei
as a function of frequency, showing structural parameters in panel (a) and
peak brightness temperatures in (b).
We compiled measurements by ourselves and others in the range of 5--700 GHz.
Table \ref{t.contShape} summarizes the geometrical parameters of the continuum nuclei.

In panel (a), the vertical axis is the source size in FWHM
or the geometrical mean of the major and minor axis FWHM for data from elliptical-model fitting.
For the data obtained by fitting a single elliptical Gaussian to a nucleus, 
our plotting symbols are ellipses similar to the fitted ellipses; they share the position angles and axial ratios.

Our compilation includes parameters of the two Arp 220 nuclei
for their 33 GHz continuum emission 
and the distribution of cm-wave compact sources within the nuclei.
The former emission is mostly from diffuse plasma,
and the latter discrete sources are mostly supernovae and supernova remnants
detected through VLBI observations.
To characterize the VLBI source distribution, 
we assumed that the spatial probability function of the radio sources in each nucleus was an elliptical Gaussian.
We then used the maximum-likelihood method
to estimate the Gaussian parameters 
from the 97 VLBI source positions reported by \citet{Varenius19}.
We disregarded the brightness of the individual sources and any spatial variation of source detectability. 
Sources closer to the center of each nucleus may be more likely to be missed by blending or absorption;
the free-free opacity has been estimated to be $\lesssim 1$ at 18 cm \citep{Parra07,Anantharamaiah00}.
However, the geometrical parameters we are most concerned with at centimeter wavelengths 
are axial ratio and position angle. 
It is the off-center sources in each nucleus that most constrain these parameters.
The VLBI source distribution and the best-fit distribution functions are plotted in Figure~\ref{f.a220VLBIfit}.

%%%%%%%%%%%%%%%%%%%%
\subsection{Size Variation with Frequency} \label{s.spatial.size}
It is evident in Figure\ref{f.contParams}(a) that the apparent continuum size of the three nuclei 
increases with an increase in frequency when dust emission dominates. 
Arp 220 has more data points. 
Both of the nuclei of Arp 220 show a gradual increase in size at a higher frequency between 100 and 700 GHz. 
The western nucleus has a steeper rising slope.
The continuum nucleus of NGC 4418 also appears larger at \about700 GHz than in 200--400 GHz.
Dust emission dominates the continuum at $\gtrsim 200$ GHz in the three nuclei.
Its fractional contributions are 0.1--0.4 and 0.5 at 100 GHz, respectively, for the nuclei of Arp 220 and NGC 4418
and about 0.95 at 300 GHz for them all
\citep[\S \ref{s.model.tauAndColumn.decomposition}, and Table \ref{t.contSpectralAnalysis}]{Sakamoto17}.
The transition between plasma and dust dominance is also seen as the sharp change of peak brightness temperature
around 100 GHz in Fig.~\ref{f.contParams}(b).
Minor plasma emission cannot be the reason for the size variation at $\gtrsim 200$ GHz.
Therefore, Figure\ref{f.contParams}(a) demonstrates that the three buried galactic nuclei have larger apparent sizes 
at higher frequencies in their dust-continuum emission.

The observed size variations with frequency are 
qualitatively consistent with the expectation outlined in Section \ref{s.spatial.basic} for dust emission of moderately opaque sources.
They are also semi-quantitatively consistent with the higher dust opacity of Arp 220W than Arp 220E, 
which we estimated in Section \ref{s.model.tauAndColumn.opacities} from their spectral indices.
In addition, the magnitude of size variation is roughly consistent with that in the BGN model.
The apparent size of Arp 220W increases by about a factor of 3 from 100 GHz to 700 GHz.
The FWHM size of the BGN model in Fig.\ref{f.contModelTbProfile}
increases by about a factor of 5 from 115 to 691 GHz.
The model has a column density about 0.5 dex higher than our estimate for Arp 220W.
Analysis of the frequency-dependent continuum structure like this one, with even better data and models, should provide 
more insights into the properties of our three dust-obscured nuclei and similar sources.

%%%%%%%%%%%%%%%%%%%%
\subsection{Variation of Radial Profile with Frequency and a Possible Distinct Core of Arp 220W} 
\label{s.spatial.2G}
The radial intensity profiles of the continuum emission that we characterized by decomposing to Gaussians 
are also consistent with the saturation of dust emission at higher frequencies in our target nuclei. 
We observed in Section \ref{s.cont.sizeShape.r2Gfit} that our target nuclei are better fitted with two concentric Gaussians 
than with a single Gaussian in most of our data.
We also observed that the smaller component tends to contribute more to the total flux at a lower frequency
in each nucleus.
Therefore, the observed frequency-dependence of the 1G size (i.e., FWHM in single-Gaussian fitting) is 
not because each nucleus maintains a Gaussian shape and has a larger FWHM at a higher frequency. 
Instead,  it is because the radial profile of continuum intensity is not a Gaussian in general and 
becomes more centrally peaked at a lower frequency.
This intensity profile is expected in moderately opaque nuclei, as explained in Section \ref{s.spatial.basic}, and
agrees with the BGN simulations in Fig.~\ref{f.contModelTbProfile}.

In this regard, it is interesting to note that, in Band 9, the central component is still discernible in NGC 4418 
(as well as in Arp 220E) but not in Arp 220W
although we estimated from spectral slopes similar dust opacities for the two, 
with NGC 4418 being less than twice more opaque than \arpW.
Since the sharp central peak should disappear at frequencies where the source is highly opaque, 
this difference implies that \arpW\ is more opaque than NGC 4418 (as well as \arpE).
Another observation that may be due to a much higher opacity toward \arpW\ than the other two is
the 3 mm spectral index $\alpha_\mathrm{3\,mm}\sim2$ for the core component of Arp 220 \citep{Sakamoto17}. 
If this compact component is predominantly dust emission, which is possible if most of the 3 mm dust emission from Arp 220W
is from this component, then the spectral index
implies $\tau_\mathrm{d,\, 3\, mm} \sim 5$, $\log(\NHH/\persquare{cm}) \sim 26.5$,
and a mean \HH\ density of $\nHH \sim 10^{6.7}$ \percubic{cm} for the 20 pc core. 
This column density is an order of magnitude larger than that from the \about1 mm spectral slope.
Admittedly, the 3 mm spectral index of \about2 may be alternatively due to a combination of
optically thin dust emission ($\alpha >2$) and plasma emission with negative $\alpha$.
It may be also partly due to optically thick plasma emission.
Still, this possible distinct core of very high column and volume densities in \arpW\ 
deserves critical examination in a separate work.

Barring errors in the opacity estimates from spectral slopes,
the example of \arpW\
implies the presence of factors other than the total dust opacity  to affect the appearance of a dusty nucleus.
They may be the small-scale spatial structures in the dust and luminosity-source distributions,
the flatness and inclination of a disk-like dust distribution, 
and 
the contribution of dust emission from an outflow.
In other words, structural analysis over a range of frequencies can potentially 
verify the opacity estimate from the continuum spectrum and further tell us other properties of the nucleus.

%%%%%%%%%%%%%%%%%%%%
\subsection{Brightness Temperatures of Dust Continuum}
\label{s.spatial.peakTb}
The high peak brightness temperatures of the dust continuum, 
as high as \about500~K in Figure \ref{f.contParams}(b) at \about350~GHz,
also indicate the presence of opaque dust. 
A peak brightness temperature of dust emission, the beam-deconvolved one obtained through visibility fitting, is 
the photospheric dust temperature in the case of optically thick emission. 
It is the dust temperature multiplied by the dust opacity in the optically thin cases.
Therefore, the peak \Tb\ of the dust continuum sets a lower limit to the physical temperature of the dust.
If the temperature is not uniform, then the limit would be to the temperature of the dust 
that contributes most to the continuum emission.
The deconvolved peak continuum brightness temperatures that we obtained are 
as high as about a third of the dust sublimation temperature. 
We attributed most of this bright continuum to dust emission in Section \ref{s.model.tauAndColumn.decomposition}.
Thus, our peak intensity observations alone suggest a high dust opacity, $\tau_{1\;\rm{mm}} \gtrsim 0.3$, for both 
the nuclei of NGC 4418 and Arp 220W. 
Needless to say, if the dust temperature is about the observed peak \Tb\ and not at the sublimation temperature, 
then the opacity constraint is $\tau_{1\;\rm{mm}} \gtrsim 1$.

Higher opacities of NGC 4418 and Arp 220W than Arp 220E are also evident in the frequency dependence 
of their peak brightness temperatures.
Figure \ref{f.contParams}(b) shows that the peak \Tb\ in the 1G fitting (filled symbols) increases by a factor of 3
from 26 K at 240 GHz to 73 K at 670 GHz for Arp 220E
while the temperature remains about the same at 150--200 K in the same frequency range for Arp 220W and NGC 4418.
These distinct trends are consistent with the higher optical depths 
and saturation of dust emission (i.e., $\tau_\mathrm{dust} \gtrsim1$) 
in the latter two nuclei.

To summarize, our continuum-based evidence for the high dust opacities 
of $\tau_{\rm d, 1 mm} \sim 1$
for the nuclei of NGC 4418 and Arp 220W includes
their spectral indices at  $\lambda \sim 1$ mm,
their size variation with frequency, 
the way the intensity profiles within the individual nuclei vary with frequency, 
the high peak brightness temperatures, 
and the lack of variation in their peak brightness temperatures with frequency. 

%%%%%%%%%%%%%%%%%%%%%%%%%%%%%%%%%%%%%%%%%%%%%%%%%%%%%%%%%%%%
\section{Structures of the Three Nuclei}
\label{s.contStructure}
We use our compilation of high-quality spatial information of continuum emission
to describe the structure of the three nuclei with more precision.

%%%%%%%%%%%%%%%%%%%%
\subsection{Oval Shapes of the Three Nuclei} 
\label{s.contStructure.shape}
Figure~\ref{f.contParams}(a) has shown that continuum emission from the three nuclei
has consistent shapes across the observed frequency range. 
In 100--700 GHz, the standard deviations of the major-axis position angle in the 1G-deconvolution are
only 2, 6, and 10 degrees for NGC 4418, Arp 220E, and Arp 220W, respectively. 
(See Table \ref{t.contShape} for the mean values.)
The major-to-minor axial ratios for single-Gaussian fitting are also consistently measured,
with the standard deviations of only 0.06--0.08 for the three nuclei.
The consistent parameters in independent observations attest to the reliability of the measurements
and the continuum shapes.

There are three important implications about the nuclei from these apparent shapes in the continuum.
First, the oval (rather than circular) shapes of the nuclei on the sky prove 
that oblate spheroids (or disks if flat enough) approximate their three-dimensional shapes better than spheres. 
For simplicity, we hereafter use the term `disk' for both an oblate spheroid
and a disk whose thickness is much smaller than its diameter.
The distinction between a sphere and a disk matters to the thermal structure of the nuclei
since, for the same total amount of obscuring material, spheres are most effective in trapping photons 
while disks (e.g., an axisymmetric disk) tend to leak photons in their polar direction.
Consequently, a disk tends to have a lower internal temperature than a sphere for the
same internal heating source.

Second, the elongated continuum shapes indicate the presence of rotation-dominated structures (i.e., rotating disks)
of molecular gas and dust in all three nuclei.
The major axes of the continuum nuclei are aligned with the kinematical major axes in these nuclei.
The gas velocity gradients across these nuclei have been attributed to rotation 
by \citet{Sakamoto99, Sakamoto13} and many ALMA observations, including \citest{Paper2}.
Radiation pressure may contribute to the disk thickness or even outflows, 
but the rotation is the most likely reason for the observed alignment. 
The shapes of the disks are better constrained in the continuum emission than through molecular line emission
since the latter is affected by spatially variable line excitation and chemical composition and the more complex radiative transfer.

Lastly, the connection between the rotating nuclei and their active star formation is evident in the data.
Namely, both nuclei in Arp 220 have matching shapes in dust and plasma emission, 
except for the minor differences in Arp 220W (see Section \ref{s.contStructure.a220w}).
Moreover, the distribution of compact radio sources in each nucleus shares the major-axis
with the dust emission (in its single-Gaussian fitting), as already noticed in \citet{Sakamoto08, Sakamoto17}.
This configuration indicates that star formation has been active in the rotating, gaseous, nuclear disks.

%%%%%%%%%%%%%%%%%%%%
\subsection{Internal Structures of the Three Nuclei} 
The consistency of the continuum shapes across frequency further improves when we decompose each nucleus
into a few structural components. 
The decomposition also provides us with cleaner pictures of the nuclei. 
Table \ref{t.contShape} has our best estimates for the shapes of the three nuclear disks and the bipolar outflow in Arp 220W.

%%%%%%%%%%%%%%%%%%%%
\subsubsection{Nuclear Disk and Bipolar Outflow of Arp 220W} 
\label{s.contStructure.a220w}
The high-quality data in Figure~\ref{f.contParams}(a) reveals
that Arp 220W, unlike the eastern nucleus, has slightly different shapes
at $\leq 33$ GHz and $\gtrsim100$ GHz.
Although the major axis is roughly in the east-west direction in both frequency ranges, 
the position angle is about 80\degr\ in the former and \about100\degr\ in the latter.
The continuum shape of the western nucleus is also consistently rounder at $\gtrsim100$ GHz  than 
at lower frequencies.

We attribute this minor but persistent discrepancy in the 1G parameters to
a significant contribution of the bipolar outflow to continuum emission at $\gtrsim100$ GHz.
Indeed, our multi-component visibility fitting to the \about700 and \about100 GHz data in Section \ref{s.contB9visfit} found 
a major-axis position angle of $83\degr\pm1\degr$ for the nuclear disk, 
in good agreement with $79\degr\pm2\degr$ at 33 GHz and $83\degr\pm5\degr$ at \about5 GHz 
(see Table \ref{t.cont_model-r2G_parameters}). 
We, therefore, suggest 83\degr\ for the major-axis position angle of the western nuclear disk in Arp 220.

We estimate the minor-to-major axial ratio of the nuclear disk in Arp 220W to be about 0.5 from the Band 3 and \about5 GHz data. 
After separating the outflow through our multi-component visibility fitting, 
the axial ratio decreases
in Band 9 from 0.83 in the 1G fit to 0.68 for the disk component.
In Band 3, the axial ratio again decreases from 0.77 in the 1G fitting to 0.50 for the nuclear disk alone. 
The 1G axial ratio in the 33 GHz continuum, 0.60, should also be affected by the plasma outflow seen in Band 3,
while the outflow is less likely to affect the stellar component traced by the \about5 GHz VLBI sources. 
The slightly larger ratio for the disk component in Band 9, 0.68, may indicate mild saturation of the dust emission
on the line of nodes or high-latitude dust due to the disk starburst.

The bipolar outflow is estimated to have a major axis position angle of 165\degr\ from
the multi-component visibility fitting at \about700 and \about100 GHz in Section \ref{s.contB9visfit}.
It is nearly perpendicular to the major axis of the nuclear disk of 83\degr.
The deviation from a right angle can still be non-zero. But it is smaller than before, i.e., $<$10\degr.
(The major axis of the 1G fitting at 100--350 GHz makes an angle of \about65\degr\ with the outflow; 
hence the deviation was about 25\degr.)
\citest{Paper2} presents line images of this outflow and discusses its geometry more in detail.

%%%%%%%%%%%%%%%%%%%%
\subsubsection{Nuclear Disks in Arp 220E and NGC 4418}  
\label{s.contStructure.n4418a220e}
For Arp 220E, we estimated the apparent shape of the nuclear disk from the
multi-component visibility fitting at \about700 and \about100 GHz in Section \ref{s.contB9visfit} 
and the 1G fitting of the \about5 GHz stellar sources. 
The nuclear disk parameters are similar to the parameters of the entire nucleus
since there is no prominent bipolar (outflow) emission in our continuum data.
The disk major axis is aligned with the direction of the velocity gradient of molecular gas.

The nuclear disks proper in the two nuclei of Arp 220 are similar in that
both are inclined disks of gas, dust, and young stars without visible misalignment among these components.

For NGC 4418, we only have the multi-component visibility fitting at \about700 GHz
to single out the nuclear disk and decide its parameters. 
Therefore, we regard the disk parameters as less robust than for the other nuclear disks.
The axial ratio of this nuclear disk is smaller (0.4) than for the entire nucleus (0.61) 
because the visibility fitting identified a misaligned central core and a faint extended component rounder than the nuclear disk.

%%%%%%%%%%%%%%%%%%%%
\subsubsection{Inclinations of the Three Nuclear Disks}
It is suggested from the axial ratios of about 0.4--0.5 that the three nuclear disks have inclination angles of \about60\degr\ or larger.
The lower limit of about 60\degr\ is for a thin disk, and a nearly edge-on configuration is allowed
for a thick disk or an oblate spheroid of axial ratio \about0.5.
In \citest{Paper2}, we discuss lopsided line absorption toward the nuclei. 
If they are due to collimated outflows, 
then edge-on nuclear disks are only possible for the outflows not perpendicular to the nuclear disks.
The continuum-based axial ratios certainly disfavor nearly face-on configurations and near-spherical shapes of the nuclear disks.

%%%%%%%%%%%%%%%%%%%%%%%%%%%%%%%%%%%%%%%%%%%%%%%%%%%%%%%%%%%%
\section{Discussion} \label{s.discussion}
Our continuum analysis has the following possible future directions to overcome current limitations. 

%%%%%%%%%%%%%%%%%%%%
\subsection{Spectral Index Survey}
\label{s.discuss.SpixForSurvey}
We have presented procedures and examples to estimate the obscuring column densities in buried galactic nuclei 
through the spectral indices of their (sub)millimeter continuum emission from dust.
In particular, the spectral index at $\lambda \sim 1$ mm, $\alpha_1$, is suitable to diagnose
buried galactic nuclei similar to those in NGC 4418 and Arp 220, 
or those having the obscuring column densities of $\log(\NHH/\persquarecm) \sim $ 25--26. 
One can use $\alpha_1$ to identify sources with even higher column densities,  
though it needs $\alpha$ at a less-saturated, longer wavelength to determine \NHH\ for such sources.
This BGN diagnostic has no selection bias against $\log(\NHH/\persquarecm) \gtrsim $ 25 
and does not need an AGN in the first place, unlike X-ray diagnostics.
The continuum spectral index $\alpha_1$ is sometimes easier to obtain than the information of specific lines, e.g., 
from archival data. 
Therefore, this diagnostic enables a census of deeply obscured nuclei, i.e., BGN, 
on whose population we currently have only limited knowledge.

In light of the circumstances above, we propose a simple criterion of
\begin{equation}
\alpha_1 \leq 3
\quad
\Longrightarrow
\quad
\log(\NHH/\persquarecm) \gtrsim 25
\end{equation}
to find the most deeply buried nuclei among radio-quiet sources.
Both NGC 4418 and Arp 220W satisfy this condition. 
There are two practical merits in this diagnostic at $\lambda \approx 1$ mm.
First, continuum emission there is predominantly dust emission unless the nucleus is radio-loud.
Therefore, one can omit the subtraction of plasma emission in the initial search.
Second, the spectral index is easy to measure with such telescopes as ALMA and SMA.
Their high-angular resolution is essential to isolate the nuclear emission and constrain its properties.
Their spectroscopic capability also allows us
to remove lines that contaminate the continuum and bias the spectral slope.
One could simultaneously observe diagnostic lines such as those from the vibrationally excited HCN.

%%%%%%%%%%%%%%%%%%%%
\subsection{BGN Model Refinement} 
\label{s.discuss.BGNconsistency}
\citet{GS19} simulated  in their BGN modeling 
not only continuum radiation but also lines of vibrationally excited HCN for various column densities.
One could invert the results to infer the column densities from HCN observations.
By comparing such HCN-based estimates with the continuum-based ones from the same BGN models,
we expect to refine the BGN modeling and its assumptions.

For example, the BGN simulations indicate that the J=4--3-to-J=3--2 ratio of the HCN($v_2$=1, $l$=1f) line
follows virtually the same relation with \NHH\ for various luminosity surface densities of the nucleus
\citep[see Fig. 12(b)]{GS19}. 
The line-flux ratio decreases from about 2.7 to 1.7 when $\log (\NHHprime /\persquarecm)$ increases from 24 to 25
because the optically thicker J=4--3 line causes more continuum absorption.
Therefore, the line ratio is a potential indicator of the column density.
However, our three nuclei have about the same HCN line ratios of \about1.9--2.0 
corresponding to $\log (\NHH /\persquarecm) \sim 24.7$
even though \arpE\ must have a lower degree of obscuration than the other two nuclei according to our continuum analysis.
This example indicates room to refine the BGN modeling 
and its assumptions, such as the structure of the model nucleus and the HCN abundance per velocity width.

%%%%%%%%%%%%%%%%%%%%
\subsection{Non-Spherical Shapes and Anisotropic Extinction}
\label{s.discuss.nonSphericalModel}
While we used a simple spherical model, buried galactic nuclei are complex systems having internal structures.
We have seen that three of them have unambiguously non-spherical shapes at 10--100 pc scales.
They most likely have anisotropic obscuration, i.e., direction-dependent column densities 
from the center to the surface of the obscuring structure. 
By analogy, we infer non-spherical shapes and anisotropic obscuration at the scale of tens of parsecs 
in a significant fraction of deeply obscured galactic nuclei.

Anisotropic extinction by a non-spherical shroud affects our specific analysis and BGN studies in general, 
as in the following examples. 
First, the sharp central peak in the sub/millimeter continuum emission may be 
partly due to the anisotropic extinction. 
We explained in Section \ref{s.spatial} that such features are in accord with the greenhouse effect 
and higher dust temperature toward the center in the BGN dust sphere. 
However, as noted in Section \ref{s.contStructure.shape}, 
a disk-like structure having less obscuration in its polar direction will leak more radiation in that direction 
from the warm interior. 
A torus structure, e.g., caused by a bipolar outflow, would have even more polar leakage. 
Such leaked radiation may contribute to the sharp central peak of the sub/millimeter continuum 
for a non-spherical BGN. 
\citet{Soifer99}  suggested such hot-dust emission as a source of infrared radiation 
at $\lambda \lesssim 15$ \micron\ from the nuclei of Arp 220. 
Second, to hide an energetically significant AGN, 
it should have both a high column density along our sightline to block the direct radiation ($\NH \gtrsim 10^{25}$ \persquarecm) 
and a high covering factor of \about1 to block indirect radiation \citep{Iwasawa01, Maiolino03, Teng15}.
Given the highly Compton-thick column densities based on our spherical analysis, 
it is likely that at least NGC 4418 and Arp 220W have the needed column densities 
in most of the directions through their compact dust cores, nuclear disks, and high-latitude material 
such as their outflows. 
It requires observational verification.

An obvious next step is imaging at a higher angular resolution, higher sensitivity, 
and higher spatial dynamic range. 
It should improve on our simple models that describe each nucleus with only a few Gaussians.
Our spectral scan guides the choice of observing frequency to minimize line contamination.
A non-spherical absorber has multiple parameters for the anisotropic extinction; 
a single column density does not fully describe it.
Therefore, on the theory side, more advanced modeling is necessary 
to obtain the set of parameters from high-quality observations.
The first targets of such study should include our three nuclei, given the abundance of their observational information.

%%%%%%%%%%%%%%%%%%%%%%%%%%%%%%%%%%%%%%%%%%%%%%%%%%%%%%%%%%%%
\section{Summary}
\label{s.summary}
Three heavily obscured nuclei in the infrared-luminous galaxies NGC 4418 and Arp 220 have been observed with ALMA
at the geometrical-mean resolution of 0\farcs14--0\farcs28 over bandwidths of 67 GHz between \frest=215 and 697 GHz.
The three are prototypes of 
highly obscured (\NH $\gtrsim 10^{25}$ \persquarecm), 
compact ($\lesssim 100$ pc),
and luminous ($\Lbol \gtrsim 10^{11}$ \Lsun) galactic nuclei.
Their internal properties, luminosity sources, and ongoing evolution have been of our interest.
This paper is for the overview of our high-resolution imaging spectroscopy observations, 
data reduction and presentation, and continuum data modeling and analysis.
A companion paper reports the line data, line identification, and line analysis \citesp{Paper2}.

We first presented the parameters of our observations and our data reduction techniques
devised for wide-band spectral scans with ALMA.
These techniques include flux self-calibration to reduce the relative amplitude error among adjacent tunings to \about1\% 
and 
detection and correction of apparently systematic (in addition to random) errors in astrometry.
These improve the accuracy of our continuum spectra and line ratios.
In addition to the conventional image reconstruction, we employed visibility-based analysis to measure 
the emission parameters through visibility fitting in every spectral channel.
We made models of the continuum emission using the continuum-dominated channels found in the line forest.
They were used for our model-based continuum subtraction from line data
for consistent continuum subtraction in a line-filled spectral scan.

We next presented models to be compared with millimeter-submillimeter continuum observations of 
deeply dust-enshrouded and luminous sources such as our three galactic nuclei.
We pointed out that the radio spectral index $\alpha$ of continuum emission at $\lambda \sim 1$ mm 
is a good indicator of the dust opacity and hence the total column density 
in the range of $\log(\NHH/\persquarecm)$ \about\ 25--26 for both physical and practical reasons.
Plasma contribution to the \about1 mm continuum is small and can be easily subtracted using our simple formalism.
We provided the $\alpha$--$\tau$ relation for dust emission from the BGN simulations of \citet{GS19}
to convert an observed 1 mm spectral index $\alpha_\mathrm{1\,mm}$ to a dust optical depth.
Combining that with an adopted $\tau$--$\NHH$ relation, the continuum spectral index can be converted 
to the total column density of the nucleus.

We further explored the frequency-dependent variation of continuum structures using both models and observations.
There are particular structural variations over frequency expected in dust continuum images. 
These frequency-dependent structures are in both BGN simulations and our BGN observations. 
They also constrain the dust opacity and hence the total column density of a nucleus.

Our main observations and their model-based interpretation are the following
for the continuum emission of the three nuclei:

%%%%%%%%%%%%%%%%%%%%%%%%%%
\begin{enumerate}
\item 
All three nuclei are bright, compact, and single-peaked emitters at our \about0\farcs2 (30, 80 pc) resolution
in our observing frequencies. 
They all have dense forests of lines but less so in the ALMA Band 9 (\frest\ \about\ 680 GHz).

\item 
The continuum nuclei are spatially well resolved in the \uv\ domain despite their compactness.
In the single elliptical Gaussian (1G) fitting, they have the major-axis FWHM of 0\farcs10--0\farcs26 (17--110 pc) 
in Bands 6 and 7 (\frest\ = 210--370 GHz) and  0\farcs17--0\farcs33 (30 -- 140 pc) in Band 9.
We found that the 1G size increases toward higher frequencies between 100 and 700 GHz,
with more variations in \arpW\ and NGC 4418 and least in \arpE.
These size variations with frequency are evidence that all three nuclei are moderately opaque in dust emission
and that the former two have higher opacities than the last.

\item
The peak brightness temperature of continuum emission in the 1G deconvolution  
increases by a factor of 3 from \about200 GHz to \about700 GHz in \arpE\
while it stays about constant in \arpW\ and NGC 4418 in the same frequency range.
This behavior is another piece of evidence that NGC 4418 and \arpW\ have higher opacities than \arpE\
and that the former emission is saturated, i.e., $\tau > 1$, while the latter is not.
The high opacities of the former are in line with their substantial brightness temperatures 
compared to the dust sublimation temperature.

\item
In every nucleus, the 1G shape (i.e., axial ratio and major-axis position angle) largely remains constant over frequency, 
and the major axis of the continuum structures align with the known nuclear disks of molecular gas.
Such alignment indicates that the primary factor deciding the dust distributions at 10--100 pc scale is 
the rotation of the nuclear gas disks.

\item  Further visibility analysis 
found that all three nuclei are better modeled with a sharper central peak and a broader envelope than a single Gaussian. 
The core component has a higher fractional flux in NGC 4418 than in the nuclei of Arp 220.
The same is true at lower frequencies for every nucleus.
The cores have FWHM in the range of 0\farcs05--0\farcs10 (8--40 pc)  
and peak brightness temperatures in 60--530 K. 
The core-envelope structure in Arp 220 is also visible at 100 GHz \citep{Sakamoto17}.

\item  We obtained the $\nu \sim 300$ GHz spectral indices of 
$\alpha=2.35 \pm 0.08$ for NGC 4418, 
$2.67 \pm 0.10$ for \arpW\, and 
$3.28 \pm 0.09$ for \arpE\
from ALMA measurements in Bands 6 and 7.
We combined these with our 100 GHz continuum measurements of the three nuclei \citep{Costagliola15, Sakamoto17}
to estimate the contribution of plasma emission of only 3--9\% at $\nu \sim 300$ GHz.

\item
Our plasma-subtracted spectral slopes and new $\alpha_\mathrm{dust}$--$\tau$ relation 
based on the BGN model lead to 
the dust optical depths, measured from the center to the surface of a spherical source, 
of $\tau_\mathrm{dust,\,1\,mm} = 2.2 \pm 0.3$ for NGC 4418, $1.2\pm0.2$ for \arpW, and $\leq 0.1$ for \arpE.
(The opacity of \arpE\ can be \about0.4 for a dust $\beta$ of 1.8.)
These dust opacities correspond to the obscuring column density $\log(\NHH/\squarecm)$ of 
$25.65 \pm 0.06$ in NGC 4418, 
$25.4\pm0.1$ in \arpW, and 
$\leq 24.4$ (or \about25.0 for $\beta=1.8$) in \arpE,
although they could be up to about 0.5 dex larger depending on the choice of a $\tau_\mathrm{dust}$--\NHH\ relation.

The higher opacities of NGC 4418 and \arpW\ than \arpE, as well as the opaqueness of the former 
(i.e., $\tau_\mathrm{\,1\,mm} \gtrsim 1$), are supported by the frequency dependence of source size, 
shape, and peak brightness temperature of their continuum emission as listed above in 2, 3 and 5.

\item
The core component of \arpW\ that \citet{Sakamoto17} measured at $\lambda = 3$ mm 
to have a size \about20 pc and a spectral index of $\approx$2 could be a distinct component
of high opacity and density if the 3 mm core emission is predominantly from dust.
Our BGN continuum analysis leads to 
$\tau_\mathrm{d,\, 3\, mm} \sim 5$, a column density of $\log(\NHH/\persquare{cm}) \sim 26.5$,
and a mean \HH\ density of $\nHH \sim 10^{6.7}$ \percubic{cm}.

\item 
Our Band 9 continuum images ($\lambda_\mathrm{rest} = 0.44$ mm) 
have the highest angular resolution  (\about0\farcs2) reported so far for the three nuclei in this band
and have high continuum sensitivity ($\sigma$ = 1--3 mJy \perbeam\ or 70--200 mK), 
thanks to less discernible lines and the resulting large bandwidth for continuum. 
We detected the bipolar continuum structure of \arpW\ for the first time at a submillimeter wavelength,
in the same position angle of about $165$\degr\ as seen at 3 mm in \citet{Sakamoto17}.
No bipolar feature is evident in our continuum data of \arpE\ and NGC 4418 though both have extended emission.

\item
We determined the shapes of the three nuclear disks with our multi-component visibility fitting.
We also fitted the distribution of radio supernovae (and remnants) in Arp 220.
The consistent shapes in different emissions indicate that molecular and ionized gas, dust, and young stars coexist in
the nuclear disks.
The three nuclear disks have axial ratios of 0.4--0.5 and hence inclinations of $\gtrsim$60\degr.
The nuclear disk in Arp 220W has a major-axis position angle of 83\degr, 
and therefore the bipolar outflow is nearly perpendicular to the nuclear disk at least on the sky plane.

\item 
On the basis of our experience with the three nuclei, we propose a simple diagnostic 
to identify deeply buried compact galactic nuclei using a continuum spectral index at $\lambda \sim 1$ mm from 
high-resolution observations:
\[ 
 \alpha_\mathrm{1\,mm} \leq 3 \quad \Longrightarrow \quad \log(\NHH/\persquarecm) \gtrsim 25 
 \]
for sources as radio-quiet as our three prototypes.
It should work together with the multi-transition observations and ratio analysis of molecular lines.

\item
A major uncertainty in the current analysis is due to the spherical modeling.
The estimated column densities also have uncertainty arising from the dust opacity law and the gas-to-dust ratio.
Non-spherical modeling should clarify 
whether the compact and bright continuum cores in the three nuclei are solely due to
the high inner temperature of the photon-trapping nuclei
or partly due to their disk structure with less obscuration in the polar direction.
Another remaining issue is how much the core/disk structures of the \about 1 mm continuum emission 
reflect multiple luminosity sources in each nucleus, i.e., the starburst in the nuclear disk and a central compact
source such as an AGN.

\end{enumerate}
%%%%%%%%%%%%%%%

We demonstrated in this work  
the considerable utility of millimeter-to-submillimeter continuum emission
in characterizing heavily obscured galactic nuclei of large luminosities.
For this purpose, one needs high-resolution and wide-band imaging spectroscopy
with close attention to calibration.
The continuum analysis also needs to include the effect of the source opaqueness on its thermal structure
(i.e., the greenhouse effect).
ALMA now provides the necessary observations, and we already have the thermal model to start.
Therefore, we expect similar observations and improved analyses for more of the buried galactic nuclei.

%%%%%%%%%%%%%%%%%%%%%%%%%%%%%%%%%%%%%%%%%%%%%%%%%%%%%%%%%%%%
%\clearpage
%%%%%%%%%%%%%%%%%%%%%%%%%%%%%%%%%%%%%%%%%%%%%%%%%%%%%%%%%%%%
\vspace{5mm}
%\begin{acknowledgements}
\acknowledgements
We are grateful to the ALMA Observatory and its staff members for realizing the observations used here,
and to the global ALMA team members for conducting the initial data calibration and for answering our questions
on data reduction.
This paper makes use of the following ALMA data: \\
ADS/JAO.ALMA\#2012.1.00377.S, \\
ADS/JAO.ALMA\#2012.1.00317.S,  \\
and \\
ADS/JAO.ALMA\#2012.1.00453.S. \\
ALMA is a partnership of ESO (representing its member states), NSF (USA) and NINS (Japan), 
together with NRC (Canada), MOST and ASIAA (Taiwan), and KASI (Republic of Korea), 
in cooperation with the Republic of Chile. 
The Joint ALMA Observatory is operated by ESO, AUI/NRAO and NAOJ.
This research has made use of NASA's Astrophysics Data System Bibliographic Services.
This research has also made use of the NASA/IPAC Extragalactic Database (NED), 
which is operated by the Jet Propulsion Laboratory, California Institute of Technology, 
under contract with the National Aeronautics and Space Administration.
This research has made use of the NASA/IPAC Infrared Science Archive, which is funded by the National Aeronautics and Space Administration and operated by the California Institute of Technology.
Herschel is an ESA space observatory with science instruments provided by European-led Principal Investigator consortia and with important participation from NASA.
KS is supported by grants MOST 108-2112-M-001-015 and 109-2112-M-001-020 from the Ministry of Science and Technology, Taiwan.
EG-A is a Research Associate at the Harvard-Smithsonian
Center for Astrophysics, and thanks the Spanish
Ministerio de Econom\'{\i}a y Competitividad for support under projects
ESP2017-86582-C4-1-R and PID2019-105552RB-C41.
Finally, we thank our reviewer for constructive criticism that greatly helped us to clarify this paper.
%\end{acknowledgements}

\facility{ALMA, SMA, NED, IRSA, Herschel}
\software{CASA \citep{CASA07}, 
mpfit \citep{More77,More93,Markwardt09}, 
uvmultifit \citep{uvmultifit14}}

%%%%%%%%%%%%%%%%%%%%%%%%%%%%%%%%%%%%%%%%%%%%%%%%%%%%%%%%%%%%%
\clearpage
\appendix
%%%%%%%%%%
\section{Spectral Indices of Calibrators} \label{a.calSpix}
Tables \ref{t.calibratorSpix4n4418} and \ref{t.calibratorSpix4arp220} list the spectral indices of the quasars
that we used to calibrate our observations.
They are either from the flux measurements in our observations or from observations on adjacent dates
in the ALMA Calibrator Source Catalogue (ACSC).

The spectral index of 3C273 in Table \ref{t.calibratorSpix4n4418}
 deviates from its usual value of about $-0.8$ in our B6--1.a and B6--2 observations of NGC 4418.
It appears anomalous, but we adopted the listed values for the following reasons.
The two were consecutive observations on Aug.~18th, 2014. 
ALMA observed B6--2 first with the primary calibrator Pallas and then B6--1.a with Ceres.
We measured the spectral index of 3C273 in the two observations against the primary calibrators known to be blackbody. 
Their absolute flux densities do not matter in these measurements.
For comparison,  ACSC has the Aug.~17th measurements of 3C273 in Band 3 (\about100 GHz) and Band 7 (343.5 GHz),
and the spectral index between them is $-0.43$.
From the ACSC data and our measurements, it looks as though 3C273 had a short flare at higher frequencies, making its spectral slope shallower than usual. 
Fortunately, whether to adopt $\alpha = -0.8$ or $-0.5$ makes an amplitude difference of only $\pm1$\% at the highest and lowest frequencies in these tunings around 260 GHz.

%%%%%%%%%%
\section{Notes on Band 9 Flux Calibration} \label{a.b9fluxcal}
ALMA has a set of primary flux calibrators, i.e., sources such as solar-system objects that have models for flux densities as a function of frequency and the date and time of the observations.
However, some ALMA observations do not contain integrations on any primary flux calibrator (because, for example,
no suitable primary calibrator was at high enough elevation during the observations.)
Those observations are supposed to be flux calibrated through secondary calibrators, i.e., quasars that the observatory monitors for their time-variable flux densities and publish their measurements in ACSC.  
This monitoring is conducted mainly in Band 3 (\about100 GHz) and Band 7 (\about340 GHz)
and occasionally in Band 6 (\about230 GHz). 
Apparently, the monitoring observations are exceedingly rare in Band 8 and above (i.e., $>$ 400 GHz). 
The flux densities of secondary calibrators at such high frequencies 
are supposed to be estimated with the power-law extrapolation from the monitoring observations at lower frequencies.

%%% 
\subsection{For Arp~220}
Our two tunings for Arp~220 were observed consecutively on June 9th, 2015.
Both used the same calibrators, 3C~279 and J1550+0527, and had no primary calibrator.
Figure \ref{f.B9fluxcal} shows twelve measurements of 3C~279 found in the ACSC within ten days of our observations.
The best-fit power law to these measurements, the dotted line in Figure \ref{f.B9fluxcal}, has
a spectral index of $-0.619 \pm 0.013$ and flux density of 3.73 Jy at 670 GHz.
(A fit using only three measurements within five days from our observations gives 3.82 Jy and $-0.60$.) 
Among the twelve measurements is one in Band 9 on May 31st or nine days before our observations. 
It is one of the only two Band 9 measurements of 3C~279 in the entire ACSC as of this writing,
and its entry is $3.36\pm 0.34$ at 373 GHz, 
which is 1$\sigma$ below the best-fit power law.

We adopted a spectral index of $0.62$ and flux density of 3.5 Jy at 670 GHz for 3C~279 
to flux calibrate both of our two datasets. 
As expected, the two tunings matched very well, within 2\%, in the overlapping channels of their Arp~220 spectra. 
For comparison, the data delivered to us from the observatory were calibrated by the ALMA staff 
using a flux scale where 3C~279 was at about 3.1--3.2 Jy around 670 GHz. 
If that calibration were correct, the spectrum of 3C~279 must have a bend 
to deviate from a power law between Band 3 and 9. 
To check the sub/millimeter spectrum of 3C~279, 
we took a set of its ACSC data measured on May 31st, 2015, in Bands 3, 6, 7, and 9. 
We took another dataset in Bands 3, 7, 9, and 10 (861 GHz) obtained between Nov.~27th and Dec.~2nd in 2018. 
On both occasions, the measurements are consistent with a power-law spectrum without a bend. 
We were therefore not compelled to adopt the lower flux scale.

The elevation-dependent atmospheric attenuation also significantly affects the final flux scale of the 
science data product. 
Our flux standard 3C~279 and science target Arp~220 were at mean elevations of 69\degr (58\degr) and 40\degr (44\degr), respectively, in our B9--1 (B9--2) observations. 
Precipitable water vapor during our observations was 0.37 mm for B9--1 and 0.34 mm for B9--2, and the zenith opacity was about 0.65 according to the ALMA atmosphere model\footnote{\sf https://almascience.nao.ac.jp/about-alma/atmosphere-model}.
Therefore, without the correction for the differential atmospheric attenuation between 3C~279 and Arp~220, 
Arp 220 at lower elevations would have been about 40\% (20\%) dimmer than it should be in B9--1 (B9--2).
The correction is applied using system temperatures in the standard ALMA procedure in CASA. 
We are not aware of how accurate this correction is supposed to be. 
However, we note again that the flux of Arp~220 agreed to 2\% in B9--1 and B9--2 at their overlaps.
It is despite the different magnitudes of their attenuation corrections.
Therefore, we do not expect a significant elevation-dependent error in the amplitude of our calibrated data.

%%% 
\subsection{For NGC~4418} 
NGC~4418 was observed three times in Band 9; twice for our B9--1 tuning and once for B9--2.
The calibrators were 3C~279, 3C~273, Ceres, and Titan.
We consulted the flux calibration against the primary calibrators and lower-frequency measurements in the ACSC, 
which has no 3C~273 data in Band 9 (or above 400 GHz), 
to decide the flux densities and the spectral indices of the quasars.  
The latter are in Tables \ref{t.calibratorSpix4n4418} and \ref{t.calibratorSpix4arp220}.
For flux densities at 670 GHz, we adopted 4.0 Jy for 3C~279 and 1.05 Jy for 3C~273 in B9--1a, 
3.5 Jy for 3C~279 and 2.55 Jy for 3C~273 in B9--1b, and 2.50 Jy for 3C~273 in B9--2. 
These are within 8\% of those used in the observatory-provided calibrations.
It turned out that the resulting flux densities of NGC~4418 agree to 1\% 
between the two B9--1 observations that are eleven months apart from each other. 
The total B9--1 data and the B9--2 data taken two weeks after the second B9--1 observations
 agreed to 1.5\% on the flux density at their overlapping channels.

%%%%%%%%%%
\section{Spectra in Larger Apertures} \label{a.lowResSpectra}
Figures \ref{f.spec_with_cont.largeAperture.ylog.N4418}
and \ref{f.spec_with_cont.largeAperture.ylog.A220} show the spectra of NGC~4418 and Arp~220 sampled 
from our 0\farcs35 image cubes in circular apertures of diameter 1\arcsec\ (NGC~4418) and 2\arcsec\ (Arp~220). 
We centered the former aperture at the nucleus and the latter at the midpoint of the two nuclei.
The major lines are labeled.
Many lines are less evident in these spectra than in smaller apertures.
It is because these large apertures dilute lines having small emitting areas.
It is also because they can contain circumnuclear emission to fill absorption against the compact continuum nuclei.
Lines having extended emitting regions appear more prominently in these spectra.
In Arp~220, lines from around the two nuclei blend 
since their systemic velocities differ by only 100 \kms\ from each other \citesp{Paper2}. 
Notable features that \citest{Paper2} addresses include: 
the lines from vibrationally-excited molecules; 
their contrast between HCN and \HCOplus\ noticed by Sakamoto et al. (2010); 
the different \thirteenCO-to-\CeighteenO\ ratios in NGC 4418 and Arp 220. 

%%%%%%%%%%%%%%%%%%%%%%%%%%%%%%%%%%%%%%%%%%%%%%%%%%%%%%%%%%%%%
\clearpage
% References

%%%%%%%%%%%%%%%%%%%%%%%%%%
%%%%%%%%%%%%%%%%%%%%%%%%%%%%%%%%%%%%%%%%%%%%%%%%%%%%%%%%
%%%%% Tables %%%%%
\clearpage
%%%%%%%%%%%%%%%%%%%%%%%%%%%%%%%%%%%%%%%%%%%%%%%%%%%%%%%%%%%%%%%%%%%%%%%%
% Table 1: Science Goals (one line for each freq_setup)
\begin{deluxetable}{lccccccccll}
\tabletypesize{\scriptsize}
\tablewidth{0pt}
\tablecaption{List of Tunings  and Major Lines in each \label{t.tunings} }
\tablehead{ 
       \colhead{name} &
	\multicolumn{4}{c}{\fobs/GHz} & 
	\multicolumn{4}{c}{\frest/GHz} & 
	\multicolumn{2}{c}{Major Molecules with Lines}
	\\
	\colhead{}  &
	\multicolumn{2}{c}{LSB} &
	\multicolumn{2}{c}{USB} &	
	\multicolumn{2}{c}{LSB} &
	\multicolumn{2}{c}{USB} &	
	\colhead{LSB} &
	\colhead{USB} 	
}
\colnumbers
\startdata
\sidehead{NGC 4418} 
B9--1 & 676.844 & 680.406 & 688.594 & 692.156 &   681.618 & 685.206 & 693.451 & 697.039 &            & SiO \\
B9--2 & 673.558 & 677.078 & 685.247 & 688.809 &   678.309 & 681.855 & 690.081 & 693.668 & CN       & CO, \HthirteenCN \\
B7--1 & 348.194 & 351.756 & 360.444 & 364.006 &   350.650 & 354.238 & 362.986 & 366.574 & \HtwoCO & \HtwoCO, HNC\vib, \HCthreeN\vib \\
B7--2 & 344.968 & 348.518 & 357.102 & 360.652 &   347.401 & 350.976 & 359.621 & 363.196 & \HNthirteenC, \CtwoH & \CHthreeOH, HNC \\
B7--3 & 341.755 & 345.276 & 353.777 & 357.319 &   344.166 & 347.711 & 356.272 & 359.839 & \HthirteenCN, CO, \HthirteenCOplus, SiO & \HCOplus, \HOCplus \\
B7--4 & 338.551 & 342.056 & 350.474 & 353.979 &   340.939 & 344.469 & 352.946 & 356.476 & CS & HCN, HCN\vib \\
B6--1 & 250.182 & 253.743 & 266.057 & 269.618 &   251.947 & 255.533 & 267.934 & 271.520 & c-C3H2, \HCthreeN & \HCOplus\vib \\
B6--2 & 247.034 & 250.581 & 262.717 & 266.264 &   248.777 & 252.349 & 264.570 & 268.142 &  & HCN, HCN\vib, \HCOplus \\
B6--3 & 217.333 & 220.708 & 230.833 & 234.442 &   218.866 & 222.265 & 232.461 & 236.096 & \thirteenCO, \CeighteenO & \\
B6--4 & 214.122 & 217.451 & 227.619 & 230.948 &   215.633 & 218.985 & 229.225 & 232.577 & \HtwoS, SiO, \HtwoCO, \HCthreeN & CO, \thirteenCS, \NtwoDplus \\
\sidehead{Arp 220} 
B9--1 & 660.969 & 664.531 & 680.469 & 684.031 &   672.979 & 676.606 & 692.833 & 696.460 & \HtwoCO & SiO \\
B9--2 & 657.724 & 661.286 & 677.119 & 680.681 &   669.675 & 673.302 & 689.423 & 693.049 &                & CO, \HthirteenCN\\
B7--1 & 344.694 & 348.256 & 356.944 & 360.506 &   350.957 & 354.584 & 363.429 & 367.057 & \HtwoCO & \HtwoCO \\
B7--2 & 341.466 & 345.015 & 353.599 & 357.148 &   347.670 & 351.284 & 360.023 & 363.637 & \HNthirteenC, CCH & HNC \\
B7--3 & 338.250 & 341.770 & 350.270 & 353.812 &   344.396 & 347.980 & 356.635 & 360.241 & \HthirteenCN, CO, \HthirteenCOplus, SiO & \HOCplus, \HCOplus \\
B7--4 & 335.044 & 338.547 & 346.965 & 350.467 &   341.132 & 344.698 & 353.269 & 356.835 & CS, \HCfifteenN & HCN, HCN\vib, \HCOplus \\
B6--1 & 245.992 & 249.553 & 261.867 & 265.428 &   250.461 & 254.088 & 266.625 & 270.251 &       & HCN\vib, \HCOplus, \HOCplus \\
B6--2 & 242.846 & 246.394 & 258.527 & 262.075 &   247.258 & 250.871 & 263.224 & 266.837 &       & \HCthreeN, HCN \\
B6--3 & 224.515 & 228.068 & 239.504 & 243.057 &   228.595 & 232.212 & 243.856 & 247.473 & CO & CS, \HCthreeN \\
B6--4 & 214.370 & 217.702 & 227.870 & 231.202 &   218.265 & 221.658 & 232.010 & 235.403 & \thirteenCO, \CeighteenO & \\
\enddata
\tablecomments{
(1) Name of the frequency setup. 
(2) and (3) are the minimum and maximum frequencies, respectively, in the LSB; (4) and (5) are the same for the USB.
The total sky-frequency coverage (excluding overlaps) is 66.2 GHz for both galaxies.
(6)--(9) are the same as (2)--(5) but at the rest frame of the target galaxy. 
We used $V$(radio, LSRK)=2100 \kms\ and 5350 \kms\ for NGC 4418 and Arp 220, respectively.
The total coverage in rest frequency is 66.7 and 67.4 GHz for NGC 4418 and Arp 220, respectively. 
(10) and (11) are notable molecules whose lines appear in LSB and USB, respectively. 
Vibrationally-excited species are marked with an asterisk in superscript.
}
\end{deluxetable}

%%%%%%%%%%%%%%%%%%%%%%%%%%%%%%%%%%%%%%%%%%%%%%%%%%%%%%%%%%%%%%%%%%%%%%%%		% Table 1: List of tunings
	%\label{t.tunings} 
	
%%%%%%%%%%%%%%%%%%%%%%%%%%%%%%%%%%%%%%%%%%%%%%%%%%%%%%%%%%%%%%%%%%%%%%%%
% Table 2: Log of observations (i.e., Execution Blocks)
\begin{deluxetable}{lccclllrrrrr}
\tablecolumns{11}
\tabletypesize{\scriptsize}
\tablewidth{0pt}
\tablecaption{Observation Log  \label{t.obslog} }
\tablehead{ 
       \colhead{name} & 
       \colhead{UT date} &
       \colhead{$T_{\rm tel}$} &       
       \colhead{$N_{\rm ant}$} &
       \colhead{flux cal.} &
       \colhead{bandpass cal.} &
       \colhead{gain cal.} &
       \colhead{$T_{\rm on}$} &  
       \colhead{$\langle T_{\rm sys} \rangle$} &   
	\multicolumn{2}{c}{uv range} &
	\colhead{$\theta_{\rm MRS}$}
	\\
	\colhead{}  &	
	\colhead{yyyy-mm-dd} &
	\colhead{min} &	
	\colhead{}  &	
	\colhead{} &
	\colhead{} &
	\colhead{} &
	\colhead{min} &
	\colhead{K} &
	\multicolumn{2}{c}{m} &
	\colhead{arcsec} 
}
\colnumbers
\startdata
%%%%%%%%%%%%%%%%
\sidehead{NGC 4418}
B9--1.a & 2014--06--17 & 39.9 & 28 & Ceres  & 3C279 & 3C273 & 8.2 & 1037 & 19.1 & 618.3 & 2.8 \\
B9--1.b & 2015--05--19 & 51.9 & 34 & Titan   & 3C279 & 3C273 & 7.6 & 704  & 16.3 & 472.5 & 3.3 \\
B9--2   & 2015--06--02 & 49.6 & 37 & 3C273 & 3C279 & 3C273 & 9.2 & 998  &  21.3 & 718.2 & 2.5 \\
B7--1.a & 2013--06--01 & 28.0 & 26 & Titan   & 3C279 & 3C273 & 7.2 &  175 &14.6 & 1266.4 & 7.0\\
B7--1.b & 2015--07--24 & 36.7 & 42 & 3C273 & 3C273 & 3C273  & 7.2 &  169 &13.5 & 1503.4 & 7.6\\
B7--2 &  2015--07--24  & 37.7 & 41 & 3C273 &  3C273 &  3C273 & 8.4 &  179 &13.9 & 1562.8 & 7.4\\  
B7--3 &  2015--07--24  & 37.7 & 42 & 3C273 &  3C273 &  3C273 & 8.4 & 157 & 13.2 &  1398.4 & 7.9\\   
B7--4 &  2015--06--29  & 37.7 & 42 & 3C273 &  3C279 &  3C273 & 8.4 & 202 & 38.5 & 1487.1 & 2.7\\                                           
B6--1.a &   2014--08--17 & 24.3 & 30 &    (Ceres)    &  3C273 &  3C273 &   5.8  & 82 & 18.1 &1132.2 & 7.6 \\   
B6--1.b &   2015--07--18 & 25.2 & 39 &    3C273     &  3C279 &  3C273 &   5.8  & 73 & 15.0 &1442.8 & 9.2 \\   
B6--2 &     2014--08--17 & 23.7 & 31 &    Pallas         &  3C273 &  3C273 &   6.4  & 88 & 17.9 &1082.7 & 7.8 \\   
B6--3 &     2015--07--18 & 24.6 & 38 &     3C273      &  3C279 &  3C273 &   5.3  & 66 & 15.1 & 1387.9 & 10.5 \\   
B6--4 &     2015--07--18 & 24.6 & 39 & J1058+015 &  3C279 &  3C273 &   5.3  & 65 & 15.0 & 1549.2 & 10.7 \\   
%%%%%%%%%%%%%%%%
\sidehead{Arp 220}
B9--1   &  2015--06--09 & 57.9 & 34 & 3C279        & 3C279            & J1550+0527 & 7.6 & 1024 & 20.1 & 543.2 & 2.7 \\
B9--2   &  2015--06--09 & 62.1 & 34 & 3C279        & 3C279            & J1550+0527 & 8.5 &  939 & 18.1 & 598.2 & 3.0 \\
B7--1   &  2015--07--17 & 40.2 & 38 & J1550+054 & J1337$-$125 & J1516+193 & 9.8 & 168 & 19.0 & 1572.8 & 5.4 \\
B7--2   &  2015--06--28 & 42.3 & 40 & J1550+054 & J1751+093 & J1516+193 & 9.8 & 204 & 30.6 & 1520.6 & 3.4 \\
B7--3   &  2015--06--27 & 41.7 & 39 & J1550+054 & J1751+093 & J1516+193 & 9.3 & 146 & 30.6 & 1557.8 & 3.4 \\
B7--4   &  2015--06--28 & 45.0 & 39 & J1550+054 & J1337$-$125 & J1516+193 & 9.8 & 188 & 34.8 & 1570.5 & 3.0 \\
B6--1   &  2015--06--30 & 33.1 & 41 & Titan     & J1550+052 & J1550+052 & 8.6 & 101 & 33.0 & 1556.6 & 4.2 \\
B6--2   &  2015--06--30 & 30.4 & 41 & Titan     & J1550+052 & J1550+052 & 8.2 & 100 & 38.3 & 1569.6 & 3.7 \\
B6--3.a &  2015--06--30 & 29.8 & 41 & Titan     & J1550+052 & J1550+052 & 7.7 &  86 & 36.3 & 1574.1 & 4.2 \\
B6--3.b &  2015--08--04 & 35.2 & 36 & Titan     & J1550+052 & J1550+052 & 7.7 &  87 & 39.1 & 1572.9 & 3.9 \\
B6--4   &  2015--06--27 & 34.7 & 40 & Titan     & J1550+052 & J1550+052 & 7.2 &  70 & 28.8 & 1479.6 & 5.6 \\
\enddata
\tablecomments{
(1) Tuning name and a, b, ..., for the first, second, ..., useable executions when there are multiple executions.
(3) Duration of the observations. 
(4) Number of antennas in the array, excluding those entirely flagged.
(5), (6), (7) Flux, bandpass, and gain calibrators, respectively. 
The B6--1a data of NGC 4418 were calibrated without referring to Ceres but using nearby quasar measurements.
(8) On-source integration time for the target galaxy.
(9) Median system temperature.  
(10) and (11) The minimum and maximum projected baseline lengths for the target galaxy.
(12) Maximum recoverable scale calculated with $0.6 \lambda / L_{\rm min}$ (ALMA Cycle 2 Technical Handbook) for the highest frequency in each tuning.
}
\end{deluxetable}

%%%%%%%%%%%%%%%%%%%%%%%%%%%%%%%%%%%%%%%%%%%%%%%%%%%%%%%%%%%%%%%%%%%%%%%%              % Table 2: Obs. log
	% \label{t.obslog}

%%%%%%%%%%%%%%%%%%%%%%%%%%%%%%%%%%%%%%%%%%%%%%%%%%%%%%%%%%%%%%%%%%%%%%%%
% Table 3 : Data parameters (rms). (one line for each freq_setup)
\begin{deluxetable}{ccCCRRRR} 
\tabletypesize{\scriptsize}
\tablewidth{0pt}
\tablecaption{Data parameters  \label{t.dataparams} }
\tablehead{ 
       \colhead{name} &
       \colhead{SB} &       
       \colhead{maj} &
       \colhead{min} &
       \colhead{p.a.} &
	\multicolumn{3}{c}{$\sigma_\mathrm{20\,MHz}$}
	\\
	\colhead{ }  &
	\colhead{ }  &
	\colhead{ }  &
	\colhead{ }  &
	\colhead{ }  &
	\colhead{native}  &		                     
	\colhead{0\farcs2 } & 
       \colhead{0\farcs35 }      
       \\
	\colhead{ }  &
	\colhead{ }  &
	\colhead{\asec} &
	\colhead{\asec} &
	\colhead{\arcdeg} &
    \multicolumn{3}{c}{mJy \perbeam}  		
}
\colnumbers
\startdata
\sidehead{NGC 4418}
B9--1 & U & 0.25 & 0.19 &  79 & 14.5 & \nd & 16.8 \\
B9--1 & L & 0.26 & 0.19 & -62 & 12.2 & \nd & 14.1 \\
B9--2 & U & 0.20 & 0.15 &  81 & 26.3 & \nd & 34.1 \\
B9--2 & L & 0.20 & 0.15 &  80  & 22.9 & \nd & 28.9 \\
B7--1 & U & 0.19 & 0.15 &  74 & 1.8 & 1.9 & 3.1 \\
B7--1 & L & 0.18 & 0.15 &  69 & 1.3 & 1.3 & 2.0 \\
B7--2 & U & 0.16 & 0.13 &  60 & 2.0 & 2.1 & 3.2 \\
B7--2 & L & 0.16 & 0.14 &  59 & 1.5 & 1.6 & 2.4 \\
B7--3 & U & 0.17 & 0.14 &  77 & 1.7 & 1.7 & 2.5 \\
B7--3 & L & 0.17 & 0.14 & \m90 & 1.4 & 1.4 & 2.1 \\
B7--4 & U & 0.15 & 0.14 & \m67 & 2.5 & 2.6 & 4.0 \\
B7--4 & L & 0.16 & 0.14 & \m67 & 2.0 & 2.1 & 3.2 \\
B6--1 & U & 0.24 & 0.20 &  54 & 0.8 & \nd & 0.9 \\
B6--1 & L & 0.24 & 0.20 &  53 & 0.6 & \nd & 0.8 \\
B6--2 & U & 0.26 & 0.24 & \m80 & 1.1 & \nd & 1.3 \\
B6--2 & L & 0.28 & 0.26 & \m77 & 1.1 & \nd & 1.2 \\
B6--3 & U & 0.31 & 0.20 &  52 & 0.9 & \nd & 1.0 \\
B6--3 & L & 0.32 & 0.22 &  53 & 0.8 & \nd & 0.9 \\
B6--4 & U & 0.34 & 0.21 &  55 & 0.8 & \nd & 0.9 \\
B6--4 & L & 0.35 & 0.22 &  55 & 0.8 & \nd & 0.9 \\
\sidehead{Arp 220}
B9--1 & U & 0.21 & 0.16 & 38 & 21.0 & \nd & 27.0 \\
B9--1 & L & 0.22 & 0.18 & 35 & 17.3 & \nd & 22.0 \\
B9--2 & U & 0.20 & 0.17 & -12 & 17.4 & \nd & 25.1 \\
B9--2 & L & 0.21 & 0.18 & -11 & 15.6 & \nd & 20.3 \\
B7--1 & U & 0.20 & 0.14 &   21 & 2.0 & 2.5 & 3.0 \\
B7--1 & L & 0.20 & 0.14 &   13 & 1.5 & 1.8 & 2.4 \\
B7--2 & U & 0.19 & 0.14 &  -22 & 2.2 & 3.0 & 3.5 \\
B7--2 & L & 0.20 & 0.14 &  -23 & 1.8 & 2.3 & 2.6 \\
B7--3 & U & 0.19 & 0.14 &  -13 & 1.7 & 2.3 & 2.8 \\
B7--3 & L & 0.20 & 0.15 &   -9 & 1.4 & 1.8 & 2.1 \\ 
B7--4 & U & 0.18 & 0.14 & -179 & 1.9 & 2.5 & 3.2 \\
B7--4 & L & 0.19 & 0.14 &   -6 & 1.7 & 2.2 & 2.7 \\
B6--1 & U & 0.25 & 0.18 &  -10 & 1.1 & \nd & 1.4 \\
B6--1 & L & 0.27 & 0.19 &  -11 & 1.0 & \nd & 1.1 \\
B6--2 & U & 0.26 & 0.18 &   12 & 1.0 & \nd & 1.2 \\
B6--2 & L & 0.28 & 0.19 &   11 & 0.9 & \nd & 1.0 \\
B6--3 & U & 0.26 & 0.19 &   11 & 0.7 & \nd & 0.8 \\
B6--3 & L & 0.29 & 0.21 &   13 & 0.6 & \nd & 0.7 \\ 
B6--4 & U & 0.33 & 0.22 &  -30 & 0.7 & \nd & 0.8 \\
B6--4 & L & 0.34 & 0.22 &  -29 & 0.7 & \nd & 0.8 \\
\enddata
\tablecomments{
(1) Tuning name.
(2) Sideband.
(3)--(5) Synthesized beam major and minor axes and position angle of the major axis measured counterclockwise from the north for
the {\tt robust} weighting parameter of  0.5.
(6) Noise rms per 20 MHz channel.  Channels containing a CO line are excluded from measurement.
(7) The same as (6) but for the data convolved to 0\farcs2 resolution. (For this, we imaged Arp 220 with {\tt robust = 0}.)
(8) The same as (6) but for the data convolved to 0\farcs35.
}
\end{deluxetable}

%%%%%%%%%%%%%%%%%%%%%%%%%%%%%%%%%%%%%%%%%%%%%%%%%%%%%%%%%%%%%%%%%%%%%%%%	% Table 3: data cube parameters
	% \label{t.dataparams}
	
%%%%%%%%%%%%%%%%%%%%%%%%%%%%%%%%%%%%%%%%%%%%%%%%%%%%%%%%%%%%%%%%%%%%%%%%
% Table : Continuum model-1G
\begin{deluxetable}{ccccccrcrc}
\tabletypesize{\scriptsize}
\tablewidth{0pt}
\tablecaption{Continuum Model : 1 Gaussian \label{t.cont_model-1G_parameters} }
\tablehead{
       \colhead{freq. range} &
       \colhead{$\nu_{\rm ref}$} &
       \colhead{$I_{\rm ref}$} &
       \colhead{$\alpha$} &   
       \colhead{$\theta_{\rm maj}$} &             
        \colhead{$\theta_{\rm min}$} &  	          
 	\colhead{P.A.} & 
       \colhead{$\eta_{\rm 1G}$} &              
       \colhead{$S_{\rm ref}^{\rm (1G)}$} &
       \colhead{$T_{{\rm b, ref}}^{\rm (1G)}$}
       \\
	\colhead{GHz}  &
	\colhead{GHz} &
	\colhead{mJy \perbeam} &
	\colhead{} &	
	\colhead{mas} &
	\colhead{mas} &		
	\colhead{deg.} &
	\colhead{} &	
	\colhead{mJy} &	
       \colhead{K} 
}
\colnumbers
\startdata 
\sidehead{NGC 4418}  
673--693 &  680 &   845.1       & 2.18  & 172.7       & \phn93.8 & \phn46.0   &  0.866   & 974.8      & 159 \\
337--364 &  350 &   131.7       & 2.37  & \phn92.3  & \phn63.7 & \phn48.5   &  0.951   & 138.4      & 235 \\ 
247--270 &  260 &   \phn62.9  & 2.19  & \phn95.5  & \phn61.6 & \phn47.2   &  0.950   & \phn66.2  & 203 \\ 
214--235 &  225 &   \phn47.2 & 1.81   & 107.9       & \phn60.1 & \phn47.4   &  0.942   & \phn50.1  & 187 \\ 
\sidehead{\arpE}  
657--685 & 670  &  1053         & 3.32  & 332.9       & 185.3 & \phn48.8 & 0.640 & 1644        & \phn73 \\
335--361 & 350  & 127.2         & 2.84  & 258.0       & 145.2 & \phn51.0 & 0.743 & 171.1       & \phn46  \\ 
214--266 & 240  & \phn36.7   & 2.90  & 255.1        & 156.1 & \phn47.2 & 0.738 & \phn49.7 & \phn26 \\ 
\sidehead{\arpW}  
657--685 & 670  &  1750        & 2.80  & 204.1        & 169.2 & \phn92.6 & 0.778 & 2250         & 177 \\
335--361 & 350  & 317.7        & 1.70  & 149.3        & 120.2 & \phn89.8 & 0.870 & 365.2       & 203 \\ 
214--266 & 240  & 115.4        & 2.37  & 144.0        & 127.5 & 101.5      & 0.869 & 132.8       & 153 \\ 
\enddata
\tablecomments{
Parameters of our continuum models.
(1) Frequency range for the model.
(2) Reference frequency for the power-law spectrum.
(3) and (4) Parameters describing our model continuum spectra for a 0\farcs35 beam.
They are displayed with many digits to precisely describe the models used, e.g., for continuum subtraction. 
It does not imply their accuracy.
The flux density in (3) is precise in relation to our dataset, but not accurate to the fourth digits in the absolute scale.
Likewise, the local spectral indices here have large uncertainties due to thermal noise, residual errors of flux self-calibration,
line contamination, and residual errors in bandpass calibration.
The signal-to-noise ratio of the continuum emission in a channel is 
about 40, 50, and 100, respectively, for NGC 4418, Arp 220E, and Arp 220W 
accross all our observed bands.
Combining them with the averaging of a few channels around each local minima and the fractional bandwidths
of individual frequency sections, 
we estimate
the uncertainties of 0.4, 0.3, and 0.2, respectively, for the spectral indices in Band 7, upper Band 6, and lower Band 6 
of NGC 4418; thermal noise dominates in these uncertainties.
Arp 220E should have errors of about 0.4 and 0.15, respectively, in Band 7 and 6.
The errors for the twice brighter Arp 220W should be less than these, but the reduction should be less than a factor of 2 
because of calibration errors at the 1\% level.
We obtain better estimates of the spectral indices in Section \ref{s.cont.spix}.
(5) and (6) Source FWHM along the major and minor axes.
(7) Position angle of the major axis, measured counter-clockwise from the north.
(8) Coupling efficiency of the model source with our 0\farcs35 beam.
(9) Total flux density of the model source at the reference frequency, corrected for the source-beam coupling (i.e., $= I_{\rm ref}/\eta_{\rm 1G}$).
(10) Peak Rayleigh-Jeans brightness temperature of the model.
See text for more model descriptions.
}
\end{deluxetable}

%%%%%%%%%%%%%%%%%%%%%%%%%%%%%%%%%%%%%%%%%%%%%%%%%%%%%%%%%%%%%%%%%%%%%%%%       % Table 4: 1G model parameters          
	% \label{t.cont_model-1G_parameters}

%%%%%%%%%%%%%%%%%%%%%%%%%%%%%%%%%%%%%%%%%%%%%%%%%%%%%%%%%%%%%%%%%%%%%%%%
% Table : Continuum parameters (from 1-Gaussian visibility fitting)
\begin{deluxetable}{ccccccccc}
\tabletypesize{\scriptsize}
\tablewidth{0pt}
\tablecaption{Continuum Visibility Fits : 1 Gaussian  \label{t.cont_visfit_1G_params} }
\tablehead{ 
       \colhead{$\nu_{\rm obs}$} &
       \colhead{$N_{\rm ch}$} &
       \colhead{$\theta_{\rm maj}$} &
       \colhead{$\theta_{\rm min}$} &
	\colhead{P.A.} &                            
       \colhead{$\frac{\theta_{\rm min}}{\theta_{\rm maj}}$} &       
	\colhead{$S_{\nu}^{(0)}$} &
	\colhead{$T_{{\rm b}, \nu}^{(0)}$} &	
	\colhead{$\langle \chi^2 \rangle$/d.o.f}        
       \\
	\colhead{GHz}  &
	\colhead{} &
	\colhead{mas} &
	\colhead{mas} &
	\colhead{\arcdeg} &
	\colhead{} &		
	\colhead{mJy} &
	\colhead{K} &	
	\colhead{}
	\\
	\colhead{(1)}  &
	\colhead{(2)}  &
	\colhead{(3)} &
	\colhead{(4)} &
	\colhead{(5)} &	
	\colhead{(6)} &	
	\colhead{(7)} &
	\colhead{(8)} &	
	\colhead{(9)}
}
\startdata
\sidehead{NGC 4418}  
690.288    &   $\phn70$    &   $171.4 \pm 1.0$          &   $\phn91.2 \pm  1.6$   &   $46.5 \pm  0.6$  &   $0.522 \pm 0.010$  & $999.0 \pm 3.0$       & $156.4 \pm 2.6$ & 1.2 \\ 
677.761    &   $130$         &   $173.3 \pm 0.7$          &   $\phn95.0 \pm  1.0$   &   $45.8 \pm  0.4$  &   $0.537 \pm 0.006$  & $972.0 \pm 2.0$       & $149.1 \pm 1.5$ & 1.3 \\ 
352.686    &   $\phn32$    &   $\phn92.3 \pm  0.8$    &   $\phn63.7 \pm  0.9$   &   $48.5 \pm  1.1$  &   $0.685 \pm 0.012$  & $142.6 \pm 1.4$       & $231.7 \pm 3.4$ & 1.2 \\
268.788    &   $\phn33$    &   $\phn94.9 \pm  1.2$    &   $\phn62.2 \pm  1.1$   &   $45.6 \pm  1.3$  &   $0.647 \pm 0.014$  & $\phn73.1 \pm 0.2$ & $206.9 \pm 4.0$ & 1.3 \\
250.750    &   $\phn38$    &   $\phn96.1 \pm  1.2$    &   $\phn60.9 \pm  1.2$   &   $48.7 \pm  1.3$  &   $0.625 \pm 0.015$  & $\phn62.8 \pm 0.2$ & $202.3 \pm 4.4$ & 1.3 \\
232.662    &   $\phn78$    &   $104.1 \pm  1.3$         &   $\phn59.5 \pm  1.1$   &   $48.5 \pm  0.9$  &   $0.551 \pm 0.013$  & $\phn54.4 \pm 0.1$ & $185.2 \pm 3.9$ & 1.2 \\
214.975    &   $\phn26$    &   $122.2 \pm  2.6$         &   $\phn62.7 \pm  2.3$   &   $45.0 \pm  1.4$  &   $0.481 \pm 0.022$  & $\phn48.0 \pm 0.4$ & $144.0 \pm 6.0$ & 1.3 \\
\sidehead{\arpE} 
680.921    &   $\phn83$    &   $335.8  \pm 1.3$    &  $184.0 \pm  \phn0.9$    &  $49.3 \pm  0.3$   &  $0.546 \pm 0.003$    &  $1761.4 \pm 6.0$     & $\phn74.4 \pm 0.4$ & 1.3 \\ 
660.893    &   $172$         &   $331.8  \pm 0.8$    &  $185.9 \pm  \phn0.6$    &  $48.5 \pm  0.2$   &  $0.558 \pm 0.002$    &  $1589.0 \pm 3.0$     & $\phn71.6 \pm 0.2$ & 1.3 \\ 
347.046   &   $218$             & $258.0 \pm  0.5$   &  $145.2 \pm  \phn0.5$    &  $51.0 \pm  0.2$   &  $0.560 \pm 0.002$    &  $172.4 \pm 0.8$       & $\phn45.9 \pm 0.1$ & 1.2 \\
264.145   &   $\phn40$        & $257.4 \pm  2.0$   &  $162.6 \pm  \phn2.0$    &  $46.5 \pm  0.8$   &  $0.609 \pm 0.009$    &  $\phn68.7 \pm 0.6$  & $\phn27.7 \pm 0.3$ & 1.2 \\
246.253   &   $\phn26$        & $266.0 \pm  2.4$   &  $152.1 \pm  \phn2.3$    &  $46.2 \pm  0.8$   &  $0.564 \pm 0.010$    &  $\phn56.7 \pm 0.5$  & $\phn27.4 \pm 0.4$ & 1.2 \\
229.703   &   $\phn75$        & $250.8 \pm  1.3$   &  $153.6 \pm  \phn1.8$    &  $48.0 \pm  0.6$   &  $0.604 \pm 0.008$    &  $\phn44.8 \pm 0.2$  & $\phn26.1 \pm 0.3$ & 1.3 \\
217.570   &   $\phn\phn4$   & $276.5 \pm  9.6$   &  $130.9 \pm 15.6$          &  $50.0 \pm  3.0$   &  $0.470 \pm 0.059$    &  $\phn38.8 \pm 0.3$  & $\phn26.4 \pm 3.0$ & 2.2 \\
\sidehead{\arpW} 
681.543   &   $\phn49$   & $206.5 \pm 0.8$ & $168.9 \pm 0.8$ & $\phn91.2 \pm 0.9$  & $0.815 \pm 0.005$ & $2400.2 \pm 6.0$ & $179.7 \pm \phn0.8$ & 1.2 \\ 
661.197   &   $158$        & $203.6 \pm 0.3$ & $169.2 \pm 0.4$ & $\phn93.0 \pm 0.4$  & $0.829 \pm 0.002$ & $2206.7 \pm 2.7$ & $178.1 \pm \phn0.4$ & 1.3 \\ 
345.545   &   $156$        & $149.3 \pm 0.2$ & $120.2 \pm 0.2$ & $\phn89.8 \pm 0.3$  & $0.804 \pm 0.002$ & $367.4 \pm 1.2$   & $207.3 \pm \phn0.4$ & 1.2 \\       
264.122   &   $\phn17$   & $142.9 \pm 0.8$ & $128.3 \pm 1.0$ & $103.3 \pm  2.4$      & $0.892 \pm 0.009$ & $173.9 \pm 0.5$   & $165.2 \pm \phn1.4$ & 1.2 \\
246.091   &   $\phn44$   & $143.9 \pm 0.5$ & $126.9 \pm 0.6$ & $\phn94.6 \pm  1.2$ & $0.876 \pm 0.005$ & $146.9 \pm 0.9$   & $160.7 \pm \phn0.9$ & 1.3 \\
230.210   &   $\phn40$   & $144.8 \pm 0.6$ & $127.8 \pm 0.6$ & $110.6 \pm  1.5$      & $0.875 \pm 0.006$ & $124.7 \pm 0.4$   & $154.3 \pm \phn0.9$ & 1.4 \\ 
\enddata
\tablecomments{
Parameters of the single-gaussian (1G) fitting to the visibilities in the continuum-dominated channels.
For Arp 220, the two nuclei were simultaneously fitted using one Gaussian for each. 
Emission in the channels used here is dominated by continuum emission, having less than 5\% (Band 6 and 7) or 1\% (Band 9)
contribution of line emission in our 0\farcs35 spectrum.
Those channels were individually fitted, and the errors of the derived parameters were rescaled so that the reduced $\chi^2$ is unity.
The parameters in the continuum-dominated channels were then averaged, in each contiguous spectral segment, using the inverse-squares of their
uncertainties as weights.
The mean values are listed here with $\pm1\sigma$ errors, which do not include systematic errors such as the ones due to flux calibration, 
line contamination, and the choice of the single Gaussian model.
(1) Mean frequency of the used data.
(2) Number of the 20 MHz-wide channels used for averaging.
(3) Major-axis FWHM. 
(4) Minor-axis FWHM.
(5) Position angle of the major axis.
(6) Ratio of major- to minor-axis FWHM.
(7) Total flux density.
(8) Peak Rayleigh-Jeans brightness temperature of the fitted Gaussian.
(9) Median of the reduced $\chi^2$ of the fit in individual channels before rescaling.
}
\end{deluxetable}

%%%%%%%%%%%%%%%%%%%%%%%%%%%%%%%%%%%%%%%%%%%%%%%%%%%%%%%%%%%%%%%%%%%%%%%%            	% Table 5: 1G fit results        
	% \label{t.cont_visfit_1G_params}		

%%%%%%%%%%%%%%%%%%%%%%%%%%%%%%%%%%%%%%%%%%%%%%%%%%%%%%%%%%%%%%%%%%%%%%%%
% Table : Continuum parameters (from 2-Gaussian visibility fitting)
\begin{deluxetable}{CRRCCCRRR}
\tabletypesize{\scriptsize}
\tablewidth{0pt}
\tablecaption{Continuum Visibility Fits: restricted 2 Gaussians  \label{t.cont_visfit_r2G_params} }
\tablehead{ 
	\colhead{} &
       \colhead{} &
      	\colhead{} &
      	\colhead{} &             
       \multicolumn{2}{c}{Component 1} &
       \multicolumn{2}{c}{Component 2}            
       \\
       \colhead{$\nu_{\rm obs}$} &
       \colhead{$N_{\rm ch}$} &     
       \colhead{$S_{\nu}^{\rm (total)}$} &
       \colhead{$\frac{S_{\nu}^{(1)}}{S_{\nu}^{\rm (total)}}$} &                
       \colhead{$S_{\nu}^{(1)}$} &
       \colhead{$\theta_{\rm maj}^{(1)}$} &
       \colhead{$S_{\nu}^{(2)}$} &
       \colhead{$\theta_{\rm maj}^{(2)}$} &    
	\colhead{$\chi^2$/d.o.f} 
       \\                
	\colhead{GHz}  &
	\colhead{} &
       	\colhead{mJy} &
       \colhead{} &	
	\colhead{mJy} &
	\colhead{\arcsec} &
	\colhead{mJy} &
	\colhead{\arcsec} &
       \colhead{}               
}
\colnumbers
\startdata 
\sidehead{NGC 4418}  
682.146 &  200 & 1062.2\pm 5.9       & 0.47\pm0.03   &  498.6\pm 27.1             & 0.080\pm0.007   & 563.5\pm 27.2             &  0.319\pm0.011   & 3.6  \\
358.127 &      3 & 153.3 \pm 2.5        & 0.64 \pm 0.08 & \phn 98.0\pm 12.3        & 0.051 \pm 0.012 & 55.3 \pm 12.3              &  0.213 \pm 0.033 & 0.9  \\
342.080 &      3 & 137.5 \pm 1.8        & 0.74 \pm 0.06 & 101.8 \pm \phn 7.8       & 0.065 \pm 0.007 & 35.7 \pm  \phn 7.7       &  0.246 \pm 0.037 & 0.7  \\
268.940 &      4 & \phn 76.7 \pm 1.8  & 0.89 \pm 0.02 & \phn 68.3\pm \phn 2.5  & 0.084 \pm 0.005 & \phn 8.4 \pm \phn 1.8  &  0.648 \pm 0.165 & 1.2  \\
268.940 &      4 & \phn 71.4 \pm 0.7  & 1                    & \phn 71.4\pm \phn 0.7   & 0.094 \pm 0.004 & \nd                                        & \nd             & 1.6  \\
247.958 &    12 & \phn 57.6 \pm 0.7  &  0.84 \pm 0.11 & \phn 48.3\pm \phn 6.4  & 0.050 \pm 0.028  & \phn 9.4\pm \phn 6.4   &  0.294 \pm 0.113 & 1.6  \\
\sidehead{\arpE}  
665.667 & 121 & 1595.0\pm 20.5      & 0.11\pm0.02   & 174.3\pm 28.7             &  0.080\pm0.025  & 1420.7\pm 34.0                        &  0.367\pm0.007 &  2.4  \\ 
354.092 &     5 &  193.3 \pm  6.3       & 0.23 \pm 0.04 & \phn44.8 \pm \phn8.1  & 0.102 \pm 0.018 & 148.5 \pm \phn9.4                    &  0.356\pm0.021 & 1.3   \\
338.452 &     5 &  162.3 \pm  4.5       & 0.17 \pm 0.03 & \phn27.4 \pm \phn4.6  & 0.054 \pm 0.026 & 135.0 \pm \phn5.8                    &  0.336\pm0.014 & 1.5  \\
337.918 &     9 &  167.3 \pm  4.0       & 0.22 \pm 0.03 & \phn36.9 \pm \phn5.1  & 0.087 \pm 0.016 & 130.4 \pm \phn6.0                    &  0.346\pm0.015 & 1.6  \\
336.067 &     3 &  169.1 \pm  9.3       & 0.36 \pm 0.06 & \phn61.1 \pm 10.4       & 0.131 \pm 0.018 & 108.0 \pm 11.5                         &  0.408\pm0.043 & 1.2  \\
264.560 &     5 &  \phn73.8 \pm  4.2  & 0.28 \pm 0.05 & \phn20.5 \pm \phn4.2  & 0.079 \pm 0.036 & \phn53.3 \pm \phn5.1               &  0.396\pm0.037 & 1.3 \\
264.560 &     5 &  \phn57.4 \pm  2.4 & 0                     & \nd                               & \nd                     & \phn57.4 \pm \phn\phn2.4        &  0.225\pm0.010 & 2.7 \\
\sidehead{\arpW}  
665.667 & 121 & 2264.1\pm 47.8               & 0.27\pm0.05 &  608.5\pm122.6            & 0.366\pm0.047          & 1655.6\pm 126.8          &  0.168\pm0.006  & 4.5 \\
665.667 & 121 & 2062.0\pm 20.0               &  0                 &  \nd                               & \nd                            &  2062.0\pm \phn20.0    &  0.189\pm0.002  & 10.5 \\ 
353.591 &    7  & \phn373.9\pm \phn3.7     & 0.27\pm0.06 & 100.0\pm \phn21.7       & 0.054\pm0.016          & 273.9\pm \phn21.9       &  0.191\pm0.009  & 2.2 \\
337.703 &    6  & \phn340.8\pm \phn3.3     & 0.21\pm0.07 & \phn70.2\pm \phn25.0  & 0.031\pm0.042          & 270.6\pm \phn25.1       &  0.173\pm0.009  & 2.2 \\
231.077 &    6  & \phn115.9\pm \phn1.0     &  0                 & \nd                                & \nd                            & 115.9\pm \phn\phn1.0  &  0.130\pm0.002  & 2.2  \\
\enddata
\tablecomments{
Parameters of the restricted-two-gaussian (r2G) fitting to the visibilities in the most continuum-dominated channels. 
Each nucleus was fitted with two Gaussians sharing the position and shape, except for rows in which $S_{\nu}^{\rm (1)}/S_{\nu}^{\rm (total)}=$ 0 or 1; 
they are single-Gaussian fit for reference.
The channels used for the fitting have less than 1\% excess over the continuum in our 0\farcs35 spectrum.
Among those channels, the ones nearby were first averaged, gridded to fifty radial bins, and then fitted with the two Gaussians. See text for more detail. 
Errors are $\pm1\sigma$ and have been rescaled so that reduced $\chi^2$ is unity.
(1) Mean frequency of the used data.
(2) Number of the 20 MHz channels used for averaging.
(3) Total flux density.
(4) Flux density fraction of the compact component.
(5)--(6) Total flux density and major-axis FWHM for the 1st (compact) component.
(7)--(8) Same for the 2nd (extended) component.
(9) Reduced $\chi^2$ of the fit before the error rescaling.
}
\end{deluxetable}

%%%%%%%%%%%%%%%%%%%%%%%%%%%%%%%%%%%%%%%%%%%%%%%%%%%%%%%%%%%%%%%%%%%%%%%%          % Table 6: r2G fit results        
	% \label{t.cont_visfit_r2G_params}

%%%%%%%%%%%%%%%%%%%%%%%%%%%%%%%%%%%%%%%%%%%%%%%%%%%%%%%%%%%%%%%%%%%%%%%%
% Table : B9 continuum multi-G fitting
\begin{deluxetable}{cCCCCCCC}
\decimals
\tabletypesize{\scriptsize}
\tablewidth{0pt}
\tablecaption{Multi-Gaussian Fitting of Band 9 Continuum \label{t.B9cont.multiGfit} }
\tablehead{
       \colhead{component} &
       \colhead{R.A. Offset} &
       \colhead{Dec. Offset} &
       \colhead{$S_\nu$} &     
       \colhead{$\theta_{\rm maj}$} &
       \colhead{$\theta_{\rm min}$} &
       \colhead{p.a.} &
       \colhead{peak \Tb} 
       \\
	\colhead{}  &
	\colhead{mas} &
	\colhead{mas} &
	\colhead{Jy} &	
	\colhead{\asec} &		
	\colhead{\asec} &	
       \colhead{\degr} &
       \colhead{K}    
}
\colnumbers
\startdata 
\sidehead{NGC 4418}  
1	& +0.5\pm0.1  	&  -0.9\pm0.0  	&  0.442\pm0.014  &  0.068\pm0.004  	& 0.046\pm0.005 &   81\pm 10  	&  370 \pm49 	\\
2	& -2.3\pm0.1  	&  -1.3\pm0.1  	&  0.561\pm0.012  &  0.290\pm0.006  	& 0.115\pm0.005 &   43.2\pm 0.5  	&  44.1\pm2.9	\\
3	& -3.5\pm3.8  	& -14.2\pm4.5  	&  0.125\pm0.008  &  1.05 \pm0.08  	& 0.60  \pm0.05  	&   39  \pm 4  	&  0.52\pm0.08	\\
\sidehead{Arp 220E}  
1	& -29.5\pm1.0      &  +5.3\pm0.1  	&  0.166\pm0.007  &  0.070\pm0.006  & 0.046\pm0.010  &  131 \pm10  		&  143\pm 33 \\ 
2	& +18.6\pm0.5  	&  +0.3\pm0.0  	&  1.168\pm0.012  &  0.353\pm0.003  & 0.167\pm0.002  &   51.1\pm 0.3  	&  54.8\pm 0.7	\\
3	& -35.9\pm2.1  	&  -3.7\pm0.2  	&  0.503\pm0.012  &  0.779\pm0.017  & 0.417\pm0.008  &   10.6\pm 1.0  	&   4.3\pm 0.2 	\\
\sidehead{Arp 220W}
2	&  3.1\pm0.1 	&   0.0\pm0.1  	&  1.694\pm0.008  &  0.183\pm0.001  & 0.125\pm0.001  &   83.6\pm 0.4  	& 203.7\pm 0.9	\\
3	& -26.8\pm0.7  	&   0.6\pm0.6  	&  0.818\pm0.008  &  0.587\pm0.006  & 0.304\pm0.003  &  162.9\pm 0.4  	&  12.6\pm 0.3  	\\
\enddata
\tablecomments{
Multi-Gaussian fitting in visibilities for Band 9 continuum, using three Gaussians on NGC 4418 and five on Arp 220.
(1)  1 = core component, 2 = main component, 3 = extended/outflow component. 
Only two components were assigned to Arp 220W since the core component of this nucleus is found insignificant in our Band 9 data. 
(2)--(3) Positional offset from the positions assumed in self-calibration. Errors less than $0.05$ mas are written as 0.0 mas. 
(4) Flux density of each component. (A flat spectrum was assumed in these fitting.) Errors do not include those from absolute flux calibration.
(5)--(7) Major and minor axis FWHM and the major axis position angle. 
(8) Peak Rayleigh-Jeans brightness temperature of the component.
See text for more model descriptions.
}
\end{deluxetable}

%%%%%%%%%%%%%%%%%%%%%%%%%%%%%%%%%%%%%%%%%%%%%%%%%%%%%%%%%%%%%%%%%%%%%%%%  % Table 7 B9 multi-G fit results
	% \label{t.B9cont.multiGfit}
	
%%%%%%%%%%%%%%%%%%%%%%%%%%%%%%%%%%%%%%%%%%%%%%%%%%%%%%%%%%%%%%%%%%%%%%%%
% Table : B9 continuum multi-G fitting
\begin{deluxetable}{cCCCCCCC}
\decimals
\tabletypesize{\scriptsize}
\tablewidth{0pt}
\tablecaption{Multi-Gaussian Fitting of Band 3 Continuum \label{t.B3cont.multiGfit} }
\tablehead{
       \colhead{component} &
       \colhead{R.A. Offset} &
       \colhead{Dec. Offset} &
       \colhead{$S_\nu$} &     
       \colhead{$\theta_{\rm maj}$} &
       \colhead{$\theta_{\rm min}$} &
       \colhead{p.a.} &
       \colhead{peak \Tb}
       \\
	\colhead{}  &
	\colhead{mas} &
	\colhead{mas} &
	\colhead{mJy} &	
	\colhead{\asec} &		
	\colhead{\asec} &	
       \colhead{\degr} &
       \colhead{K}  
}
\colnumbers
\startdata
\sidehead{Arp 220E}  
1	&  -12.00\pm0.62  &   -0.53\pm0.44  &   2.33\pm 0.09  &  0.085\pm0.003  & 0.040\pm0.002  &   55.6\pm 2.1  &   74.6\pm 3.9  \\
2	&  +11.90\pm1.73  &   -5.23\pm0.50  &   7.53\pm 0.25  &  0.342\pm0.008  & 0.162\pm0.004  &   54.6\pm 0.9  &   14.7\pm 0.5  \\
3	&  -16.30\pm8.06  &  +36.06\pm2.60  &   2.83\pm 0.28  &  0.619\pm0.043  & 0.304\pm0.023  &   22.1\pm 3.4  &    1.6\pm 0.3  \\
\sidehead{Arp 220W}
1	&   +1.87\pm0.07  &   -1.89\pm0.07  &   9.82\pm 0.07  &  0.048\pm0.000  & 0.035\pm0.000  &  138.5\pm 1.5  &  626.4\pm 7.2  \\
2	&  -11.06\pm0.22  &   +3.27\pm0.18  &  11.60\pm 0.19  &  0.237\pm0.003  & 0.119\pm0.002  &   82.2\pm 0.5  &   44.9\pm 0.8  \\
3	&  -13.40\pm1.07  &  -14.32\pm1.40  &   3.16\pm 0.18  &  0.464\pm0.024  & 0.183\pm0.011  &  167.0\pm 1.5  &    4.0\pm 0.4  \\
\enddata
\tablecomments{
Multi-Gaussian fitting in visibilities for the Band 3 continuum data from \citet{Sakamoto17};
we fitted the visibilities imaged in their Fig.~4 with six Gaussians, assigning three to each nucleus.
(1)  1 = core component, 2 = main component, 3 = extended/outflow component. 
(2)--(3) Positional offset from the 1G positions in Table 1 of \citet{Sakamoto17}. 
(4) Flux density of each component. (A flat spectrum was assumed in the fitting.) Errors do not include those from absolute flux calibration.
(5)--(7) Major and minor axis FWHM and the major axis position angle. 
(8) Peak Rayleigh-Jeans brightness temperature of the component. 
}
\end{deluxetable}

%%%%%%%%%%%%%%%%%%%%%%%%%%%%%%%%%%%%%%%%%%%%%%%%%%%%%%%%%%%%%%%%%%%%%%%%  % Table 8 B3 multi-G fit results
	% \label{t.B3cont.multiGfit}	

%%%%%%%%%%%%%%%%%%%%%%%%%%%%%%%%%%%%%%%%%%%%%%%%%%%%%%%%%%%%%%%%%%%%%%%%
% Table : Continuum model
\begin{deluxetable}{cccccrccccrccc}
\decimals
\tabletypesize{\scriptsize}
\tablewidth{0pt}
\tablecaption{Continuum Model: restricted 2 Gaussians  \label{t.cont_model-r2G_parameters} }
\tablehead{
       \colhead{freq. range} &
       \colhead{$\nu_{\rm ref}$} &
       \colhead{$I_{\rm ref}$} &
       \colhead{$\alpha$} &       
        \colhead{$\frac{\theta_{\rm min}}{\theta_{\rm maj}}$} &  	          
 	\colhead{P.A.} & 
       \colhead{$\frac{S_{\nu}^{(1)}}{S_{\nu}^{\rm (r2G)}}$} &
       \colhead{$\theta_{\rm maj}^{(1)}$} &
       \colhead{$\theta_{\rm maj}^{(2)}$} &
       \colhead{$\eta_{\rm r2G}$} &
       \colhead{$S_{\rm ref}^{\rm (r2G)}$}  &
       \colhead{$T_{\rm b, ref}^{\rm (r2G)}$}  &
       \colhead{$T_{\rm b, ref}^{(1)}$}  &
       \colhead{$T_{\rm b, ref}^{(2)}$}                                                         
       \\
	\colhead{GHz}  &
	\colhead{GHz} &
	\colhead{mJy \perbeam} &
	\colhead{} &	
	\colhead{} &		
	\colhead{deg.} &	
       \colhead{} &
       \colhead{mas} &
       \colhead{\asec} &
       \colhead{} &
       \colhead{mJy} &
       \colhead{K} &
       \colhead{K} &
       \colhead{K}               
}
\colnumbers
\startdata 
\sidehead{NGC 4418}  
673--693 &  680 &   845.1       & 2.18  & 0.533 & \phn46.0  & 0.47 & 80 & 0.32 & 0.807  & 1046       & 403       & 376       & \phn27 \\
337--364 &  350 &   131.7       & 2.37  & 0.685 & \phn48.5  & 0.71 & 61 & 0.23 & 0.916  & 143.8      & 406       & 394       & \phn12 \\
\sidehead{\arpE}  
657--685 & 670  &  1053         & 3.32  & 0.555 & \phn48.8  & 0.11 & 80 & 0.37 & 0.637  & 1654       & 192       & 139       & \phn53 \\
335--361 & 350  & 127.2         & 2.84  & 0.560 & \phn51.0  & 0.22 & 97 & 0.35 & 0.691  & 184.0      & \phn97  & \phn76  & \phn21  \\ 
214--266 & 240  & \phn36.7    & 2.90  & 0.594 & \phn47.2  & 0.22 & 97 & 0.35 & 0.684  & \phn53.7 & \phn57  & \phn44  & \phn12 \\ 
\sidehead{\arpW}  
335--361 & 350  & 317.7        & 1.70   & 0.804 & \phn89.8  & 0.24 & 51 & 0.18 & 0.855  & 371.6      & 530       & 424       & 105 \\ 
214--266 & 240  & 115.4        & 2.37   & 0.878 & 101.5       & 0.54 & 93 & 0.22 & 0.852  & 135.4      & 236       & 204       & \phn32  \\
\enddata
\tablecomments{
Parameters of our r2G continuum models for individual nuclei. 
(There are no r2G models for NGC 4418 in Band 6 and Arp 220 W in Band 9 because we adopt 1G models for their continuum subtraction.) 
(1) Frequency range for which the model is built. 
(2) Reference frequency for the power-law spectrum.
(3)--(4) Parameters describing our continuum spectrum obtained with a 0\farcs35 beam.
(5)--(6) Major-to-minor axis ratio and the major axis position angle. These are from our 1G measurements. 
We assume that both Gaussians in our r2G model share these.
(7) Fractional contribution of the first Gaussian to the total flux density in the r2G model.
(8) and (9) Major-axis FWHM of the first and second Gaussians, respectively.
(10) Coupling efficiency of the r2G model source with a 0\farcs35 beam.
(11) Total flux density of the model source at the reference frequency, corrected for the source-beam coupling (i.e., $= I_{\rm ref}/\eta_{\rm r2G}$).
(12)--(14) Peak Rayleigh-Jeans brightness temperatures of the r2G model, its component 1, and component 2, respectively.
See text for more model descriptions.
}
\end{deluxetable}

%%%%%%%%%%%%%%%%%%%%%%%%%%%%%%%%%%%%%%%%%%%%%%%%%%%%%%%%%%%%%%%%%%%%%%%%  % Table 9: r2G model parameters      
	% \label{t.cont_model-r2G_parameters}

%%%%%%%%%%%%%%%%%%%%%%%%%%%%%%%%%%%%%%%%%%%%%%%%%%%%%%%%%%%%%%%%%%%%%%%%
% Table : Continuum spectral parameters
\begin{deluxetable}{lRCC}
\tablecolumns{4}
\tabletypesize{\scriptsize}
\tablewidth{0pt}
\tablecaption{Continuum Spectra ($\lambda \about 1$mm)  \label{t.contSpectra} }
\tablehead{ 
       \colhead{name}             & 
       \colhead{$S_{300}$}     &
       \colhead{$\alpha$}       &
       \colhead{$\bar{\nu}$} 
	\\
	\colhead{}           &
	\colhead{mJy}     &
	\colhead{}           &
	\colhead{GHz}                
}
\colnumbers
\startdata
NGC 4418  &   95.2 \pm 1.5   &  2.35 \pm 0.08  & 270 \\
\arpE         & 103.0 \pm 1.6   &  3.28 \pm 0.09  & 292  \\
\arpW        & 244.4 \pm 4.5   &  2.67 \pm 0.10  & 302  \\
\enddata
\tablecomments{
Parameters of the power-law spectra fitted to our ALMA measurements between 200 GHz and 400 GHz.
(For NGC 4418, we included two data points of \citet{Costagliola15} in the frequency range.)
The fitting function has a form of
$S_\nu = S_{300} (\nu / \mathrm{300\; GHz})^{\alpha} $.
We list in (4) the geometrical mean frequency of the data used for each  fitting.
In wavelength, it ranges 0.99--1.11 mm.
The fitted spectrum of NGC 4418 is shown in Figure \ref{t.contSpectra}.
}
\end{deluxetable}

%%%%%%%%%%%%%%%%%%%%%%%%%%%%%%%%%%%%%%%%%%%%%%%%%%%%%%%%%%%%%%%%%%%%%%%%

      % Table 10: Continuum spectra of the three nuclei at ~1mm
	% \label{t.contSpectra} 	

%%%%%%%%%%%%%%%%%%%%%%%%%%%%%%%%%%%%%%%%%%%%%%%%%%%%%%%%%%%%%%%%%%%%%%%%
% Table : Line Contamination
\begin{deluxetable}{cccc}
\tablecolumns{4}
\tabletypesize{\scriptsize}
\tablewidth{0pt}
\tablecaption{Line Contribution to Total Flux in our  Spectra  \label{t.lineFraction} }
\tablehead{ 
       \colhead{Band}             & 
       \colhead{NGC 4418}     &
       \colhead{\arpE}       &
       \colhead{\arpW} 
	\\
	\colhead{}         &
	\colhead{\%}     &
	\colhead{\%}      &
	\colhead{\%}                
}
\colnumbers
\startdata
   9  &  14.3  & \phn5.5  & \phn4.6  \\
   7  &  30.7  & 19.1       & 17.0 \\
   6  &  26.9  & 22.0       & 20.9 \\
\enddata
\tablecomments{
The ratio of the total flux in continuum-subtracted line emission to the total flux in line+continuum emission in our 0\farcs35
spectra in Figures \ref{f.spec_with_cont.ylog.N4418}--\ref{f.spec_with_cont.ylog.A220W}.
An absorption line has negative flux in this calculation.
Note that our spectral coverage is not the same for NGC 4418 and Arp 220 in Bands 6 and 9. 
}
\end{deluxetable}

%%%%%%%%%%%%%%%%%%%%%%%%%%%%%%%%%%%%%%%%%%%%%%%%%%%%%%%%%%%%%%%%%%%%%%%%

      % Table 11: Fraction of line in our total flux.
	% \label{t.lineFraction} 	
		
%%%%%%%%%%%%%%%%%%%%%%%%%%%%%%%%%%%%%%%%%%%%%%%%%%%%%%%%%%%%%%%%%%%%%%%%
% Table : Continuum spectral analysis to opacity
\begin{deluxetable}{lllllllll}
\tablecolumns{9}
\tabletypesize{\scriptsize}
\tablecaption{$\alpha$--$\tau$ Analysis with the BGN model  \label{t.contSpectralAnalysis} }
\tablehead{ 
       \colhead{Nucleus} & 
       \colhead{$S_{\nu}$} &
       \colhead{$\alpha_{\nu}$} &       
       \colhead{$S_{\mathrm{p}, \nu}$ } &
       \colhead{$\alpha_{\mathrm{p}, \nu}$ } & 
       \colhead{$f_{\mathrm{p}, \nu}$ } &             
       \colhead{$\alpha_{\mathrm{d}, \nu}$ } &
       \colhead{$\tau_{\mathrm{d}, \nu}$ } & 
       \colhead{$\log \NHH$}                    
	\\
	\colhead{}  &
	\colhead{mJy} &
	\nocolhead{} &
       \colhead{mJy} &
       \colhead{} &
       \colhead{\%} &       
       \colhead{} &       
       \colhead{} &
       \colhead{\persquarecm}                      	                
}
\colnumbers         
\startdata
NGC 4418  & $\phn95.2 \pm 1.5$ & $2.35 \pm 0.08$	& $3.4\pm0.7$ & $-0.30$ & $3.6\pm0.8$ & $2.45\pm0.09$ & $2.2\pm0.3$ & $25.7\pm0.1$ \\
\arpE         & $103.0 \pm 1.6$	   & $3.28 \pm 0.09$  	& $6.7\pm0.6$ & $-0.37$ & $6.5\pm0.5$ & $3.53\pm0.10$ & $\leq 0.1$         & $\leq 24.4$ \\
\arpW         & $244.4 \pm 4.5$	   & $2.67 \pm 0.10$  	& $9.1\pm0.9$ & $-0.37$ & $3.7\pm0.4$ & $2.79\pm0.10$ & $1.2\pm0.2$ & $25.4 \pm 0.1$\\
\enddata
\tablecomments{
Analysis of the continuum spectra at $\nu = 300$ GHz.
(2), (3): Flux density and spectral index from our observations, taken from Table \ref{t.contSpectra}.
(4), (5): Estimated flux density and estimated or assumed spectral index of plasma emission.
NGC 4418 has a continuum flux density of $9.9 \pm 1.0$ mJy and spectral index $\alpha \sim 1.5$ at 98 GHz  \citep{Costagliola15}. 
Assuming this index being a weighted mean of $\alphap=-0.4$ of plasma emission and $\alphad=3.3$ of slightly opaque dust emission, 
which is consistent with the final result, 
the fractional contribution of plasma is 0.49; or in the range of 0.45--0.53 for \alphap\ between $-0.1$ and $-0.7$. 
Conservatively using $0.5 \pm 0.1$ for this fraction, we estimate the plasma flux density at 98 GHz to be $5.0\pm1.1$ mJy.
Further assuming that the 98 GHz plasma emission is half synchrotron and half free-free emission, the emission at 300 GHz should have 
0.68 times the 98 GHz flux density and a spectral index of $-0.30$ (66\% in free-free emission).
The assumption on the synchrotron fraction at 98 GHz makes little difference to the final results at 300 GHz (see text). 
Arp 220 E and W respectively have at 104 GHz continuum flux densities of $11.9\pm0.9$ and $23.9\pm1.3$ mJy,
fractional contributions of plasma emission of $0.87\pm0.04$ and $0.59\pm0.05$ \citep{Sakamoto17},
and thus have $10.4\pm0.9$ and $14.1\pm1.4$ mJy of plasma emission.
Under the assumptions of synchrotron and free-free spectral indices being $-0.7$ and $-0.1$, respectively, 
the observations of $\alpha_\mathrm{6-33\,GHz} \approx -0.6$ in both nuclei \citep{Barcos-Munoz15} indicate the fractional contributions
of $f_\mathrm{syn}=5/6$ and $f_\mathrm{f-f}=1/6$ at $\sqrt{6\times33}$=14 GHz. 
With these fractions, the plasma emission at 300 GHz should have 0.65 times the 104 GHz flux densities and a composite spectral index of 
$-0.37$ (56\% in free-free emission).
(6): Fraction of the plasma component in the total flux density, $f_{\mathrm{p}, \nu}=S_{\mathrm{p}, \nu}/S_{\nu}$.
(7): Spectral index of dust emission calculated with equation (\ref{eq.alpha_d_exact}).
(8): Optical depths of dust emission calculated with the $\alpha$--$\tau$ relation of BGN models in equation (\ref{eq.bgn_spix}).
The one for Arp 220 E is a $1\sigma$ upper limit.
(9): Gas column density from the center to the surface in the fiducial BGN model, calculated from $\tau_{\mathrm{d}, \nu}$ and
the formula (\ref{eq.bgn_nhh-tau}). 
The $\pm1\sigma$ errors do not include the uncertainty in the conversion factor.
}
\end{deluxetable}

%%%%%%%%%%%%%%%%%%%%%%%%%%%%%%%%%%%%%%%%%%%%%%%%%%%%%%%%%%%%%%%%%%%%%%%%

      % Table 12: Dust Opacity and Column Density (alpha-tau analysis)
	% \label{t.contSpectralAnalysis} 		
	
%%%%%%%%%%%%%%%%%%%%%%%%%%%%%%%%%%%%%%%%%%%%%%%%%%%%%%%%%%%%%%%%%%%%%%%%
% Table : Continuum Shapes
\begin{deluxetable}{lccCLl}
\tablecolumns{5}
\tabletypesize{\scriptsize}
\tablewidth{0pt}
\tablecaption{Radio Continuum Shapes \label{t.contShape} }
\tablehead{ 
       \colhead{Name} & 
       \colhead{Component} &
       \colhead{Frequency} &
       \colhead{P.A. } &
       \colhead{$\theta_{\text{min}}/\theta_{\text{maj}}$ } &
       \colhead{Reference}
	\\
	\colhead{}  &
	\colhead{} & 
	\colhead{GHz} &
	\colhead{\degr} &
	\colhead{}  &
	\colhead{} 
}
\colnumbers   
\startdata
NGC 4418 	& whole (1G) 		& 200--700 		& \phn47  \pm 1 	& 0.61 \pm 0.07 		& 1(\S\ref{s.cont.sizeShape.1Gfit}) \\
{              }	& nuclear disk		& (680)			& \phn43		& 0.4				& 1(\S\ref{s.contB9visfit}) \\
\hline
{              }  	& whole (1G) 		& 100--700 		& \phn49  \pm 2 	& 0.57 \pm 0.07 		& 1(\S\ref{s.cont.sizeShape.1Gfit}), 2, 3 \\
{Arp 220E}  	& whole (1G) 		& 33 			& \phn56 \pm 1 	& 0.54 \pm 0.08 		& 4 \\
{              }  	& whole (1G) 		& \about5\tnm{a} 	& \phn47 \pm 5 	& 0.48 \pm 0.07 		& 5 \\
{              }	& nuclear disk		& (5,100,670)	& \phn51		& 0.47				& 1(\S\ref{s.contB9visfit}) \\
\hline
{                }  	& whole (1G) 		& 100--700	 	& 100 \pm 9       & 0.83 \pm 0.05   		& 1(\S\ref{s.cont.sizeShape.1Gfit}), 2, 3 \\
{		  }	& whole (1G) 		& 33 			& \phn79 \pm 2 	& 0.60 \pm 0.08 	 	& 4 \\
{Arp 220W}	& whole (1G) 		& \about5\tnm{a} 	& \phn83 \pm 5 	& 0.47 \pm 0.07 		& 5 \\
{              }	& nuclear disk		& (5, 100, 670)	& \phn83		& 0.5				& 1(\S\ref{s.contStructure.a220w}) \\
{              }	& outflow		& (100, 670)		& 165			& 0.45				& 1(\S\ref{s.contB9visfit}) \\
\enddata
\tablecomments{
Summary of the radio continuum shapes, either as a whole nucleus or for internal components, for the three nuclei.
The effect of the observational resolution, $\lesssim$$0\farcs2$, has been removed.
(2) We denote with ``whole (1G)'' the nucleus as a whole. Its parameters are from the fitting with a single elliptical Gaussian. 
The ``nuclear disk'' and ``outflow'' are from our decomposition of the individual nuclei.
(3) The frequency or range of frequencies for the 1G measurements of apparent shapes. 
For individual components (as entities), frequencies of the referenced data are given in the parenthesis.
(4) Major axis position angle.
The 100--700 GHz measurements are averaged and listed with their standard deviation.
For each ALMA band, we took one (averaged) value.
The \about5 GHz parameters are from the distribution of compact sources and are explained in Fig.~\ref{f.a220VLBIfit}.
The parameters for the ``nuclear disk'' and ``outflow'' are based on multi-component visibility fitting in Section \ref{s.contB9visfit};
the nuclear disk parameters for Arp 220 also include 1G fitting of the \about5 GHz data. 
(5) Same as (3) but for the minor-to-major axial ratio.
\tablenotetext{a}{The VLBI measurements are mainly at 5 GHz but some at 1.6 and 8.3 GHz.}
}
\tablerefs{
1: This work for 210--360 GHz and \about675 GHz. 
2: \citet{Sakamoto17} for 104 GHz; 
3: \citet{Wheeler20} for 428 GHz;
4: \citet{Barcos-Munoz15} for 32.5 GHz;
5: \citet{Varenius19} fitted in this work
}
\end{deluxetable}

%%%%%%%%%%%%%%%%%%%%%%%%%%%%%%%%%%%%%%%%%%%%%%%%%%%%%%%%%%%%%%%%%%%%%%%%

		% Table 13: Radio continuum shapes, summary
	% \label{t.contShape} 	

%%%%% Figures %%%%%
\clearpage

%%%%%%%%%%%%%%%%%%%%%%%%%%%%%%%%%%%%%%%%%%%%
% Fig.   Continuum images, B9

%%%%%%%%%%%%%%%%%%%%%%%%%%%%%%%%%%%%%%%%%%%% 		   % Fig. 1 Cont images of B9
	% \label{f.B9cont}

%%%%%%%%%%%%%%%%%%%%%%%%%%%%%%%%%%%%%%%%%%%%
% Fig.   Passband calibration and Flux-selfcalibraiton

%%%%%%%%%%%%%%%%%%%%%%%%%%%%%%%%%%%%%%%%%%%% 		   % Fig. 2 Calibration
	% \label{f.calib}

%%%%%%%%%%%%%%%%%%%%%%%%%%%%%%%%%%%%%%%%%%%%
% Fig.   Astrometry of Arp 220 nuclei

%%%%%%%%%%%%%%%%%%%%%%%%%%%%%%%%%%%%%%%%%%%% % Fig. 3 Arp 220 astrometry
	% \label{f.a220position}
	
%%%%%%%%%%%%%%%%%%%%%%%%%%%%%%%%%%%%%%%%%%%%
% Fig.   Spectrum Sections - ylog: N4418

%%%%%%%%%%%%%%%%%%%%%%%%%%%%%%%%%%%%%%%%%%%% % Fig. 4: spectra with continuum, N4418
	%# \label{f.spec_with_cont.ylog.N4418}, 
%%%%%%%%%%%%%%%%%%%%%%%%%%%%%%%%%%%%%%%%%%%%
% Fig.   Spectrum Sections - ylog : Arp 220E
%%%%%%%%%%%%%%%%%%%%%%%%%%%%%%%%%%%%%%%%%%%%
 % Fig. 5: spectra with continuum, A220E
	%# \label{f.spec_with_cont.ylog.A220E}, 
%%%%%%%%%%%%%%%%%%%%%%%%%%%%%%%%%%%%%%%%%%%%
% Fig.   Spectrum Sections - ylog : Arp 220W

%%%%%%%%%%%%%%%%%%%%%%%%%%%%%%%%%%%%%%%%%%%%
 % Fig. 6: spectra with continuum, A220W
	%# \label{f.spec_with_cont.ylog.A220W}	

% 1G parameters, parameter-spectum
%%%%%%%%%%%%%%%%%%%%%%%%%%%%%%%%%%%%%%%%%%%%
% Fig.   Visibility fit (1-Gaussian) 
% NGC 4418 Band 9 and 7

%%%%%%%%%%%%%%%%%%%%%%%%%%%%%%%%%%%%%%%%%%%%
% (b) NGC 4418 Band 6
\begin{figure}[t]
\figurenum{\ref{f.uvfit.1G} (b)}
\begin{center}
\includegraphics[height=65mm]{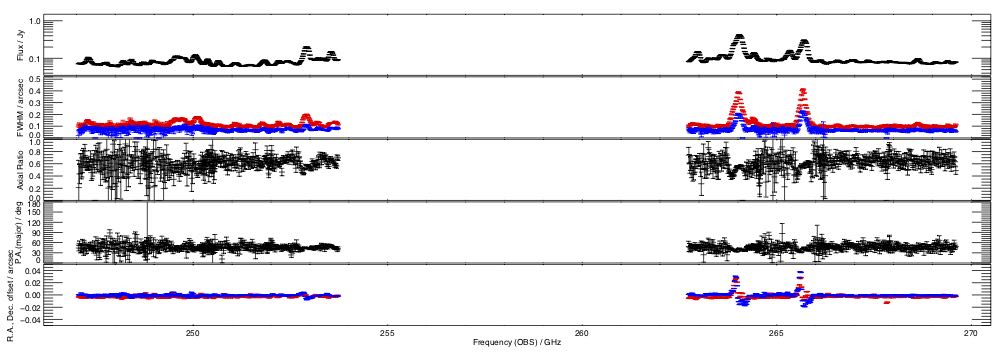} % .eps} 
\includegraphics[height=65mm]{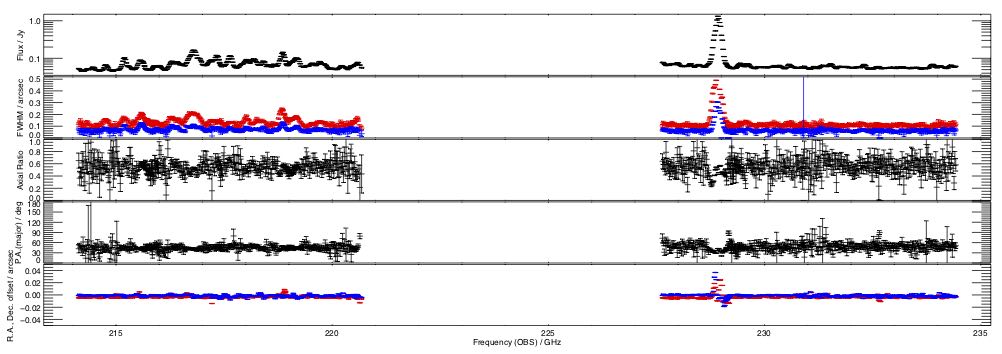} % .eps} 
\end{center}
\caption{NGC 4418 in Band 6 (Online-only)}
\end{figure}

%%%%%%%%%%%%%%%%%%%%%%%%%%%%%%%%%%%%%%%%%%%%
% (c) Arp 220E Band 9 and 7
\begin{figure}[t]
\figurenum{\ref{f.uvfit.1G} (c)}
\begin{center}
\includegraphics[height=65mm]{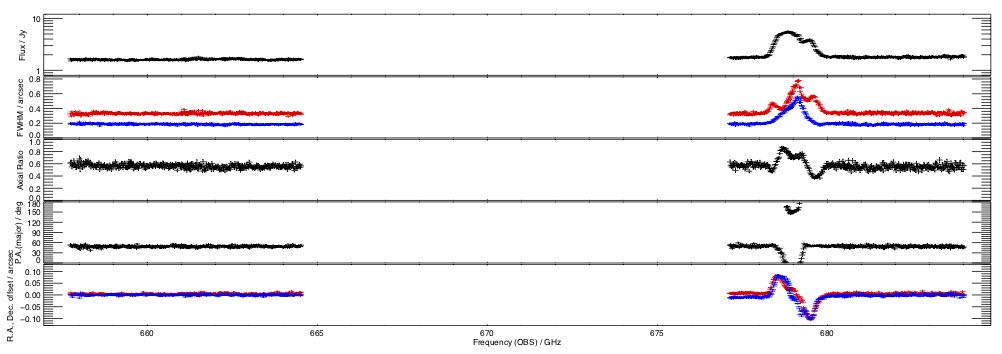} % .eps} 
\includegraphics[height=65mm]{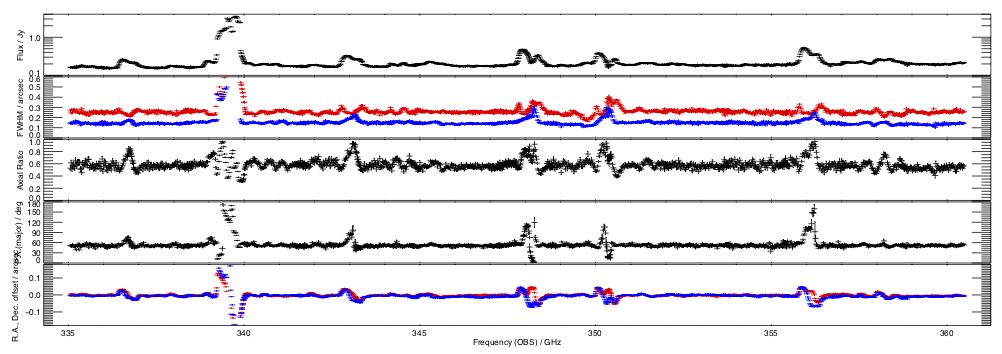} % .eps} 
\end{center}
\caption{\arpE\ in Band 9 and 7 (Online-only)}
\end{figure}
%%%%%%%%%%%%%%%%%%%%%%%%%%%%%%%%%%%%%%%%%%%%
% (d) Arp 220E Band 6
\begin{figure}[t]
\figurenum{\ref{f.uvfit.1G} (d)}
\begin{center}
\includegraphics[height=65mm]{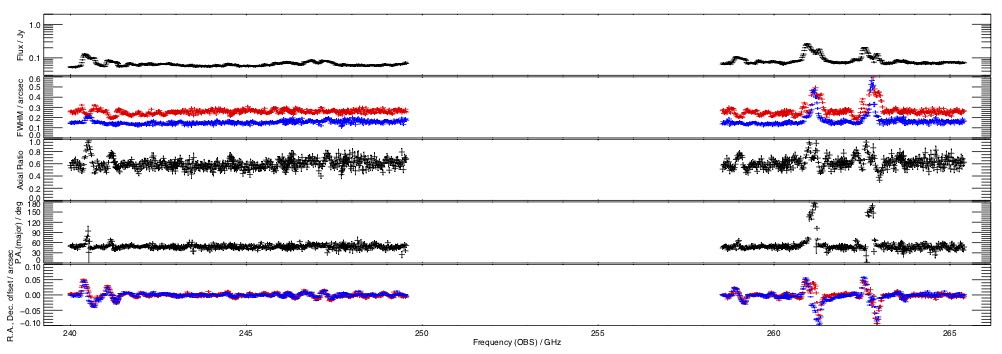} % .eps} 
\includegraphics[height=65mm]{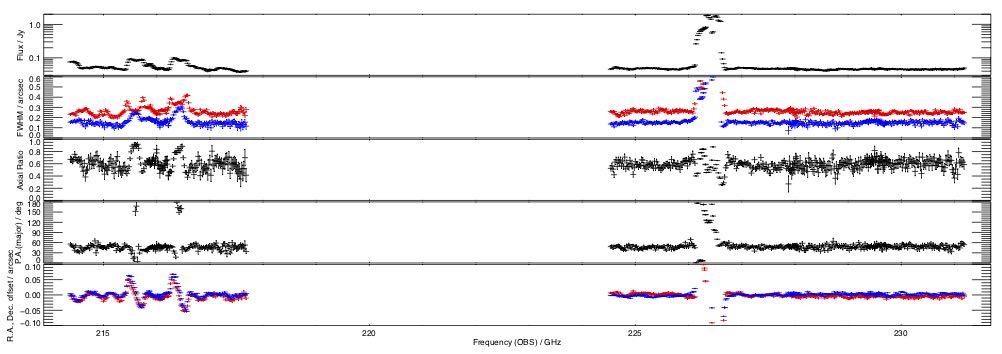} % .eps} 
\end{center}
\caption{\arpE\ in Band 6 (Online-only)}
\end{figure}

%%%%%%%%%%%%%%%%%%%%%%%%%%%%%%%%%%%%%%%%%%%%
% (e) Arp 220W Band 9 and 7
\begin{figure}[t]
\figurenum{\ref{f.uvfit.1G} (e)}
\begin{center}
\includegraphics[height=65mm]{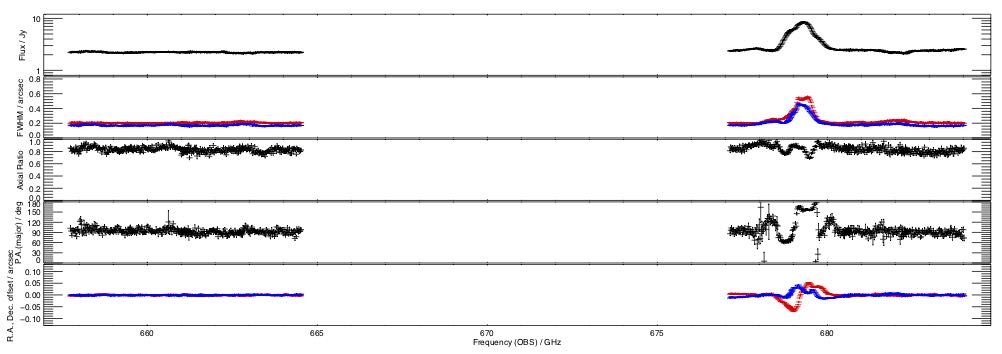} % .eps} 
\includegraphics[height=65mm]{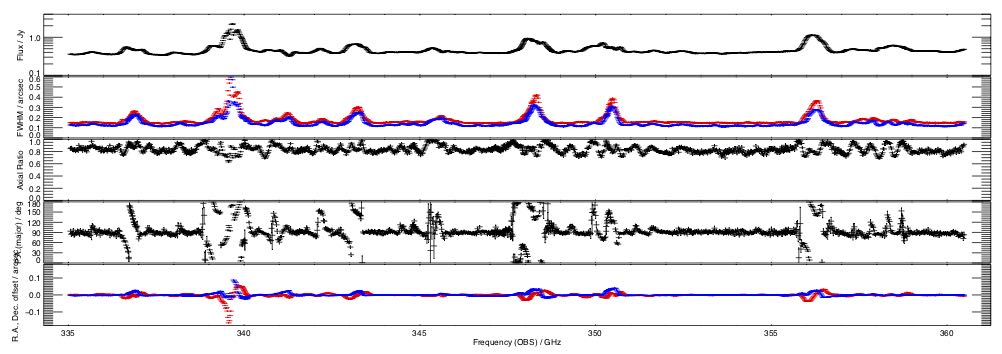} % .eps} 
\end{center}
\caption{\arpW\ in Band 9 and 7 (Online-only)}
\end{figure}
%%%%%%%%%%%%%%%%%%%%%%%%%%%%%%%%%%%%%%%%%%%%
% (f) Arp 220W Band 6
\begin{figure}[t]
\figurenum{\ref{f.uvfit.1G} (f)}
\begin{center}
\includegraphics[height=65mm]{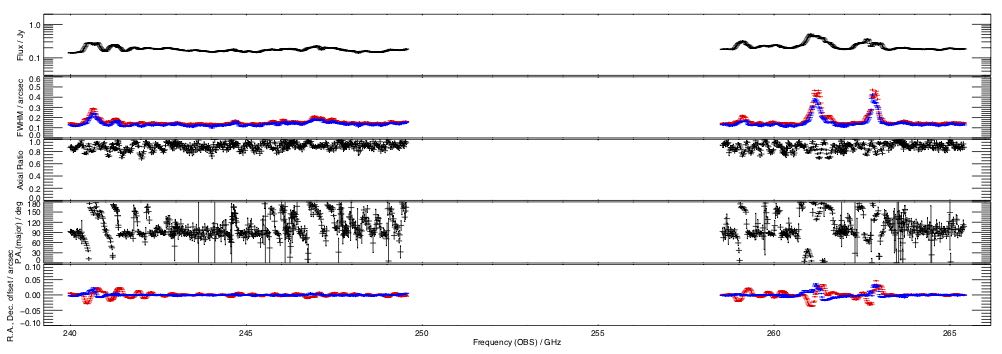} % .eps} 
\includegraphics[height=65mm]{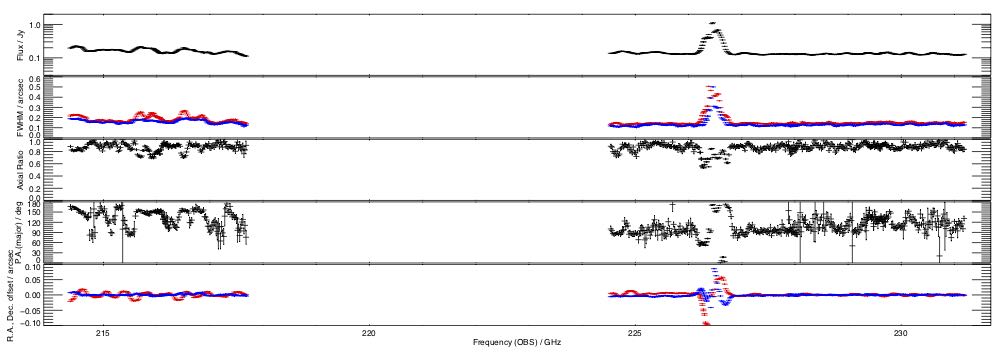} % .eps} 
\end{center}
\caption{\arpW\ in Band 6 (Online-only)}
\end{figure}

%%%%%%%%%%%%%%%%%%%%%%%%%%%%%%%%%%%%%%%%%%%%
 % Fig. 7: spectra of 1G parameters
	%# \label{f.uvfit.1G}

% vis fitting
%%%%%%%%%%%%%%%%%%%%%%%%%%%%%%%%%%%%%%%%%%%%
% Fig.   Visibility fit (2-Gaussian)
\begin{figure}[t]
\begin{center}
\includegraphics[width=58mm]{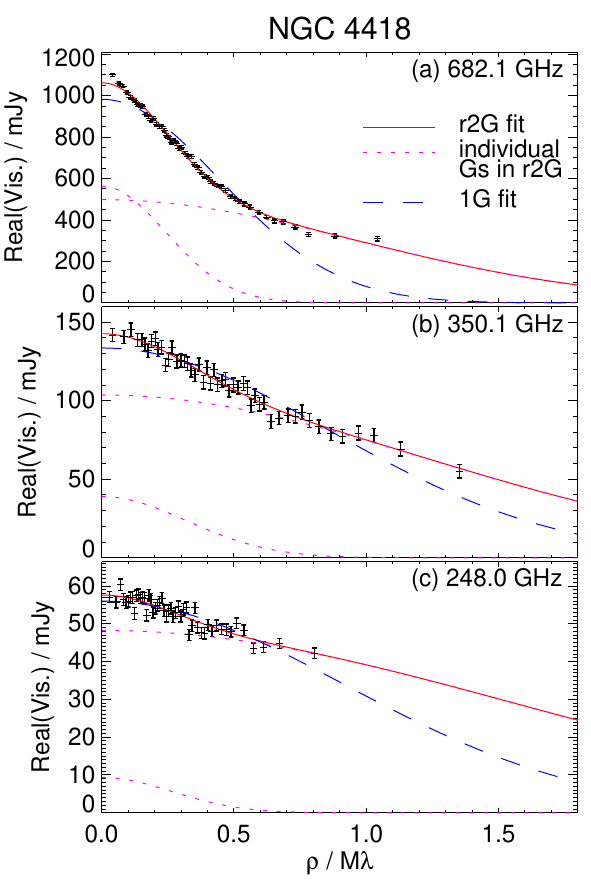} % .eps}  
\includegraphics[width=58mm]{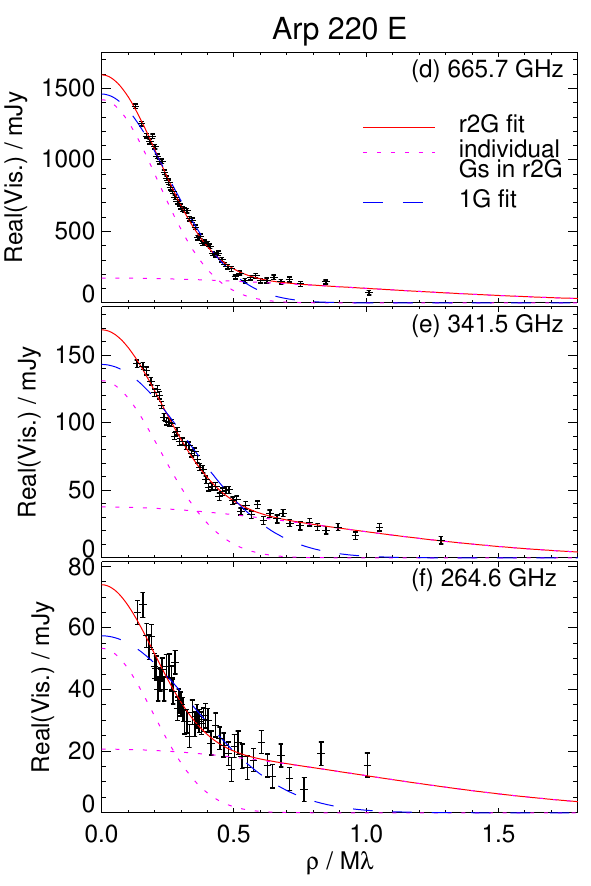} % .eps} 
\includegraphics[width=58mm]{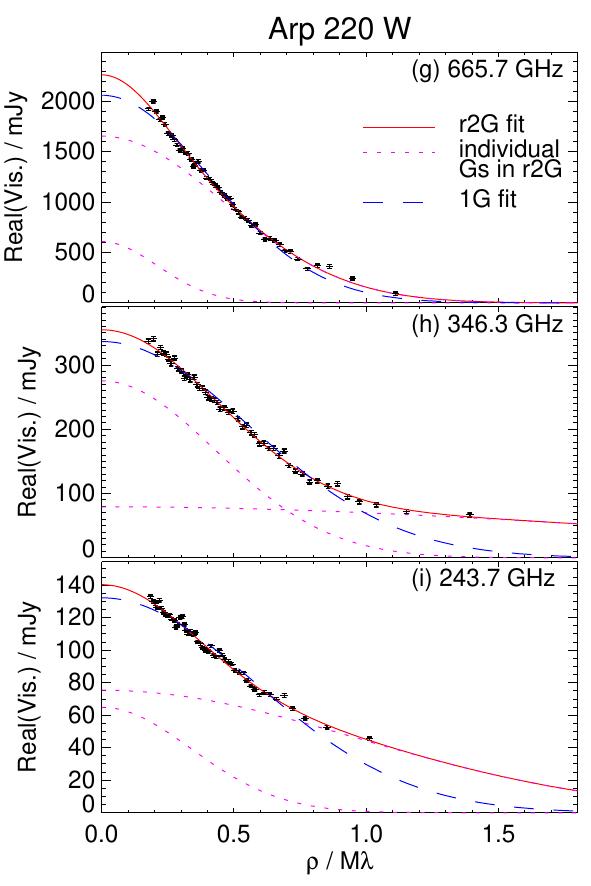} % .eps} 
\end{center}
\caption{ \label{f.uvRealFit.2Gauss}
Visibility fits of three nuclei using elliptical Gaussians at continuum-dominated channels. 
One plot each is for the ALMA Bands 9, 7, and 6. 
The mean frequency of the data is at the top-right corner of each panel.
Visibility data are plotted in black with $\pm1\sigma$ error bars.
They are corrected for the assumed position angle and the axial ratio of each nucleus and are plotted as a function of 
the semi-minor axis in an elliptic coordinate in the \uv\ domain (see text for details).
The visibility data are fitted with a pair of concentric and homologous Gaussians. 
We show each Gaussian as a magenta dotted curve and their sum as a red curve.
Also shown for comparison are single-Gaussian fits in blue dashed curves.
}
\end{figure}

%%%%%%%%%%%%%%%%%%%%%%%%%%%%%%%%%%%%%%%%%%%%                   % Fig. 8: r2G fit. real(vis)-uv_length
	%# \label{f.uvRealFit.2Gauss}

% Figure: Continuum Global Spectra
%%%%%%%%%%%%%%%%%%%%%%%%%%%%%%%%%%%%%%%%%%%%
% Fig.   mm-submm spectra of N4418, A220 continuum 
\begin{figure}[!thb]
\centering
\includegraphics{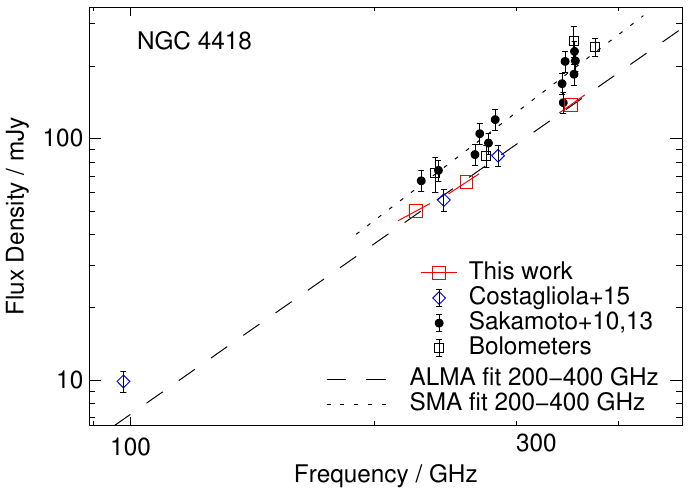} % .eps} 
\hspace{5mm}
\includegraphics{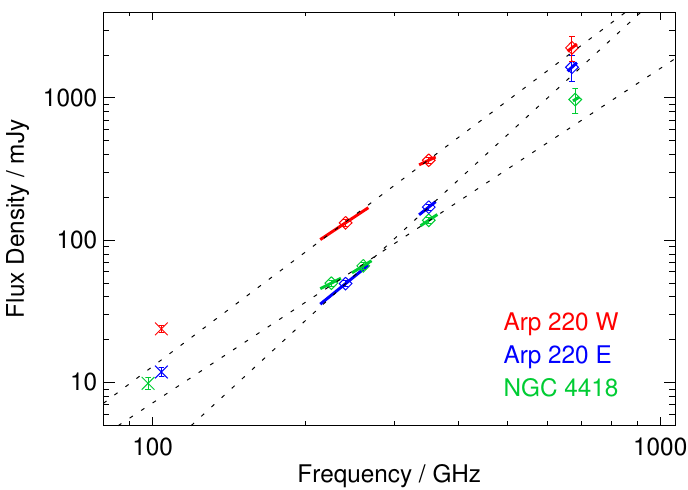} % .eps} 
\caption{ \label{f.contSpecGlobal}
Continuum spectra of the three nuclei.
(Left) Photometries of the NGC 4418 nucleus between 98 and 375 GHz.
Measurements in this work are plotted as red boxes with line segments for local spectral slopes.
Blue symbols are from our low-resolution spectral scan with ALMA \citep{Costagliola15}. 
The dashed line is a power-law fit to the two ALMA datasets combined between 200 and 400 GHz;
$S_\nu / {\rm mJy} = (95.2 \pm 1.5)\times (\nu / 300\, {\rm GHz})^{2.35\pm0.08}$.
Black-filled symbols are from our earlier measurements with the SMA, 
and the dotted line is their power-law fit having a spectral index of $2.55\pm0.18$ \citep{Sakamoto10,Sakamoto13}.
Black open symbols are bolometer measurements between 200 and 400 GHz \citep{Roche93, Dunne00}.
(Right) ALMA photometry of the three nuclei in Arp 220 and NGC 4418 between 98 and 680 GHz.
Data around 100 GHz are from \citet{Costagliola15} and \citep{Sakamoto17},
and the rest are from this work.
The thick color line segments are our local continuum spectra for each spectral section. 
The dotted lines are the power-law spectra that best fit our 200-400 GHz measurements.
Our Band 9 data (670--680 GHz) are plotted with $\pm$20\% errors.
}

\end{figure}

%%%%%%%%%%%%%%%%%%%%%%%%%%%%%%%%%%%%%%%%%%%%     % Fig. 9 N4418, A220, global cont spectrum
	% \label{f.contSpecGlobal}

\clearpage

%%%%% Section 4. Models	
	
% Figure: BGN-model: spix (slab, analytic)
%%%%%%%%%%%%%%%%%%%%%%%%%%%%%%%%%%%%%%%%%%%%
% Fig.   model spix. slab
\begin{figure}[t]
\epsscale{0.42}
\plotone{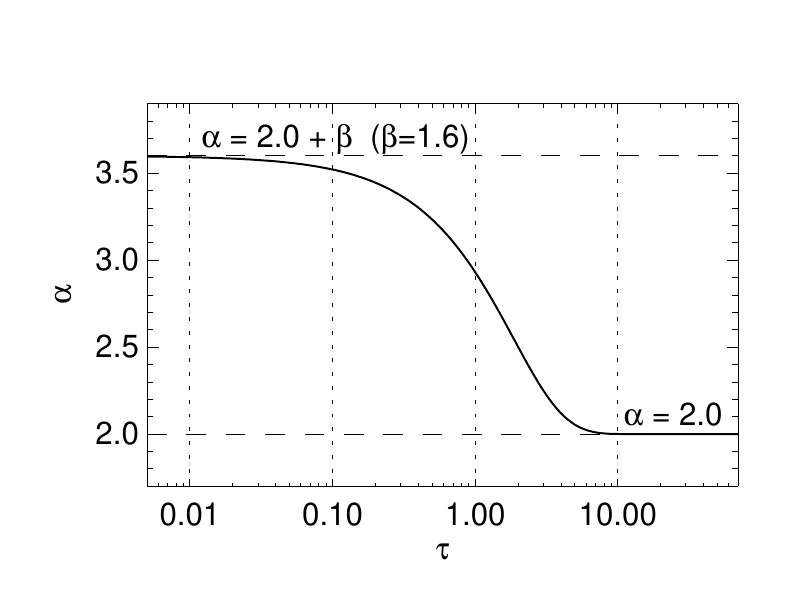} % .eps} 
\caption{ \label{f.contModelSpix_slab}
The relation between the radio spectral index $\alpha$ (for $S_\nu \propto \nu^\alpha$) and
the optical depth $\tau$ for a uniform slab of warm dust. 
See text for assumptions and (\ref{eq.simple_spix}) for the formula.
The spectral index is sensitive to the optical depth in the approximate range of $\tau=$ 0.1 -- 5.
The parameter $\beta$ is the power-law index of the dust absorption coefficient ($\kappa_\nu \propto \nu^\beta$) 
and is set to 1.6.
}
\end{figure}
%%%%%%%%%%%%%%%%%%%%%%%%%%%%%%%%%%%%%%%%%%%%     % Fig. 10 Spix-slab.
	% \label{f.contModelSpix_slab}
	
% Figure: BGN-model: spix (BGN, numerical)  
%%%%%%%%%%%%%%%%%%%%%%%%%%%%%%%%%%%%%%%%%%%%
% Fig.   model spix. BGN model
\begin{figure}[t]
\epsscale{0.84}
\plottwo{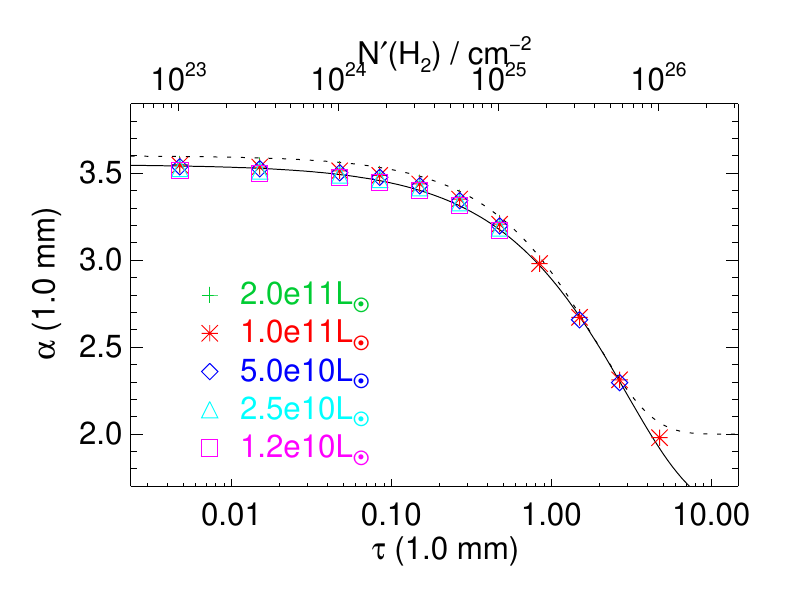}{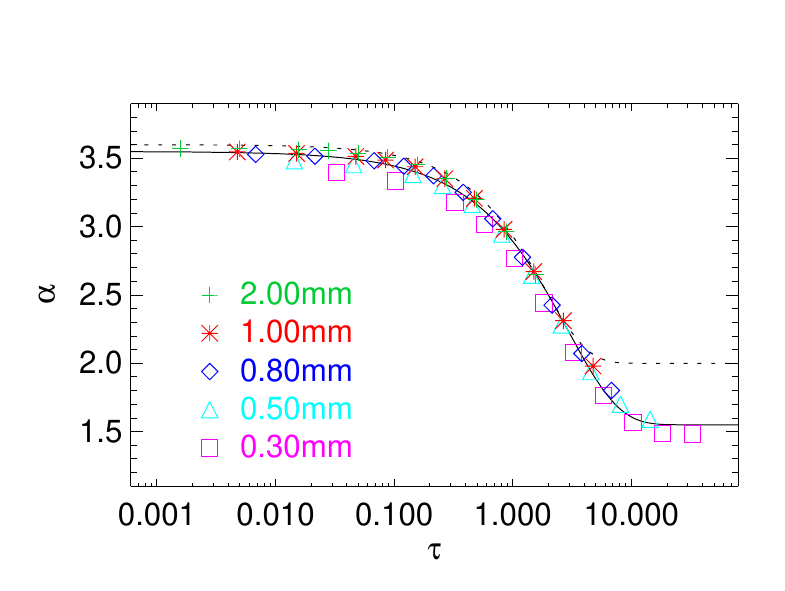} % .eps}
\caption{ \label{f.contModelSpix_BGN}
The relations between the radio spectral index $\alpha$ (for $S_\nu \propto \nu^\alpha$) and
the optical depth $\tau$ for an internally-heated spherical object in the BGN simulations of \citet{GS19}.
The optical depth, as well as the gas column density on the upper horizontal axis of the left panel, 
is that from the center to the source surface. 
The dust parameter $\beta$ is 1.6, as in Fig.~\ref{f.contModelSpix_slab}.
The dotted curve is equation (\ref{eq.simple_spix}) for the analytic slab model, 
while the solid line is a fit to the BGN simulations in equation (\ref{eq.bgn_spix}).
(Left) Both $\alpha$ and $\tau$ are for the wavelength of 1 mm.  
Simulations for various source luminosities are plotted 
to show that the $\alpha$--$\tau$ relation does not depend on luminosity.
(Right) Optical depths and spectral indices are plotted for various wavelengths. 
The model luminosity is $10^{11}$\Lsun\ though it is irrelevant as seen in the left panel.
}
\end{figure}
%%%%%%%%%%%%%%%%%%%%%%%%%%%%%%%%%%%%%%%%%%%%  % Fig. 11 Spix-BGN
	% \label{f.contModelSpix_BGN}	

%%%%%%%%%%%%%%%%%%%%%%%%%%%%%%%%%%%%%%%%%%%%
% Fig.  Various tau-NH2 relations for Arp 220
\begin{figure}[t]
\epsscale{0.45}
\plotone{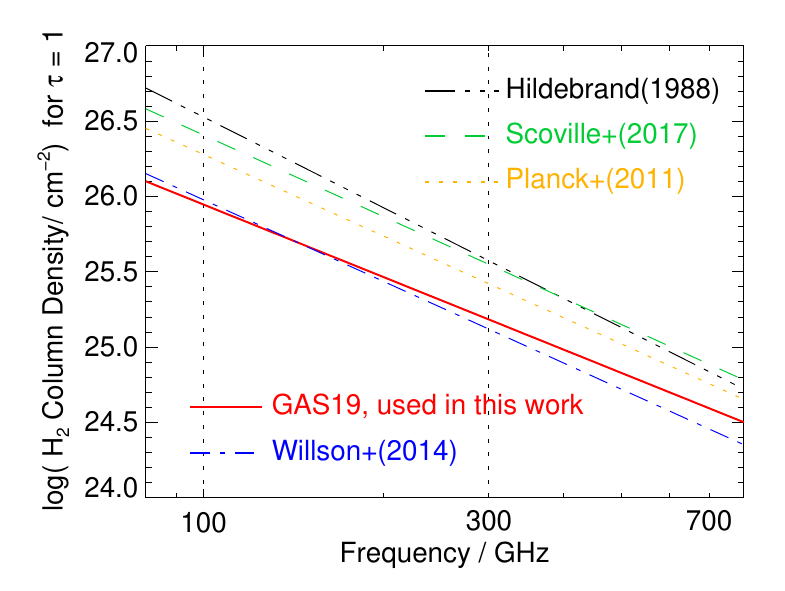} % .eps}  
\caption{ \label{f.tau2NH2}
Various relations between frequency and \HH\ column density for $\tau_{\rm dust}=1$.
The red line is the relation we adopted in this work from \citet{GS19}.
The black three-dots-dash line is the classic relation of \citet{Hildebrand83}; \citet{Sakamoto08} and \citep{Sakamoto13}
used it for Arp 220 and NGC 4418, respectively.
The green dashed line is the relation by \citet{Scoville14} and is used for Arp 220 in \citet{Scoville17}.
The yellow dotted line is the relation for nearby molecular clouds \citep{Planck11_25}.
The blue dot-dashed line is the relation adopted by \citet{Wilson14} for Arp 220. 
}
\end{figure}
%%%%%%%%%%%%%%%%%%%%%%%%%%%%%%%%%%%%%%%%%%%% % Fig. 12. Various tau-NH2 relations for Arp 220
	% \label{f.tau2NH2}

% Figure: BGN-model: Radial T_b profile	
%%%%%%%%%%%%%%%%%%%%%%%%%%%%%%%%%%%%%%%%%%%%
% Fig.  Tb profiles over impact parameter; for four freqs.
\begin{figure}[t]
\epsscale{0.45}
\plotone{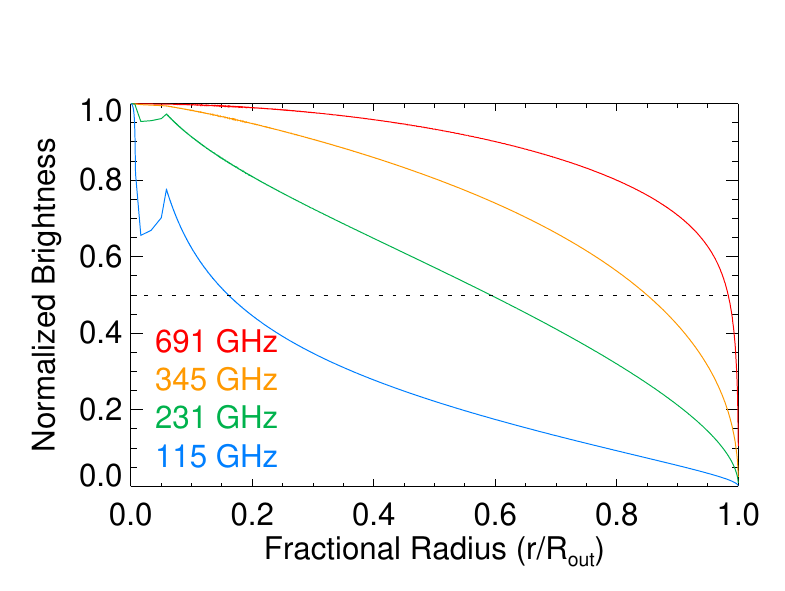} % .eps} 
\caption{ \label{f.contModelTbProfile}
Radial profiles of continuum brightness in a BGN model at four frequencies.
This particular model has an AGN of $10^{11}\Lsun$ in a spherical gas shroud 
whose radial column density is $\NHHprime = 10^{26}$ \persquarecm.
The sky-projected brightness is normalized at the center for each frequency
to show the significant variation of the half-power radius across the frequencies.
The wiggles of the 115 GHz profile near the center are a modeling artifact (due to the
central cavity and the particular implementation of the central AGN).
Since BGN models are scalable, the normalized profiles here are the same for
sources having $L/\pi R_{\rm out}^2 = 1.1\times10^{8}\Lsun \persquarepc$,
where $R_{\rm out}$ is the source radius.
}
\end{figure}
%%%%%%%%%%%%%%%%%%%%%%%%%%%%%%%%%%%%%%%%%%%%  % Fig. 13 radial Tb profiles in BGN model
	% \label{f.contModelTbProfile}

%%%%%  Section 5. Discussion

% Figure: Continuum size, shape, peak Tb
%%%%%%%%%%%%%%%%%%%%%%%%%%%%%%%%%%%%%%%%%%%%
% Fig.   Continuum size, shape, Tb.
\begin{figure}[t]
\centering
\includegraphics{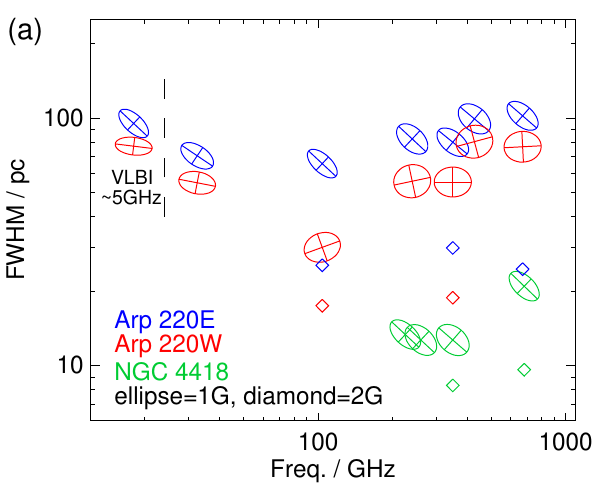} % .eps} 
\hspace{15mm}
\includegraphics{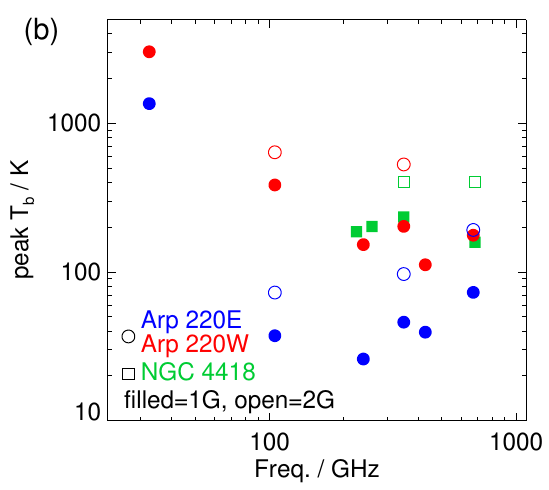} % .eps} 
\caption{ \label{f.contParams}
The three nuclei in continuum emission: (a) size and shape, (b) peak brightness temperature.
(a) 
Equivalent FWHMs (= geometrical means of the major- and minor-axis, deconvolved FWHMs) are plotted
as a function of frequency. Beam smearing has been removed through visibility fitting or image-domain deconvolution.
Symbols are ellipses for single-Gaussian models and diamonds for the smaller components in double-Gaussian models.
Each ellipse has its position angle and axial ratio from the best-fit model.
The left-most data points are from the distribution of compact radio sources in VLBI observations, mostly at 5 GHz 
(see Fig.~\ref{f.a220VLBIfit}). 
(b) 
Peak brightness temperatures of the continuum models in (a). 
Symbols (circles for Arp 220 and squares for NGC 4418) are filled and open, respectively, 
for the single- and double-Gaussian models. 
The high brightness temperatures at 33 GHz are due to plasma emission.
Data sources:
\citet{Varenius19} for VLBI sources;
\citet{Barcos-Munoz15} for 32.5 GHz;
\citet{Sakamoto17} for 104 GHz;
\citet{Wheeler20} for 428 GHz;
this work for 200--350 GHz and \about675 GHz.
}

\end{figure}

%%%%%%%%%%%%%%%%%%%%%%%%%%%%%%%%%%%%%%%%%%%%          % Fig. 14: continuum shapes
	% \label{f.contParams}
	
% Figure: VLBI sources
%%%%%%%%%%%%%%%%%%%%%%%%%%%%%%%%%%%%%%%%%%%%
% Fig.   Distribution of VLBI sources in Arp 220
\begin{figure}[t]
\centering
\includegraphics{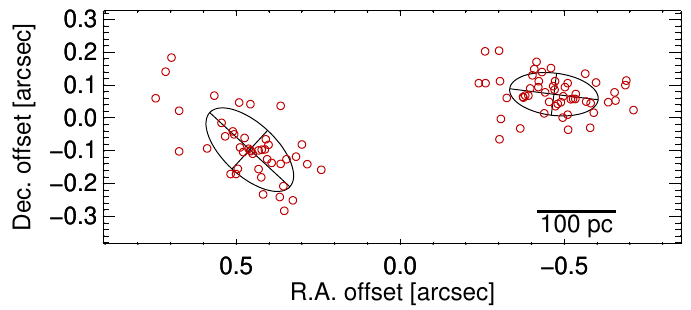} % .eps} 
\caption{ \label{f.a220VLBIfit}
Distribution of compact radio sources in Arp 220.
The red circles are the 97 VLBI sources consistently imaged and cataloged by \citet{Varenius19}; there are 45 in E and 52 in W.
Assuming that they follow a spatial probability function in the shape of an elliptical Gaussian in each nucleus,
we estimated its parameters using the maximum likelihood method.
The two ellipses are the best-fit Gaussians indicating their FWHM sizes and position angles.
Their center positions (ICRS) are 
R.A.(E) = 15\hr34\mn57\fs2904, Dec.(E)= $+23$\degr30\amin11\farcs322, with $\pm13$ mas error in each coordinate,
and 
R.A.(W) = 15\hr34\mn57\fs2230 ($\pm16$ mas), Dec.(W)= $+23$\degr30\amin11\farcs491 ($\pm8$ mas).
The FWHM of E is 
$(0\farcs33\pm0\farcs04)\times(0\farcs16\pm0\farcs02)$ with the major axis position angle of $47\degr\pm5\degr$, 
and that for W is
$(0\farcs27\pm0\farcs03)\times(0\farcs13\pm0\farcs01)$ with the major axis position angle of $83\degr\pm5\degr$.
The center positions agree within 23 mas (E) and 11 mas (W)
with the ALMA $\lambda = 3$ mm continuum positions  in \citet{Sakamoto17}.
}
\end{figure}
%%%%%%%%%%%%%%%%%%%%%%%%%%%%%%%%%%%%%%%%%%%%	% Fig. 15: VLBI sources
	% \label{f.a220VLBIfit}

%%%%%  Appendix
\clearpage 
%% Appendix A
%%%%%%%%%%%%%%%%%%%%%%%%%%%%%%%%%%%%%%%%%%%%%%%%%%%%%%%%%%%%%%%%%%%%%%%%
% Table : Spectral index of calibrators
\begin{deluxetable}{lcLL}
\tablecolumns{4}
\tabletypesize{\scriptsize}
\tablewidth{0pt}
\tablecaption{
Quasar Spectral Indices Used for NGC 4418 Calibration
\label{t.calibratorSpix4n4418} }
\tablehead{ 
       \colhead{name} & 
       \colhead{UT date} &
       \colhead{$\alpha$(3C~273)} &
       \colhead{$\alpha$(3C~279)}           
	\\
	\colhead{}  &	
	\colhead{yyyy-mm-dd} &
	\colhead{} &
	\colhead{}   
}
\colnumbers   
\startdata
B9--1.a & 2014--06--17 &  -0.87  & -0.60  \\ 
B9--1.b & 2015--05--19 & -0.77   & -0.65  \\
B9--2   & 2015--06--02 &  -0.78   & \nd    \\
B7--1.a & 2013--06--01 &  -0.95   & -0.68     \\
B7--1.b & 2015--07--24 &  -0.83   & \nd    \\
B7--2 &  2015--07--24  &  -0.83   & \nd      \\  
B7--3 &  2015--07--24  &  -0.83   & \nd     \\   
B7--4 &  2015--06--29  &  -0.84   & -0.59  \\                                           
B6--1.a &   2014--08--17 &  -0.65 &  \nd    \\   
B6--1.b &   2015--07--18 &  -0.82 &  -0.61  \\   
B6--2 &     2014--08--17 &  -0.53 &  \nd     \\   
B6--3 &     2015--07--18 &  -0.82 &  -0.61   \\   
B6--4 &     2015--07--18 &  -0.82 &  -0.61    \\   
\enddata
\tablecomments{
(1) Tuning name plus a, b, ..., for the first, second, ..., executions. 
(3) and (4) The spectral indices adopted for calibration for 3C~273 and 3C~279, respectively.
}
\end{deluxetable}

%%%%%%%%%%%%%%%%%%%%%%%%%%%%%%%%%%%%%%%%%%%%%%%%%%%%%%%%%%%%%%%%%%%%%%%%     % Table 14. assumed calibrator spix.  for N4418
	% \label{t.calibratorSpix4n4418}, 
%%%%%%%%%%%%%%%%%%%%%%%%%%%%%%%%%%%%%%%%%%%%%%%%%%%%%%%%%%%%%%%%%%%%%%%%
% Table : Spectral index of calibrators
\begin{deluxetable}{lcLLLLL}
\tablecolumns{7}
\tabletypesize{\scriptsize}
\tablewidth{0pt}
\tablecaption{
Quasar Spectral Indices Used for Arp 220 Calibration
\label{t.calibratorSpix4arp220} }
\tablehead{ 
       \colhead{name} & 
       \colhead{UT date} &
       \colhead{$\alpha$(J1550)} &
       \colhead{$\alpha$(J1516)} &
       \colhead{$\alpha$(J1337)} &
       \colhead{$\alpha$(J1751)} &
       \colhead{$\alpha$(3C~279)}            
	\\
	\colhead{}  &	
	\colhead{yyyy-mm-dd} &
	\colhead{} &
	\colhead{} &
	\colhead{} &
	\colhead{} &	
	\colhead{}   
}
\colnumbers
\startdata
B9--1.a,b & 2015--06--09 & -0.65  & \nd   & \nd      &  \nd   & -0.62 \\
B7--1    &  2015--07--17 & -0.64  & -0.44 &  -0.66  &  \nd   & \nd \\
B7--2    &  2015--06--28 & -0.64  & -0.44 &  \nd     & -0.68 & \nd \\
B7--3    &  2015--06--27 & -0.64  & -0.44 &  \nd     & -0.68 & \nd \\
B7--4    &  2015--06--28 & -0.64  & -0.44 & -0.67  &  \nd    & \nd \\
B6--1    &  2015--06--30 & -0.64  & \nd    & \nd      & \nd    & \nd  \\
B6--2.a &  2015--06--30 & -0.64  & \nd    & \nd      & \nd    & \nd  \\
B6--3.a &  2015--06--30 & -0.64   & \nd    & \nd      & \nd   & \nd \\
B6--3.b &  2015--08--04 & -0.63   & \nd    & \nd      & \nd   & \nd  \\
B6--4    &  2015--06--27 & -0.64  & \nd    & \nd      & \nd    & \nd \\
\enddata
\tablecomments{
(1) Tuning name plus a, b, ..., for the first, second, ..., executions. 
(3) to (7) The spectral indices adopted for calibration, for 
J1550+052, J1516+193, J1337$-$125, J1751+093, and 3C~279, respectively.
}

\end{deluxetable}

%%%%%%%%%%%%%%%%%%%%%%%%%%%%%%%%%%%%%%%%%%%%%%%%%%%%%%%%%%%%%%%%%%%%%%%%

     % Table 15. assumed calibrator spix. for Arp220
	% \label{t.calibratorSpix4arp220}
								    
%% Appendix B
% Figure. Band9 flux cal of 3C279 for Arp 220
%%%%%%%%%%%%%%%%%%%%%%%%%%%%%%%%%%%%%%%%%%%%
% Fig.   Band 9 flux cal for Arp 220
\begin{figure}[t]
\centering
\includegraphics{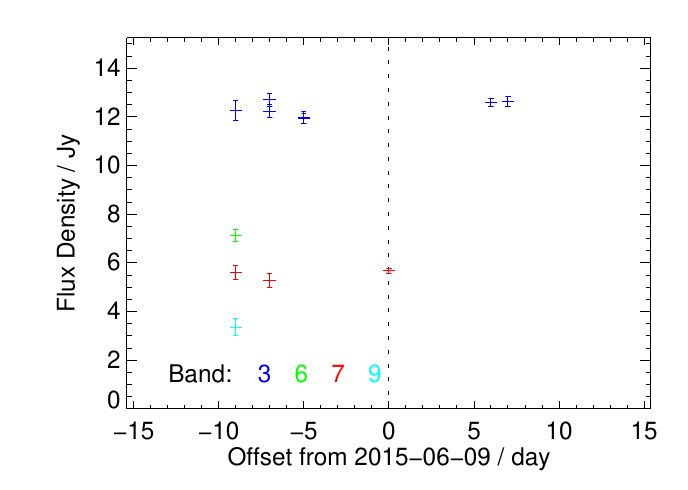} % .eps} 
\includegraphics{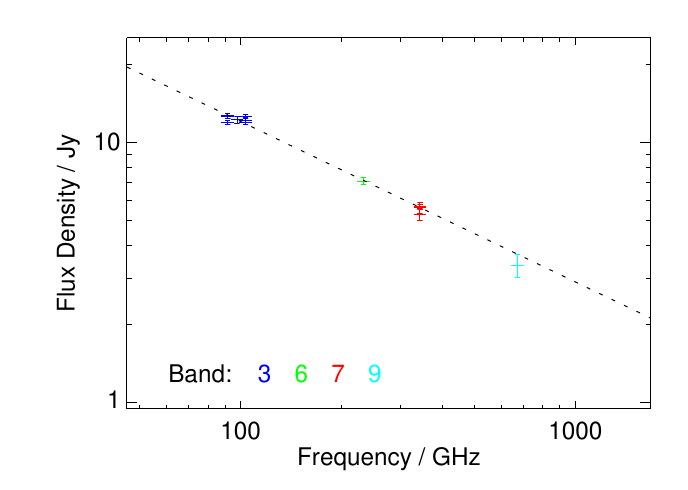} % .eps} 
\caption{ \label{f.B9fluxcal}
Flux measurements of 3C~279 in the ALMA Calibrator Source Catalog referenced for our Band 9 calibration of Arp~220. 
(Left) Measurements around June 9th, 2015, are plotted as a function of time.
(Right) The same data are plotted as a function of frequency along with the best-fit power-law line having a spectral index of 
$-0.619$.
}
\end{figure}
%%%%%%%%%%%%%%%%%%%%%%%%%%%%%%%%%%%%%%%%%%%%			% Fig. 16: 3C279 fluxcal.
	% \label{f.B9fluxcal}	

%% Appendix C
% Large Aperture Spectra
%%%%%%%%%%%%%%%%%%%%%%%%%%%%%%%%%%%%%%%%%%%%
% Fig.   N4418 spectra with continuum - ylog, 1" aperture
\begin{figure}[t]
\begin{center}
\includegraphics{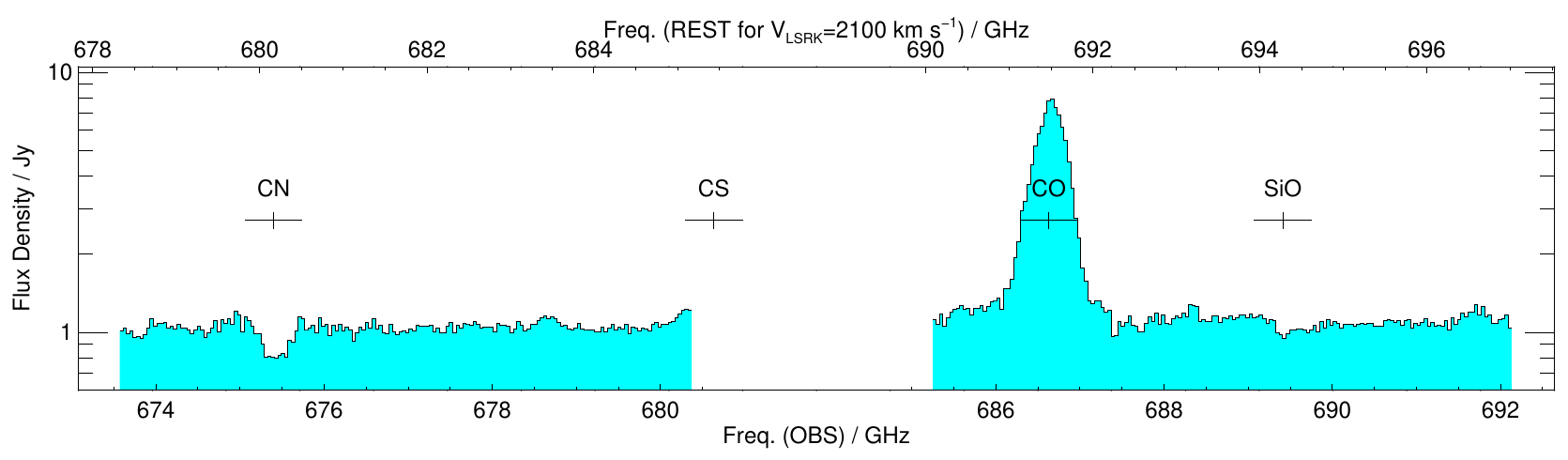} % .eps}  
\includegraphics{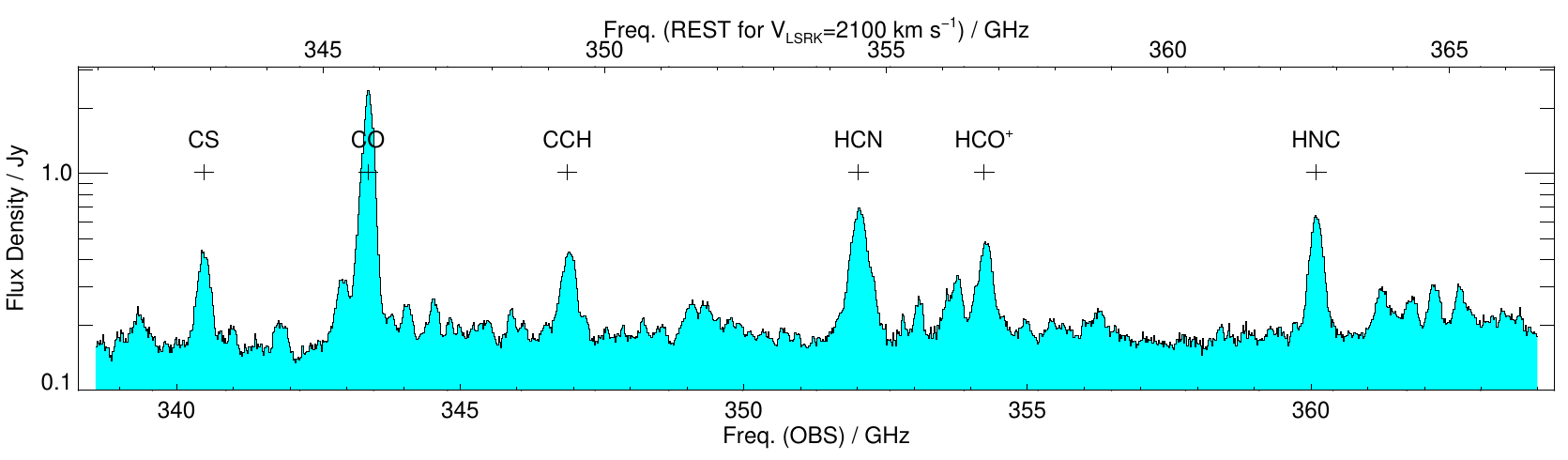} % .eps}  
\includegraphics{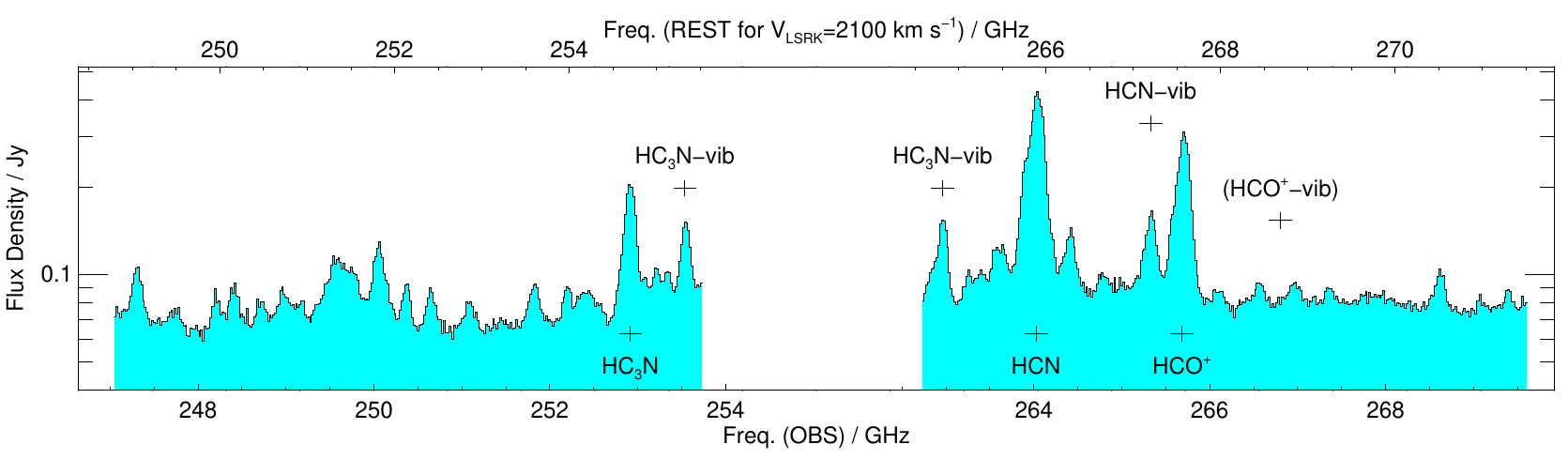} % .eps}  
\includegraphics{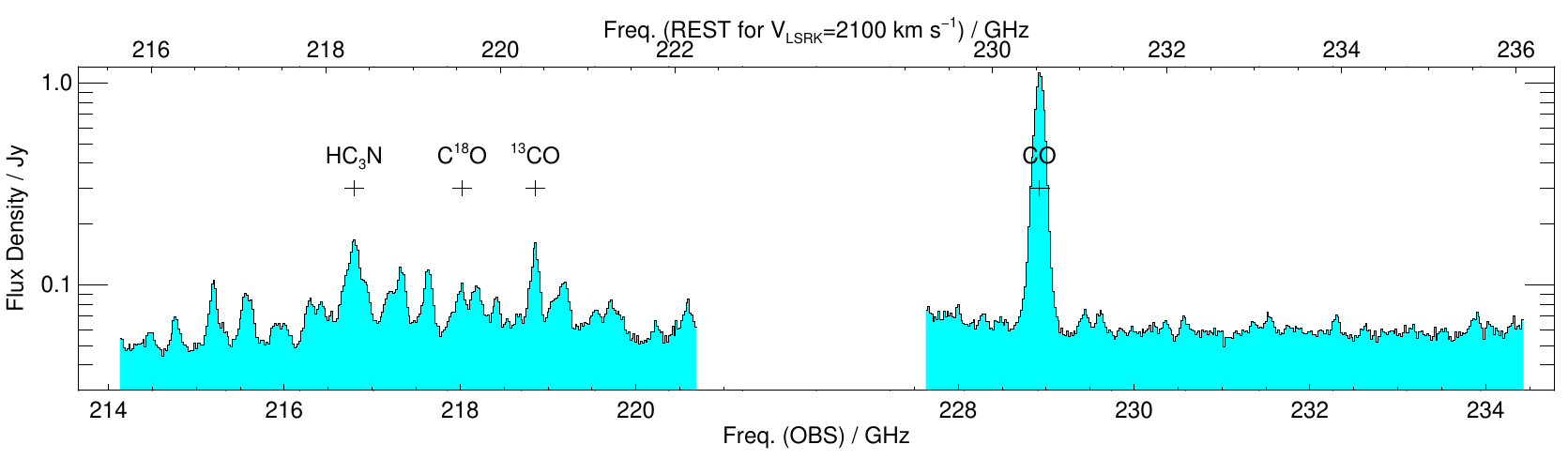} % .eps}   
\end{center}
\caption{ \label{f.spec_with_cont.largeAperture.ylog.N4418}
NGC 4418 spectra in a 1\arcsec-diameter aperture centered at the nucleus. 
The data are sampled from 0\farcs35 resolution data cubes.
The spectral resolution is 40 MHz in Band 9 and 20 MHz in the rest. 
Major lines have labels and 300 \kms-wide scale bars.
}
\end{figure}

%%%%%%%%%%%%%%%%%%%%%%%%%%%%%%%%%%%%%%%%%%%%
 	% Fig. 17 spectra in large aperture. N4418
	% \label{f.spec_with_cont.largeAperture.ylog.N4418}
%%%%%%%%%%%%%%%%%%%%%%%%%%%%%%%%%%%%%%%%%%%%
% Fig.   Spectrum Sections - ylog : Arp 220D ,2" aperture
\begin{figure}[t]
\begin{center}
\includegraphics{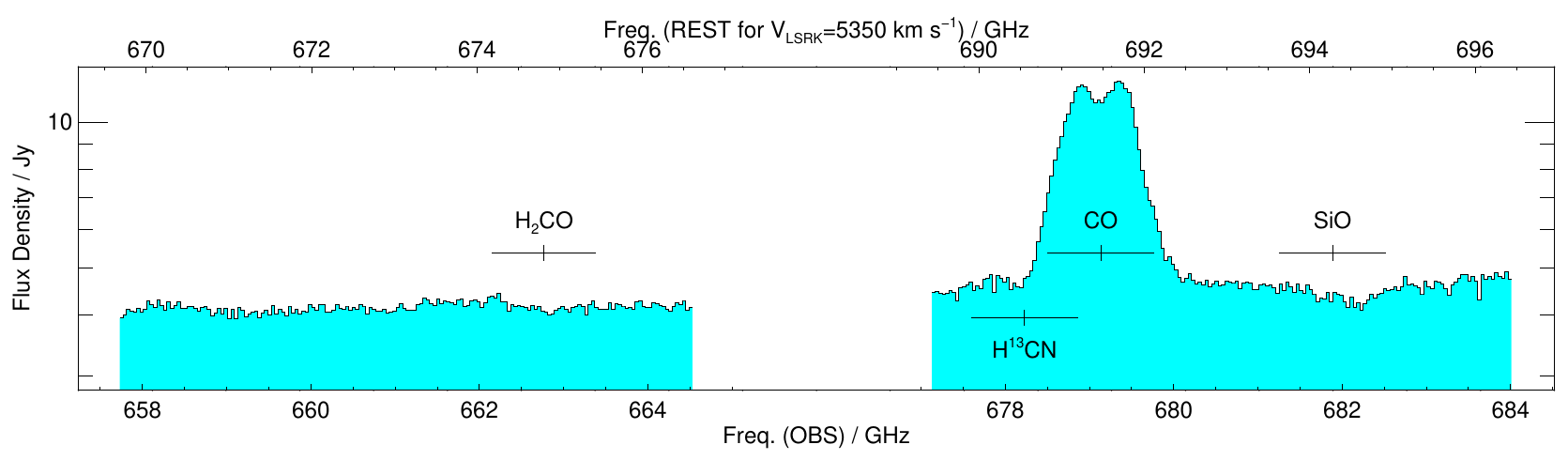} % .eps}  
\includegraphics{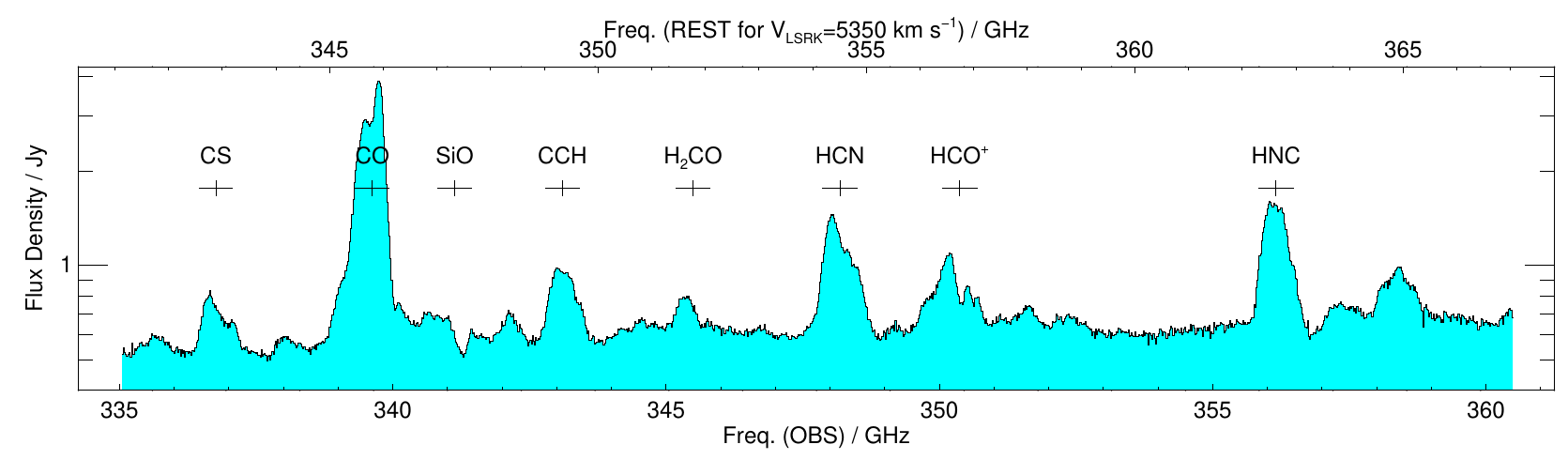} % .eps}  
\includegraphics{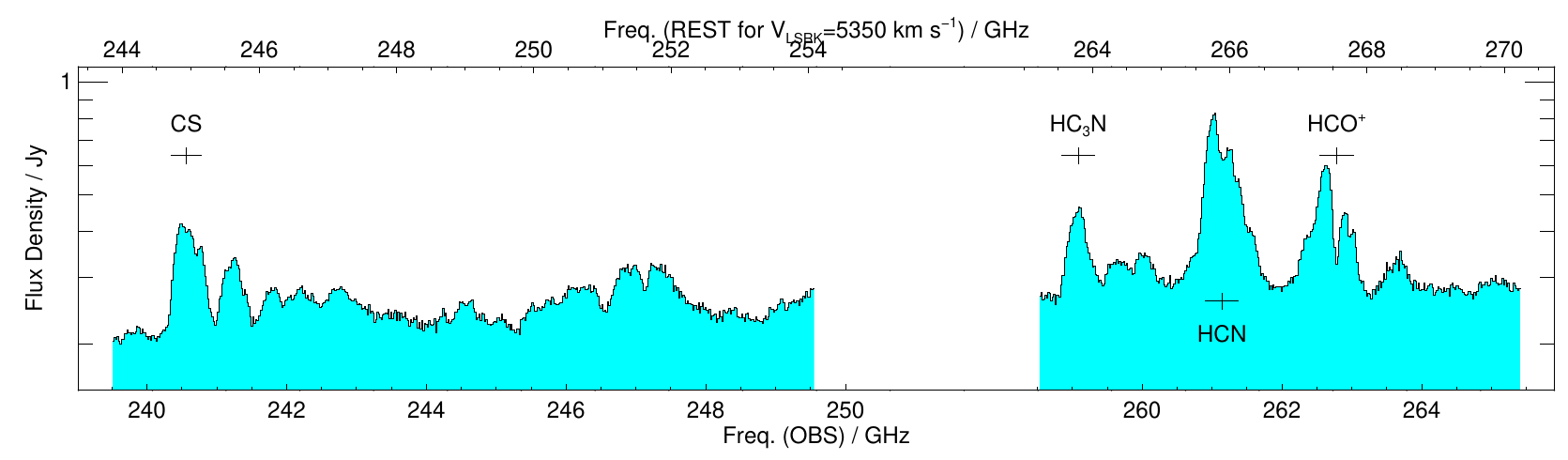} % .eps}   
\includegraphics{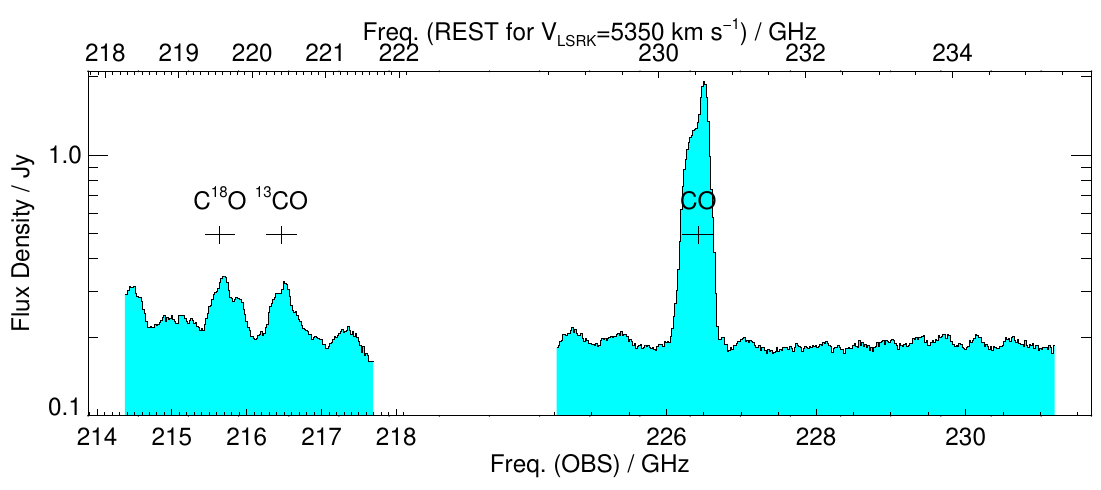} % .eps}   
\end{center}
\caption{ \label{f.spec_with_cont.largeAperture.ylog.A220}
Arp 220 spectra in a 2\arcsec-diameter aperture centered at the midpoint of the two nuclei.
The data are sampled from 0\farcs35 resolution data cubes.
The spectral resolution is 40 MHz in Band 9 and 20 MHz in the rest. 
Major lines have labels and 550 \kms-wide scale bars.
}
\end{figure}
%%%%%%%%%%%%%%%%%%%%%%%%%%%%%%%%%%%%%%%%%%%% 	% Fig. 18 spectra in large aperture. A220	
	% \label{f.spec_with_cont.largeAperture.ylog.A220}
	
%%%%%%%%%%%%%%%%%%%%%%%%%%%%%%%%%%%%%%%%%%%%%%%%%%%%%%%%

\end{document}